\documentclass[preprint,aps,prc,superscriptaddress,showpacs,nofootinbib,secnumarabic,floatfix]{revtex4-1}

\pdfoutput=1

\usepackage{graphicx}
\usepackage{bm}
\usepackage{color}
\usepackage{gensymb}
\usepackage{amsmath}
\usepackage{slashed}
\usepackage{appendix}
\usepackage{hyperref}
\usepackage{ulem}

\newcommand{\ltsim}{\protect\raisebox{-0.5ex}{$\:\stackrel{\textstyle <}
	{\sim}\:$}}

\newcommand{\bvec}[1]{\ensuremath{\boldsymbol{#1}}}

\makeatletter
\let\@makefntextOrig\@makefntext
\def\@makefntext#1{\@makefntextOrig{%
\baselineskip=14pt
\let\@finalstrut\@gobble #1}}
\makeatother
\begin{document}

\title{Effects of chiral symmetry restoration on meson and dilepton production in relativistic heavy-ion collisions}  

\newcommand{\JINR}{Joint Institute for Nuclear Research, 141980 Dubna, Russia}
\newcommand{\JLU}{Institut f\"ur Theoretische Physik, Justus-Liebig-Universit\"at, 35392 Giessen, Germany}
\newcommand{\HFHF}{Helmholtz Research Academy Hesse for FAIR (HFHF), Campus Giessen, 35392 Giessen, Germany}

\author{A.B. Larionov}
\email{Corresponding author:\\[-6pt] larionov@theor.jinr.ru}
\affiliation{\JINR}

\author{L. von Smekal}
\affiliation{\JLU}
\affiliation{\HFHF}

\begin{abstract}
We include effects of chiral symmetry and its restoration in the kinetic equations for baryon propagation and explore the consequences for $\eta$, $\pi^0$, $\rho$ and dilepton production in heavy-ion collisions at 1-2A GeV. Numerical calculations are performed using the GiBUU microscopic transport model supplemented by the parity-doublet model for the mean fields of the nucleon and the $N^*(1535)$ resonance. 
In this chiral model, a strong dropping of the Dirac mass of the $N^*(1535)$ in the high-density stage of a collision leads to a considerable enhancement in the production of this resonance as compared to the standard (non-linear) Walecka model. As the system expands, the Dirac masses of these abundant soft  $N^*(1535)$ resonances gradually increase and ultimately cross the $N \eta$ decay threshold.
As a result, an enhanced low-energy $\eta$ production is observed in the calculations with chiral mean fields.
Comparing with TAPS data on $\eta$ and $\pi^0$ production we find that the chiral model improves the agreement for 
the $m_t$-spectra of $\eta$'s at small $m_t$ in heavy colliding systems.
A similar enhancement is also observed in the soft $\rho$ production due to chiral symmetry and its partial restoration.
The resulting dilepton yields at low and intermediate invariant masses are slightly enhanced due to these chiral effects which further improve the agreement between GiBUU transport simulations and 
HADES data for C+C at 1A GeV.
\end{abstract}

\maketitle

\section{Introduction}
\label{intro}

Lattice QCD calculations predict the appearance of close in mass hadrons of the same spin
but opposite parity, i.e.~parity doubling, if chiral symmetry is restored  \cite{Glozman:2012fj,Aarts:2017rrl}.
In effective theories for the baryonic sector of QCD, parity doubling can be introduced within linear sigma models by using either so-called naive or mirror assignments for the transformations of the chiral components of two opposite-parity fermion species \cite{Jido:1998av,Jido:2001nt}. With the naive assignment, right-handed and left-handed
components of the two fermions transform alike, under chiral rotations. With the mirror assignment, originally introduced
in Ref.~\cite{Detar:1988kn}, on the other hand, the right-handed
component of the second fermion species transforms like the left-handed component of the first one and vice-versa (see Eqs.~(\ref{trans1}), (\ref{trans2}) below).
Both assignments allow for chirally-invariant mixing terms between the two fermion species. After diagonalization, however, the coupling between
the two fermions completely vanishes in the naive assignment. Moreover, both fermions become massless when chiral symmetry is restored.
In contrast, with the mirror assignment, the coupling between the two fermions cannot be removed by diagonalization
(see Eqs.~(\ref{g_piN+N+}), (\ref{g_piN-N-}), (\ref{g_piN+N-}) below), and they become degenerate in mass with chiral symmetry intact and unbroken. In this way, the parity-doublet model (PDM) with mirror assignment allows for a chirally invariant common mass term whose origin in QCD is attributed to the gluonic contribution to the scale anomaly as the main origin of the nucleon mass. Spontaneous chiral symmetry breaking then essentially generates only the mass splitting between the two fermion species of opposite parity, i.e.~between the nucleon and the $N^*(1535)$ resonance as the lowest-lying negative parity partner of the nucleon, in QCD. This PDM with mirror assignment, also referred to as mirror baryon model, is therefore used as a basis for effective hadronic theories to describe the phenomenology of a chiral phase transition inside dense baryonic matter \cite{Hatsuda:1988mv,Zschiesche:2006zj,Sasaki:2010bp,Weyrich:2015hha}.

The search for signals of such a transition from ordinary nuclear matter to
an even higher density phase with nearly restored chiral symmetry, whether this is baryonic, quark or quarkyonic matter, is an important theoretical and experimental problem. What the PDM as an effective hadronic theory can provide, to address this problem, are experimentally testable predictions from the assumed existence of chirally symmetric baryonic matter at high density. One such prediction, in qualitative agreement with the lattice studies \cite{Glozman:2012fj,Aarts:2017rrl}, is that the mass of the lower-lying parity partner such as the nucleon varies comparatively weakly with density while the higher-lying one as the $N^*(1535)$ drops considerably, especially across the chiral transition \cite{Zschiesche:2006zj,Sasaki:2010bp,Weyrich:2015hha,Tripolt:2021jtp}. The qualitative behavior of the parity-partner baryon masses therefore resembles that of the masses of the chiral-partner vector and axial-vector mesons $\rho$ and $a_1$ \cite{Jung:2016yxl,Jung:2019nnr,Tripolt:2021jtp}.
For dilepton production in heavy-ion collisions, their chiral mixing and chiral symmetry restoration were argued to manifest themselves in bumpy structures in $\omega$-$\rho$
and $\phi$ regions \cite{Sasaki:2019jyh}. A strong low-energy resonance excitation peak at around 250~MeV in both, vector and axial-vector spectral functions, due to the dropping $N^*(1535)$ resonance mass across the chiral transition, was predicted as a possible signal in heavy-ion collisions at a few GeV/nucleon from high statistics measurements of 
an increased dilepton yield at correspondingly low invariant masses \cite{Tripolt:2021jtp}.

Another possibility is to look for signals of an enhanced production and abundance of the $N^*(1535)$ resonance from $NN$ collisions. When chiral symmetry gets fully restored, the partial densities of the $N^*(1535)$ resonances and the nucleons become equal to one another.
This remains true inside nuclear matter in thermal equilibrium as a consequence of equal Dirac masses of the parity partners, assuming that the vector mean-fields of $N$ and $N^*(1535)$ are the same as well when chiral symmetry is restored.
In heavy-ion collisions thermal equilibrium is not necessarily
expected to be reached. Nevertheless, an increased $N^*(1535)$ production due to the lower production threshold in $NN \to NN^*$ collisions should be observed, when the PDM is used for the calculations of the baryonic mean fields. 

The $N^*(1535)$ resonance has a large branching ratio ($\sim 30-55\%$  \cite{Zyla:2020zbs}) for the $\eta N$ decay channel which makes this resonance
the most important channel of $\eta$ production in $\gamma$-nucleus reactions near threshold \cite{RoebigLandau:1996xa,Lehr:2003km}.
Since the PDM predicts that the difference between the $N^*(1535)$ and nucleon Dirac masses decreases with increasing baryon density, the $N^* \to \eta N$ decay channel
closes at $\rho_B \sim 0.4\,\rho_0$ according to Ref.~\cite{Jido:2008ng}
where the impact of the PDM dynamics on the coherent $\eta$-mesonic nuclei photoproduction has been discussed. This was also suggested to be the reason for the $A^{2/3}$ dependence of $\eta$-production cross section \cite{Kim:1998upa}. More recent PDM studies, taking into account $N^*$-hole loop contributions to the $\eta$ self-energy, confirmed the strong decrease of the $N^*(1535) \to N \eta$ decay width in a nuclear medium \cite{Suenaga:2017wbb}. 

In heavy-ion collisions, the compression-expansion dynamics of the bulk nuclear medium should lead to the disappearance of the mean field effects on the mass difference between $N^*$'s and nucleons and allow for $N^* \to \eta N$ decays towards the end of the time evolution, even for slow $N^*$'s in the central region of the colliding system. In contrast to the situation in a static nuclear medium, this should then result in an increased production of $\eta$'s due to the chiral mean-field effects in the PDM with chirally symmetric dense baryonic matter.

In Ref.~\cite{Zhang:2018ool} $\pi-N-\Delta$ dynamics has been studied both, in a box with periodic boundary conditions and in heavy-ion collisions below 1A GeV on the basis of the BUU model with the Skyrme energy density functional Sk$\chi m^*$ fitted to the equation of state (EOS) and effective masses from chiral two and three-body interactions. The authors of Ref.~\cite{Zhang:2018ool} have predicted a substantial enhancement of pion production, due to threshold mean field effects, although the $\pi^-/\pi^+$ ratio remains
essentially unchanged.

The aim of our work is to directly study the effects of partial restoration of chiral symmetry in heavy-ion collisions
at beam energies of 1-2A GeV. We extend the Giessen Boltzmann-Uehling-Uhlenbeck (GiBUU) transport model \cite{Buss:2011mx} by the baryonic mean fields calculated on the
basis of the PDM. The in-medium production  thresholds in GiBUU are modified to take into account a stronger in-medium mass drop of $N^*(1535)$ as compared to
the nucleon. In our numerical analysis, we compare the calculations within the chiral PDM and the non-linear and not chiral Walecka model, and demonstrate that the former
leads to a significant enhancement of low-transverse-mass $\eta$ production at midrapidity. The comparison with TAPS data on $\eta$ production shows
that the PDM improves the low-$m_t$ behavior for heavier colliding systems, Ar+Ca and Au+Au at 0.8A GeV, but leads to some overestimation at low $m_t$'s
for C+C at 0.8 and 1.0A GeV. We have also calculated dilepton production in C+C at 1A GeV where we observe increased contributions from the $\rho$-meson direct
decays at low invariant masses, and from $\eta$-Dalitz decays at intermediate invariant masses within the PDM, which improves the agreement with HADES data.

Our analysis is based on the parameterizations of the (extended) PDM from Refs.~\cite{Zschiesche:2006zj,Shin:2018axs}. In the recent Ref.~\cite{Kim:2020sjy},
the extended PDM of Ref.~\cite{Shin:2018axs} has been included in the new DJBUU transport code that has been then applied, in-particular, to the study of
the proton directed flow $v_1$ and nucleon rapidity distributions in Au+Au at 400A MeV showing a good agreement with available FOPI data on the rapidity
dependence of $v_1$, independent of the choice of the incompressibility ($K=215$ and 240 MeV) and the chirally invariant PDM mass parameter ($m_0=600-900$ MeV).
Inelastic channels were not discussed in Ref.~\cite{Shin:2018axs}, however. Note that the sophisticated collision term and the large number of degrees
of freedom in the GiBUU code \cite{Buss:2011mx} allow for studies of particle production in heavy-ion collisions in a wide range of beam energies, ranging from $\sim$ hundreds
A MeV up to $\sim$ tenths A GeV. 

The description of massive vector fields based on the Proca formalism used, in-particular, in the Walecka model is still questionable, see Ref.~\cite{Jung:2019nnr}. On the other hand, vector repulsion is certainly needed for  a realistic description of the nuclear EOS. To this end, we will apply in this work contact four-fermion interaction terms of the Nambu-Jona-Lasionio type to describe vector repulsion. 
With a proper choice of mass parameters and coupling constants for the corresponding vector Hubbard vector fields, the Walecka model description is restored for infinite nuclear matter, however.

The structure of our work is as follows: In Sec.~\ref{model}, the PDM is described starting from the basic Lagrangians.  The in-medium dispersion relations for the
nucleon and its parity partner, and the equations of motion (EOMs) for the classical $\sigma$, $\omega$ and $\rho$ fields are rederived.
The dispersion relations are then used in the kinetic equations for the particle propagation in classical meson fields including
elementary elastic and inelastic collisions as well as resonance decays. The in-medium thresholds in the collision term are explained in Sec.~\ref{CollTerm}.
In Sec.~\ref{INM} we present the calculations of the equation-of-state of nuclear matter at zero temperature and of the density dependence of the Dirac masses of the nucleon
and the $N^*(1535)$ that show a chiral phase transition at high densities. In Sec.~\ref{AuAu1AGeV}, the time evolution
for central Au+Au collision at 1A GeV is studied. It is shown that the PDM leads to a dramatic enhancement of $N^*(1535)$ production at intermediate stages of the collision, but
only to a moderate enhancement of $\eta$ and $\rho$ production. In Sec.~\ref{compExp} we present a systematic comparison with TAPS data on $\eta$ and $\pi^0$ production at 0.8-2.0 A GeV and
also provide predictions at lower beam energies, 0.6 A GeV, i.e.~far below the quasifree $\eta$ production threshold in $pp$ collisions ($E_{\rm beam}=1.255$ GeV).
Dilepton production is discussed in the end of Sec.~\ref{compExp} for the selected case of C+C at 1A GeV. We conclude and discuss some possible next steps for future extensions in Sec.~\ref{summary}.

\section{The model}
\label{model}

We apply the parity doublet model (PDM) with mirror assignment \cite{Jido:2001nt}
\begin{eqnarray}
     N_{1R} &\to& R N_{1R}, N_{1L} \to L N_{1L},  \label{trans1}\\
     N_{2R} &\to& L N_{2R}, N_{2L} \to R N_{2L}.  \label{trans2}
\end{eqnarray}
Here, $N_1$ and $N_2$ are the fields of nucleon (1) and its negative parity partner (2),
while  ``$R$'' and ``$L$'' denote the right- and left-handed components:
$N_{iR} = (1+\gamma_5)N_i/2$, $N_{iL} = (1-\gamma_5)N_i/2$, $i=1,2$. The isospin $SU(2)$ transformations
$R$ and $L$ act independently on the right- and left-handed components of the nucleon fields, Eq.(\ref{trans1})
and thus their combination belongs to the direct product $SU(2)_R \otimes SU(2)_L$ called chiral group.
The mirror assignment in Eq.~(\ref{trans2}) entails that the right-handed component of a negative parity partner transforms like a left-handed nucleon and vice versa. 
The PDM Lagrangian is written as follows:
\begin{eqnarray}
  \cal{L} &=&  \bar N_1 [i\xout{\partial} + g_1 (\sigma + i\gamma_5 \bvec{\tau}\bvec{\pi})] N_1  \nonumber \\
          &&  + \bar N_2 [i\xout{\partial} + g_2 (\sigma - i\gamma_5 \bvec{\tau}\bvec{\pi})] N_2  \nonumber \\
          &&  - m_0 (\bar N_1 \gamma_5 N_2 - \bar N_2 \gamma_5 N_1) + \cal{L}_{\rm mes} + \cal{L}_{\rm 0} + \cal{L}_{\rm 1}~,      \label{Lagr}
\end{eqnarray}
where  $\bvec{\tau}$ are the isospin Pauli matrices, and the combinations
of $\sigma$ and $\bvec{\pi}$ coupling terms to the baryons in Eq.~(\ref{Lagr}) are invariant under chiral rotations as in the original Gell-Mann-L\'evy model.
The Lagrangian (\ref{Lagr}) includes non-diagonal coupling terms between $N_1$ and $N_2$ baryon fields which
are chirally invariant:
\begin{equation}
  \bar N_1 \gamma_5 N_2 = \bar N_{1R} \gamma_5 N_{2L} + \bar N_{1L} \gamma_5 N_{2R}
  \to 
  \bar N_{1R} R^\dagger \gamma_5 R N_{2L} + \bar N_{1L} L^\dagger \gamma_5 L N_{2R} = \bar N_1 \gamma_5 N_2~, \label{chirTrans}
\end{equation}
(and similar for the $\bar N_2 \gamma_5 N_1$ term) where the mirror assignment (\ref{trans1}), (\ref{trans2}) is used.

The (pseudo-)scalar meson Lagrangian has the following form:
\begin{eqnarray}
  \cal{L}_{\rm mes} &=&  \frac{1}{2}\partial_\mu\sigma\partial^\mu\sigma + \frac{1}{2}\partial_\mu\vec{\pi}\partial^\mu\vec{\pi} \\
                  &&   + \frac{\bar{\mu}^2}{2}(\sigma^2+\bvec{\pi}^2) - \frac{\lambda}{4} (\sigma^2+\bvec{\pi}^2)^2
                       + \frac{\lambda_6}{6} (\sigma^2+\bvec{\pi}^2)^3 + \varepsilon \sigma~.  \label{Lagr_mes} \nonumber
\end{eqnarray}
The combination $\sigma^2+\bvec{\pi}^2$ is chirally invariant, and
the $\varepsilon \sigma$ term is
included for the small explicit braking of the two-flavor chiral symmetry.

Moreover, in Eq.~(\ref{Lagr})
we have included  isoscalar and isovector four-fermion Nambu-Jona-Lasinio-type interaction terms,
\begin{eqnarray}
   \cal{L}_{\rm 0} &=& -G_{\rm 0}(\bar N_1 \gamma^\mu N_1 + \bar N_2 \gamma^\mu N_2)^2~, \label{L_0} \\
   \cal{L}_{\rm 1} &=& -G_{\rm 1}[ (\bar N_1 \gamma^\mu \bvec{\tau} N_1 + \bar N_2 \gamma^\mu \bvec{\tau} N_2)^2
                      +(\bar N_1 \gamma^\mu \gamma_5 \bvec{\tau} N_1 - \bar N_2 \gamma^\mu \gamma_5 \bvec{\tau} N_2)^2 ]~. \label{L_1}
\end{eqnarray}
The form in Eq.~(\ref{L_0}) describes a local current-current interaction in the total baryon number channel, repulsive for $G_0>0$ and independent of the parity partner.  The form in Eq.~(\ref{L_1}) is determined by chiral symmetry together with parity.
It is unique, if we require it to be independent of the parity partner also.  To see this, first consider individual left- and right-handed $SU(2)$ currents $\bvec{j}_{iL}$ and $\bvec{j}_{iR}$ for both parity partners. Due to the mirror assignment, the two chirally invariant current-current interactions are then of the form $(\bvec{j}_{1L}+\bvec{j}_{2R})^2$ and $(\bvec{j}_{1R}+\bvec{j}_{2L})^2$. Their coupling strengths must be the same because of parity which exchanges the two. It is now simply a matter of defining total vector and axial-vector currents as the sum and the difference of the two currents in these bilinears, i.e.~$\bvec{j}=\bvec{j}_{1L}+\bvec{j}_{2R} + \bvec{j}_{1R}+\bvec{j}_{2L} \equiv \bvec{j}_1 +\bvec{j}_2 $ and  $\bvec{j}_A = -(\bvec{j}_{1L}+\bvec{j}_{2R}) + \bvec{j}_{1R} + \bvec{j}_{2L}  \equiv \bvec{j}_{A1} - \bvec{j}_{A2}$, where $\bvec{j}_i^\mu = \bar N_i \gamma^\mu \bvec{\tau} N_i $ and $\bvec{j}_{Ai}^\mu = \bar N_i \gamma^\mu \gamma_5 \bvec{\tau} N_i $ are the usual vector and axial-vector currents of parity partner $i=1,2$. This then leads to the form in (\ref{L_1}), and explains the relative minus sign in the axial-vector current-current interaction. 

Applying Hubbard-Stratonovich transformations to the four-fermion short-distance interaction terms (\ref{L_0}) and (\ref{L_1}) these are equivalently represented as follows:
\begin{eqnarray}
  \cal{L}_{\rm 0} &=& \frac{m_\omega^2}{2} \omega^\mu \omega_\mu - g_\omega \omega_\mu (\bar N_1 \gamma^\mu N_1 + \bar N_2 \gamma^\mu N_2)~,  \label{L_0_bos} \\
  \cal{L}_{\rm 1} &=& \frac{m_\rho^2}{2} (\bvec{\rho}^\mu \bvec{\rho}_\mu + \bvec{a}_1^\mu \bvec{a}_{1\mu}) \nonumber\\
                 &&    -g_\rho \bar N_1 (\bvec{\rho}^\mu - \gamma_5 \bvec{a}_1^\mu) \gamma_\mu \bvec{\tau} N_1
                       -g_\rho \bar N_2 (\bvec{\rho}^\mu + \gamma_5 \bvec{a}_1^\mu) \gamma_\mu \bvec{\tau} N_2~, \label{L_1_bos}
\end{eqnarray}
where only the rations $m_\omega^2/g_\omega^2 = 1/(2G_0) $ and $m_\rho^2/g_\rho^2 = 1/(2G_1) $ represent the independent model parameters, as determined by the short-range interaction strengths $G_0$ and $G_1$.

We emphasize that the Hubbard fields $\omega$, $\bvec{\rho}$ and $\bvec{a}_1$
are auxiliary fields to linearize the short-range current-current interactions whose EOMs are the constraint equations,
\begin{align}
  \omega^\mu &= \frac{\sqrt{2 G_{\rm 0}}}{m_\omega} \, \big( \bar N_1 \gamma^\mu N_1 + \bar N_2 \gamma^\mu N_2 \big)~,    \label{omega^mu}\\
  \bvec{\rho}^\mu &= \frac{\sqrt{2 G_{\rm 1}}}{m_\rho} \, \big( \bar N_1 \gamma^\mu \bvec{\tau} N_1 + \bar N_2 \gamma^\mu \bvec{\tau} N_2 \big)~, \label{rho^mu}\\  
  \bvec{a}_1^\mu &= \frac{\sqrt{2 G_{\rm 1}}}{m_\rho} \, \big( \bar N_1 \gamma^\mu \gamma_5 \bvec{\tau} N_1 - \bar N_2 \gamma^\mu \gamma_5 \bvec{\tau} N_2 \big)~. \label{a_1^mu}
\end{align}
In particular, these Hubbard fields do not themselves represent dynamical massive vector fields. 
Therefore, the parameters $m_\omega$ and $m_\rho$ in (\ref{L_0_bos}) and (\ref{L_1_bos}) do not necessarily have to represent the physical meson masses of $\omega$ and $\bvec\rho $ either. While this is only a matter of interpretation, it is important to remember that these short-range interactions are not due to boson exchanges, but really represent the contact interactions $G_0$ and $G_1$ which parameterize short distance QCD interactions beyond any effective mesonic description. Because these interactions determine only the ratios $m_\omega/g_\omega $ and $m_\rho/g_\rho $, however, we may nevertheless insert the physical $\omega$ and $\bvec \rho$ masses here  without loss, as usually done in the literature, and adjust the dimensionless couplings $g_\omega$ and $g_\rho$ accordingly.

The physical positive ($N_+$) and negative ($N_-$) parity baryon fields that have definite masses are obtained
by performing the $SO(4)$ transformation:  
\begin{equation}
   \left(\begin{array}{c}
          N_+\\
          N_-
   \end{array}\right)
 = \left(\begin{array}{cc}
          \cos\Theta         &  \gamma_5\sin\Theta \\
         -\gamma_5\sin\Theta &  \cos\Theta
         \end{array}\right)
   \left(\begin{array}{c}
          N_1\\
          N_2
   \end{array}\right)~, \label{PhysFields}
\end{equation}
where the mixing angle $\Theta$ is obtained from the condition of the diagonalization of the mass matrix \cite{Jido:2001nt}
which gives
\begin{equation}
  \tan 2\Theta = -\frac{2m_0}{\sigma(g_1+g_2)}~.    \label{Theta}
\end{equation}
The corresponding values of the masses of the positive and negative parity baryons are 
\begin{equation}
   m_\pm = \frac{1}{2}\left[\sqrt{\sigma^2(g_1+g_2)^2+4m_0^2} \pm \sigma (g_2-g_1)\right]~.   \label{m_pm}
\end{equation}
The Lagrangian (\ref{Lagr}) can be rewritten in terms of the physical baryon fields as follows:
\begin{eqnarray}
  \cal{L} &=&  \bar N_+ [i\xout{\partial} - m_+ - i g_{\pi N_+ N_+} \gamma_5 \bvec{\tau}\bvec{\pi}
                         - (g_\omega\omega^\mu + g_\rho \bvec{\tau} \bvec{\rho}^\mu  - g_{a_1} \gamma_5 \bvec{\tau} \bvec{a}_1^\mu ) \gamma_\mu ] N_+  \nonumber \\
           &&  + \bar N_- [i\xout{\partial} - m_- - i g_{\pi N_- N_-} \gamma_5 \bvec{\tau}\bvec{\pi}
    - (g_\omega\omega^\mu + g_\rho  \bvec{\tau} \bvec{\rho}^\mu + g_{a_1} \gamma_5 \bvec{\tau} \bvec{a}_1^\mu ) \gamma_\mu ] N_-  \nonumber \\
           &&  + \frac{m_\omega^2}{2} \omega^\mu \omega_\mu +  \frac{m_\rho^2}{2} (\bvec{\rho}^\mu \bvec{\rho}_\mu + \bvec{a}_1^\mu \bvec{a}_{1\mu}) \nonumber \\
           &&  - i g_{\pi N_+ N_-} \bar N_+ \bvec{\tau}\bvec{\pi} N_- + i g_{\pi N_+ N_-} \bar N_- \bvec{\tau}\bvec{\pi} N_+   \nonumber \\
           &&  + g_{a_1 N_+ N_-} \bar N_+ \gamma_\mu \bvec{\tau} \bvec{a}_1^\mu N_-   +  g_{a_1 N_+ N_-} \bar N_- \gamma_\mu \bvec{\tau} \bvec{a}_1^\mu N_+ 
               + \cal{L}_{\rm mes}~,      \label{LagrPhys}
\end{eqnarray}
where the coupling constants are
\begin{eqnarray}
  g_{\pi N_+ N_+} &=& -g_1\cos^2\Theta - g_2 \sin^2\Theta~,   \label{g_piN+N+}\\
  g_{\pi N_- N_-} &=& g_2\cos^2\Theta + g_1 \sin^2\Theta~,  \label{g_piN-N-}\\
  g_{\pi N_+ N_-} &=& \frac{g_1-g_2}{2} \sin 2\Theta~,      \label{g_piN+N-}\\
  g_{a_1} &=& g_\rho \cos 2\Theta~,        \label{g_a1}\\
  g_{a_1 N_+ N_-} &=& g_\rho \sin 2\Theta~.  \label{g_a1N+N-}
\end{eqnarray}

The $\sigma$ mean field represents the expectation value of the scalar condensate $\langle \bar q q\rangle$.
In vacuum, the Goldberger-Treiman relation yields $\sigma = f_\pi$ with $f_\pi=93$ MeV being the pion decay constant.
Below we will disregard the pion mean field $\langle \bvec{\pi} \rangle$ as it has negative parity and thus disappears
in the nuclear matter ground state. We will also disregard the isovector axial-vector Hubbard field $\bvec{a}_1^\mu$ as its expectation value also vanishes in spin-saturated nuclear matter. 

As usual, we apply Lagrange's EOMs
\begin{equation}
   \partial_\mu\left(\frac{\partial\cal{L}}{\partial\partial_\mu q}\right) - \frac{\partial\cal{L}}{\partial q} = 0  \label{LagrEOM}
\end{equation}
for the fields $q \equiv \sigma, \omega_\nu, \bvec{\rho}_\nu, \bar N_\pm$ which give:  
\begin{eqnarray}
  && \partial_\mu\partial^\mu\sigma(x)
  - \bar{\mu}^2\sigma + \lambda\sigma^3 - \lambda_6\sigma^5 - \varepsilon
  = -\sum_{i=\pm} \frac{\partial m_i}{\partial \sigma} \langle \bar N_i(x) N_i(x) \rangle~, \label{EOM_sigma} \\
  &&   \omega^\nu(x)
      = \frac{g_\omega}{m_\omega^2} \sum_{i=\pm} \langle \bar N_i(x) \gamma^\nu N_i(x) \rangle~, \label{EOM_omega} \\
  &&   \bvec{\rho}^\nu(x)
      = \frac{g_\rho}{m_\rho^2} \sum_{i=\pm} \langle \bar N_i(x) \gamma^\nu \bvec{\tau} N_i(x) \rangle~,  \label{EOM_rho} \\
  &&  [\gamma^\mu(i\partial_\mu - V_\mu) - m_\pm] N_\pm(x) = 0~, \label{EOM_N_pm}
\end{eqnarray}
where $x\equiv(t,\bvec{r})$ is Minkowski spacetime, and $\langle \ldots \rangle$ denotes averaging over the actual state of the many-body system.
$V$ stands for the vector Hubbard field matrix (in isospin space), 
\begin{equation}
  V_\mu = g_\omega \omega_\mu + g_\rho\bvec{\tau}\bvec{\rho}_\mu~.   \label{V}
\end{equation}

The EOMs for the baryons (\ref{EOM_N_pm}) have the form of Dirac equations in external vector fields, with the vacuum masses
replaced by the Dirac masses $m_\pm$. If the mesonic mean fields vary slowly in space and time, and the baryons are fast enough to adjust to these variations, 
then this equation can be solved with a plane-wave ansatz $N_\pm \propto \exp(-ipx)$:
\begin{equation}
  [\gamma^\mu p^*_\mu -  m_\pm] N_\pm = 0~,         \label{EOM_N_pm_PWA}  
\end{equation}
where $p^*_\mu \equiv p_\mu - (V_{\mu})_{I_z I_z}$ is the kinetic four-momentum of the baryon with $I_z=\pm1/2$.
This gives the dispersion relation (in-medium mass-shell condition):
\begin{equation}
  (p^*)^2 - m_\pm^2 = 0~.                   \label{massShellCond}
\end{equation}

Now we will require self-consistency, i.e. we calculate the scalar densities $\langle \bar N_\pm N_\pm \rangle$
and baryon currents $\langle \bar N_\pm \gamma^\nu N_\pm \rangle$, $\langle \bar N_\pm \gamma^\nu \bvec{\tau} N_\pm \rangle$,
in Eqs.~(\ref{EOM_sigma}), (\ref{EOM_omega}), (\ref{EOM_rho}),  assuming
that the baryons occupy certain states in momentum space while their Dirac spinors satisfy Eq.~(\ref{EOM_N_pm_PWA}).
This gives the following expressions:
\begin{eqnarray}
   \langle \bar N_\pm(x) N_\pm(x) \rangle &=& g_s\! \sum_{I_z=\pm1/2} \int \frac{d^3p}{(2\pi)^3}\, \frac{m_\pm}{p^{*0}} \, f_{\pm, I_z}(x,\bvec{p})~,   \label{rho_s}\\
   \langle \bar N_\pm(x) \gamma^\nu N_\pm(x) \rangle &=& g_s\! \sum_{I_z=\pm1/2} \int \frac{ d^3p}{(2\pi)^3} \,  \frac{p^{*\nu}}{p^{*0}} \, f_{\pm, I_z}(x,\bvec{p})~, \label{j_B^nu}\\
   \langle \bar N_\pm(x) \gamma^\nu \tau^3 N_\pm(x) \rangle &=& g_s\! \sum_{I_z=\pm1/2} \int \frac{d^3p}{(2\pi)^3} \, \frac{p^{*\nu}}{p^{*0}}\, \tau^3_{I_z I_z} \, f_{\pm, I_z}(x,\bvec{p})~, \label{j_I^nu}
\end{eqnarray}
where $f_{\pm, I_z}(x,\bvec{p})$ are the phase-space distribution functions (occupation numbers)
that depend on the isospin projection $I_z$, and $g_s=2$ is the spin degeneracy.
We furthermore assume that the state of the system is characterized by certain numbers of protons or their parity partners ($I_z=+1/2$)
and neutrons or their parity partners ($I_z=-1/2$), without any isospin-mixed states, so that only the third isospin component of the baryon current in Eq.~(\ref{j_I^nu}) is non-vanishing, and hence only the third isospin component of the $\bvec{\rho}$ field in Eq.~(\ref{EOM_rho}), likewise.

The distribution functions are normalized such that $g_s f_{\pm, I_z}(x,\bvec{p})  \frac{d^3rd^3p}{(2\pi)^3}=$(number of particles in the phase space element $d^3rd^3p$).
For simplicity, we did not include the contributions from
antibaryons in Eqs.~(\ref{rho_s})-(\ref{j_I^nu}) (although this can be readily done following Ref.~\cite{Larionov:2008wy}). Below, where it does not cause confusion, we will drop the baryon type ($\pm$) and isospin ($I_z$) indices for brevity.

Applying Liouville's theorem, we can now write the kinetic equation for the baryons:
\begin{equation}
  \left( \frac{\partial}{\partial t} + \frac{\partial p^0}{\partial \bvec{p}} \frac{\partial}{\partial \bvec{r}}
       - \frac{\partial p^0}{\partial \bvec{r}} \frac{\partial}{\partial \bvec{p}} \right) f(x,\bvec{p}) = I_{\rm coll}[\{f\}]~,   \label{kinEq}     
\end{equation}
where
\begin{equation}
    p^{0} = \sqrt{m_\pm^2+(\bvec{p}^*)^2} + V^0    \label{p^0}
\end{equation}
is the single-particle energy. The collision term $I_{\rm coll}[\{f\}]$ in the right-hand-side (r.h.s.) of Eq.~(\ref{kinEq}) is a functional of the phase-space
distribution functions of the various particle species. The Vlasov equation is obtained in the limit $I_{\rm coll} \to 0$.

If the $\sigma$ and $\omega$ fields are momentum-independent (the more general case is discussed, e.g.~in Ref.~\cite{Blaettel:1993uz}) one can simplify their calculation
by introducing the distribution functions $f^*(x,\bvec{p}^*)$ in kinetic phase space. They are defined such that $g_s f^*(x,\bvec{p}^*) \frac{d^3rd^3p^*}{(2\pi)^3}=$
(number of particles in the kinetic phase space element $d^3rd^3p^*$). With $d^3p^*=d^3p$, we then have $f^*(x,\bvec{p}^*)=f(x,\bvec{p})$. After the variable transformation
$\bvec{p} \to \bvec{p}^*$ in Eq.~(\ref{kinEq}), one obtains the following equation:
\begin{equation}
  (p_0^*)^{-1}
 \left [ p^*_\mu \partial^\mu + (p_\mu^* {\cal F}^{\alpha\mu} 
                                   + m_\pm \partial^\alpha   m_\pm)
    \frac{\partial}{\partial p^{*\alpha}} \right] f^*(x,\bvec{p}^*) = I_{\rm coll}[\{f^*\}]~,        \label{kinEqStar}
\end{equation}
where $\alpha=1,2,3$ and $\mu=0,1,2,3$. ${\cal F}^{\mu\nu}=\partial^\mu V^\nu-\partial^\nu V^\mu$ is the field-strengths tensor obtained from the vector Hubbard field.

Eqs.~(\ref{kinEq}), (\ref{kinEqStar}) combined with the field EOMs (\ref{EOM_sigma})-(\ref{EOM_rho})
obey local energy-momentum conservation, i.e.~the energy-momentum tensor $T^{\mu\nu}$ satisfies the continuity equation,
\begin{equation}
   \partial_\nu T^{\mu\nu} = 0~.      \label{ContEq4T}
\end{equation}
Eq.~(\ref{ContEq4T}) can be directly proven if one takes into account that collisions conserve the four-momentum density which is expressed as
\begin{equation}
  \sum \int d^3p\,  p^\mu \, I_{\rm coll}[\{f\}] = 0~,     \label{4momDensCons}
\end{equation}
where the sum is taken over all particle species. The explicit form of the energy-momentum tensor is given as follows:
\begin{eqnarray}
  T^{\mu \nu} &=&    g_s\sum_{i=\pm} \sum_{I_z=\pm1/2}  \int\frac{d^3p}{(2\pi)^3} \,\frac{p^\mu p^{*\,\nu}}{p^{*\,0}}\, f_{i, I_z}(x,\bvec{p})
   + \partial^\mu\sigma \partial^\nu\sigma \nonumber \\
  && - g^{\mu\nu} \left[ \frac{1}{2}\partial_\lambda\sigma\partial^\lambda\sigma 
                       + \frac{\bar{\mu}^2}{2}\sigma^2 - \frac{\lambda}{4}\sigma^4
                       + \frac{\lambda_6}{6}\sigma^6 + \varepsilon \sigma 
                       + \frac{1}{2} m_\omega^2 \omega_\kappa \omega^\kappa
                       + \frac{1}{2} m_\rho^2 \rho^3_\kappa\rho^{3,\kappa} \right]~.   \label{T^numu}
\end{eqnarray}
In the actual calculations we neglect the space-time derivatives in the EOM (\ref{EOM_sigma}) for the $\sigma$ field which corresponds to a static treatment of the meson Lagrangian.
 The energy-momentum tensor is then given by Eq.~(\ref{T^numu})
with space-time derivatives of the $\sigma$ field removed.
In particular, in this approximation the three-momentum densities 
$T^{i 0}$, $i =1,2,3$ are then mean-field independent.

Note that the form of the Lagrangian (\ref{LagrPhys}) corresponds to the Lagrangians
used in previous PDM based studies as long as the vector mean fields are assumed to be given by constant values, i.e.~independent of space-time variables. In order to obtain a reasonable description  of the EOS of infinite nuclear matter, we can therefore directly employ the model parameters of these previous studies here as well.
The two sets of model parameters that we have used in our present calculations are adopted from Ref.~\cite{Zschiesche:2006zj}
and Ref.~\cite{Shin:2018axs}. They are referred to as Set P3 and Set 2, respectively, in the following, and listed explicitly in Table~\ref{tab:par}.
\begin{table}[htb]
  \caption{\label{tab:par}
    The sets of parameters of the PDM.}
  \begin{center}
    \begin{tabular}{lllll}
    \hline
    \hline
    \hspace{3cm}          &  Set P3 \cite{Zschiesche:2006zj}  &  Set 2 \cite{Shin:2018axs} \\
    \hline
    $m_0$ (MeV)            &  790    &      700 \\
    $m_\sigma$ (MeV)        &  370.63  &      384.428 \\
    $m_\omega$ (MeV)        &  783    &      783 \\
    $m_\rho$ (MeV)         &   ---      &     776 \\       
    $g_\omega$             &  6.79     &      7.05508 \\
    $g_\rho$              &   0        &     4.07986 \\ 
    $g_1$                 &  13.00    &      14.1708 \\ 
    $g_2$                 &   6.97    &      7.76222 \\
 $\lambda_6 f_\pi^2$       &   0       &      15.7393 \\      
    $m_+$ (MeV)            &  939    &      939   \\
    $m_-$ (MeV)            &  1500    &      1535   \\
    $K$ (MeV)              &  510.57  &      215     \\    
    \hline
    \hline
    \end{tabular}
  \end{center}
\end{table}
It is assumed in both sets that the negative parity partner of the nucleon is the $N^*(1535)$ resonance, although with somewhat different
values of its pole mass. The parameters $\bar{\mu}$ and $\lambda$ of the meson Lagrangian (\ref{Lagr_mes})
are related to the $\pi$ and $\sigma$-meson masses as follows:
\begin{eqnarray}
  \bar{\mu}^2   &=& \frac{m_\sigma^2-3m_\pi^2}{2} + \lambda_6f_\pi^4~,    \label{mu} \\
  \lambda &=& \frac{m_\sigma^2-m_\pi^2}{2f_\pi^2} + 2\lambda_6f_\pi^2~. \label{lambda}
\end{eqnarray}
These relations can be obtained by decomposing the non-linear self-interaction terms of the meson Lagrangian
into  powers of $\bvec{\pi}$ and $\Delta\sigma = \sigma - f_\pi$ and identifying the corresponding mass terms, $=-m_\pi^2\bvec{\pi}^2/2 -m_\sigma^2\Delta\sigma^2/2$, for pions and the $\sigma$-meson. Substituting Eqs.~(\ref{mu}), (\ref{lambda}) in the
EOM of the $\sigma$-field (\ref{EOM_sigma}) then determines 
the parameter in the symmetry breaking term for the uniform vacuum solution,
\begin{equation}
  \varepsilon = m_\pi^2 f_\pi~.          \label{varepsilon}
\end{equation}

\subsection{Collision term}
\label{CollTerm}

The collision term $I_{\rm coll}[\{f^*\}]$ describes two- and three-body collisions and resonance decays.
Its detailed description can be found in Ref.~\cite{Buss:2011mx} and, with a focus on the dilepton production channels,
in the recent Ref.~\cite{Larionov:2020fnu}.
Since the GiBUU transport model includes a long list of baryon resonances, it must be specified which mean field potential is acting on them.
For simplicity, we assume that the scalar and vector fields acting on all baryons except the $N^*(1535)$ are identical to the nucleon
scalar and vector fields, respectively.
It was assumed in previous GiBUU calculations with the relativistic mean field
model that the scalar potentials acting on the incoming and the outgoing baryons in two-body scatterings
$B_1 B_2 \to B_3 B_4$ and resonance production and decays $B M \leftrightarrow B^\prime$ are always the same.
This allowed to express the so-called ``free'' invariant collision energy $\sqrt{s}_{\rm free}$ -- that governs the corresponding
cross sections and decay widths -- in terms of incoming particles. This substantially simplifies calculations, since 
the final channel is unknown apriori and is sampled by Monte-Carlo. Such an assumption is, however,
no longer true if the PDM is applied for the calculations of the nucleon and $N^*(1535)$ potentials. In the inelastic production
channel $N_1 N_2 \to N_3 N_4^*$ the scalar potentials of the nucleons are $S_1=S_2=S_3=m_+ - m_N$ while the scalar potential
of the outgoing resonance is $S_4=m_- - m_{N^*}$ where $m_N$ and $m_{N^*}$ are the vacuum masses of the nucleon and $N^*(1535)$,
respectively. Therefore, we define the free invariant collision energy as
\begin{equation}
   \sqrt{s}_{\rm free} = \sqrt{s^*} - S_3 - S_4~,    \label{sqrts_free}
\end{equation}
where $s^* = (p_1^*+p_2^*)^2$ is the in-medium center-of-mass (c.m.) collision energy squared. Note that we still assume that the vector potentials
of all baryons are equal which leads to the kinetic four-momentum conservation: $p_1^*+p_2^*=p_3^*+p_4^*$. Eq.~(\ref{sqrts_free}) correctly
matches the vacuum and in-medium thresholds, i.e.~the condition $\sqrt{s}_{\rm free} \geq m_N + m_{N^*}$ (vacuum) is equivalent to the
condition $\sqrt{s^*} \geq m_+ + m_-$ (in-medium). Note that, in the special case where the sums of the vector self-energies of incoming and outgoing particles are the same, our in-medium threshold condition coincides with those from Refs.~\cite{Zhang:2018ool,Ferrini:2005jw}. 

  Some more details are in order for the dilepton spectra.
  It is commonly accepted that the broadening of the $\rho$-meson plays a key role in
  the description of the dilepton spectra from heavy-ion collisions
  \cite{NA60:2006ymb,vanHees:2006ng,Bratkovskaya:2007jk,Endres:2015fna,Staudenmaier:2017vtq,Larionov:2020fnu,Schmidt:2021hhs}.
  In the present calculations, we take into account
  the collisional broadening of the $\rho$-meson spectral function and apply the
  off-shell potential ansatz in the propagation of $\rho$-mesons according to Ref.~\cite{Larionov:2020fnu}.
  The $pn \to pn  e^+ e^-$ and  $pp \to pp  e^+ e^-$ bremsstrahlung cross sections based on the boson exchange model
  of Ref.~\cite{Shyam:2010vr} are included. The correction (enhancement) factor of the $pn$ bremsstrahlung cross section
  according to Eq.(63) of Ref.~\cite{Larionov:2020fnu} that is tuned to describe the dilepton invariant mass spectrum from reaction $dp \to e^+ e^- p_{fast} X$
  at 1.25A GeV measured by HADES \cite{Agakishiev:2009yf} is taken into account.

\subsection{Infinite nuclear matter}
\label{INM}

This section deals with infinite nuclear matter at zero temperature. To avoid misunderstanding we note that the equations and numerical results of this section serve for the qualitative
purposes only and do not influence our transport simulations directly.

Both sets of parameters listed in Table~\ref{tab:par} are adjusted to reproduce the saturation properties of nuclear matter,
i.e.
\begin{eqnarray}
 & &  \frac{\partial {\cal E}(\rho_B)/\rho_B}{\partial \rho_B}_{|\rho_B=\rho_0} = 0~,      \label{satur1}\\ 
 & &  \frac{{\cal E}(\rho_0)}{\rho_0} \simeq -16~\mbox{MeV}~,      \label{satur2}
\end{eqnarray}
where
\begin{equation}
  {\cal E}(\rho_B) \equiv T^{00}(\rho_B) - T^{00}(0)  - m_N\rho_B~,   \label{EnDensNR}
\end{equation}
is the non-relativistic energy density,
and $\rho_0=0.16$ fm$^{-3}$ is the nuclear matter density at saturation.
However, they predict very different values of the incompressibility of infinite nuclear matter, cf.~Tab.~\ref{tab:par}, which is defined as
\begin{equation}
  K=9\rho_0^2 \frac{\partial^2 {\cal E}(\rho_B)/\rho_B}{\partial \rho_B^2}_{|\rho_B=\rho_0}~.   \label{K}
\end{equation}
The value of $K$ can be determined most accurately from the isoscalar giant monopole resonance (ISGMR) centroid energies in heavy
nuclei measured by inelastic $\alpha$ scattering. The reviews of theoretical methods and experimental results on ISGMR
are given in Refs.~\cite{Blaizot:1980tw,Garg:2018uam} and on giant resonances in general -- in Ref.~\cite{HW01} .
By analysing ISGMR in doubly-magic nuclei, like $^{208}$Pb, the authors of Ref.~\cite{Garg:2018uam} concluded $K=240\pm20$ MeV
where the uncertainty comes from the concrete form of the energy-density functional.
At the same time, open-shell nuclei are typically associated with lower values of $K$.
Recent Skyrme-Hartree-Fock RPA calculations of ISGMR for large sets of nuclei \cite{Bonasera:2018onp,Bonasera:2020twz} 
concluded $K=210-240$ MeV although larger values up to $\sim260$ MeV (from $^{68}$Ni) seem also to be possible. 
Heavy-ion flow data analyses allow for a wider range, $K=200-380$ MeV,  \cite{Danielewicz:2002pu}.
The neutron star observables seem to be consistent with $K=200-300$ MeV \cite{Ghosh:2021bvw}.
  Set P3 is thus certainly at the extreme upper end,\footnote{We include calculations with Set P3 rather to demonstrate insensitivity to the value of $K$ for our purposes.} while Set 2 yields an EOS which is one of those with comparatively small $K$ values in agreement with
ISGMR frequencies \cite{Gaitanos:2010fd}.
In addition, Set 2 includes the coupling to the $\rho$-meson which leads to a
nuclear symmetry energy, $E_{\rm sym}(\rho_0)=31$~MeV \cite{Shin:2018axs}, in agreement with other phenomenological models (c.f. Ref.~\cite{Zhang:2021xdt} and refs. therein).

The total energy density in uniform and isospin-symmetric nuclear matter at zero temperature is expressed as follows:
\begin{equation}
   T^{00}(\rho_B) = \frac{2}{\pi^2} \sum_{i=\pm} p_{Fi}^4  g\left(\frac{m_i}{p_{Fi}}\right) - \frac{\bar{\mu}^2}{2}\sigma^2
                   + \frac{\lambda}{4}\sigma^4 - \frac{\lambda_6}{6}\sigma^6 - \varepsilon\sigma
                   + g_\omega\omega^0\rho_B - \frac{m_\omega^2}{2}(\omega^0)^2~,    \label{T^00_NM}
\end{equation}
where $p_{Fi}$ are the Fermi momenta of the nucleons ($i=+$) and their negative parity partners ($i=-$),
\begin{equation}
  \rho_B = \frac{2}{3\pi^2} \sum_{i=\pm} p_{Fi}^3    \label{rho_B}
\end{equation}
is the baryon density, and
\begin{equation}
  g(a) \equiv \int\limits_0^1 dx x^2 \sqrt{x^2+a^2} 
  = \frac{1}{8}\left[ (1+a^2)^{3/2} + \sqrt{1+a^2}
                      -\frac{1}{2}(1+a^2)^2\log\left(\frac{\sqrt{1+a^2}+1}{\sqrt{1+a^2}-1}\right) \right]~.
         \label{g_vs_a}
\end{equation}

The scalar field $\sigma$ and the Fermi momenta $p_{Fi}$ are calculated by solving
the static and uniform version of Eq.~(\ref{EOM_sigma}), i.e.
\begin{equation}
  \sum_{i=\pm} \frac{\partial m_i}{\partial \sigma} \, \rho_{si}
  - \bar{\mu}^2\sigma + \lambda\sigma^3 - \lambda_6\sigma^5 - \varepsilon  = 0~, \label{EOM_sigma_NM}
\end{equation}
where the partial scalar densities $\rho_{si} \equiv \langle \bar N_i(x) N_i(x) \rangle$, see Eq.~(\ref{rho_s}), are expressed as
\begin{equation}
  \rho_{si} = \frac{2p_{Fi}^3}{3\pi^2} \,f\left(\frac{m_i}{p_{Fi}}\right)~,   \label{rho_s_NM}
\end{equation}
with
\begin{equation}
  f(a) = 3a\int\limits_0^1 \frac{dx x^2}{\sqrt{x^2+a^2}}
  = \frac{3}{2} a \left[ \sqrt{1+a^2}
    - \frac{a^2}{2}  \log\left(\frac{\sqrt{1+a^2}+1}{\sqrt{1+a^2}-1}\right) \right]~.
     \label{f_vs_a}
\end {equation} 
Requiring local chemical equilibrium, the chemical potentials of the nucleons and the negative parity baryons should both be equal to the baryon chemical potential $\mu_B$, i.e.
\begin{equation}
    \mu_+ = \mu_- = \mu_B~,~~~\mu_i = \sqrt{p_{Fi}^2 + m_i^2} + g_\omega \omega^0~.      \label{chemEquil}
\end{equation}
Since the vector field does not depend on the sort of baryon, this allows to express the Fermi momenta of the baryons as functions of the  $\sigma$ field,
\begin{equation}
    p_{Fi}(\sigma) = \sqrt{\max(0,\mu_B^{*\,2}-m_i^2(\sigma))}~,       \label{p_Fi}
\end{equation}
where $\mu_B^* =\mu_B - g_\omega \omega^0$ is the effective chemical potential (c.f. Ref.~\cite{Zschiesche:2006zj}),
and the dependence of the masses on the $\sigma$ field is given by Eq.~(\ref{m_pm}). 

Eq.~(\ref{EOM_sigma_NM}) was solved numerically with respect to the $\sigma$ field for different values of $\mu_B^*$ treated as a free parameter. After this, the  zero-component of the vector field
($\bvec{\omega}=0$ in the rest frame of nuclear matter due to isotropy) simply follows from the spacetime independent version of Eq.~(\ref{EOM_omega}):
\begin{equation} 
   \omega^0 = \frac{g_\omega}{m_\omega^2} \rho_B~.       \label{EOM_omega_NM}
\end{equation}

\begin{figure}
  \includegraphics[scale = 0.40]{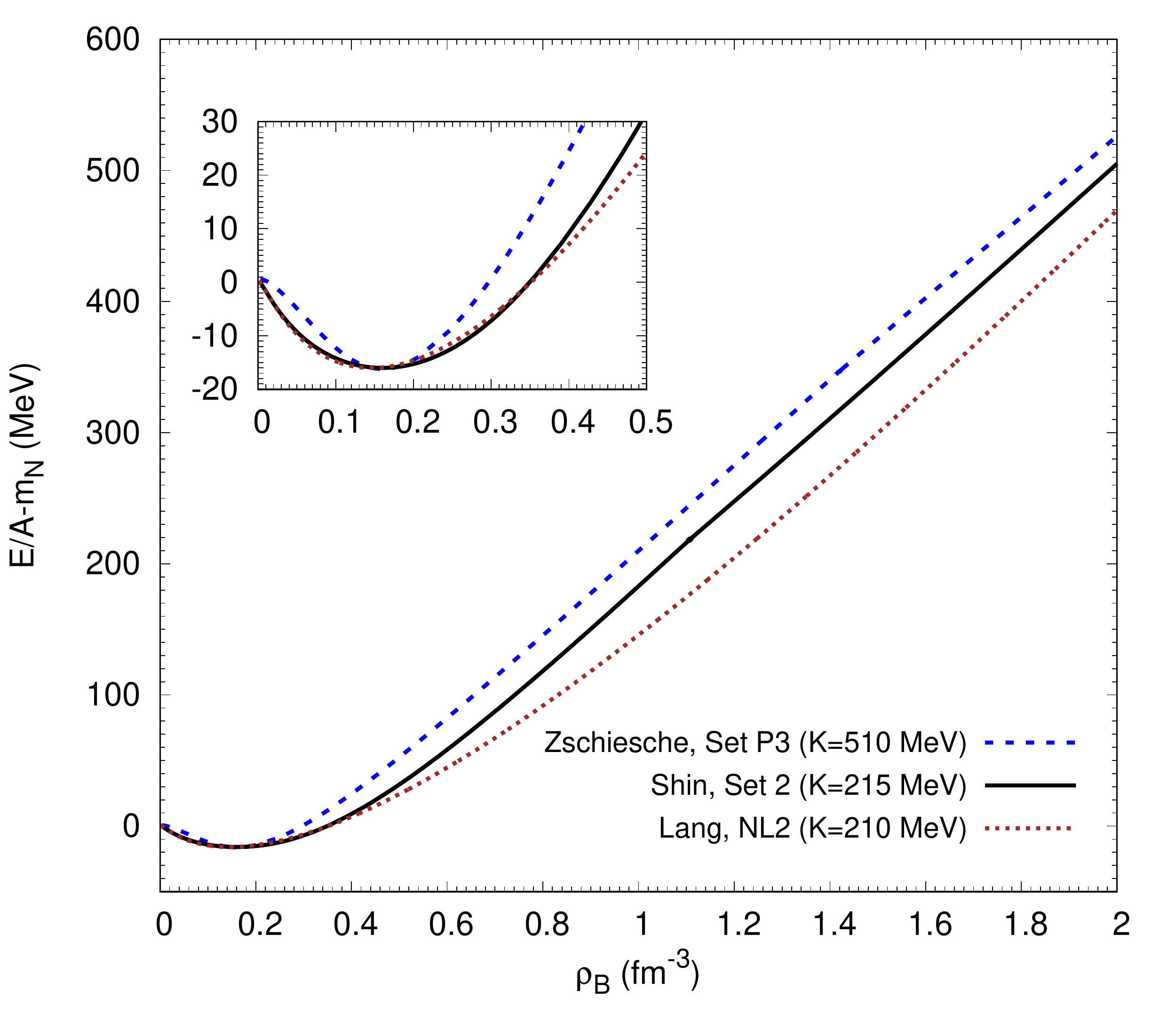}
  \caption{Energy per baryon as a function of the baryon density calculated for Set P3 (dashed line), Set 2 (solid line), and NL2 (dotted line). The inset shows the region near the normal
    nuclear matter density.}
    \label{fig:eos} 
\end{figure}

Fig.~\ref{fig:eos} shows the equation of state for the two parameterizations of the PDM from Table~\ref{tab:par} and, for comparison, for the non-linear Walecka model parameterization NL2 of Ref.~\cite{Lang:1992jz}.
Set 2 and NL2 produce  very similar EOS's for baryon densities up to  $\rho_B < 3 \rho_0$.
In contrast, as a consequence of the large bulk modulus, Set P3 predicts a very different EOS around $\rho_0$, in particular at subnormal densities.

\begin{figure}
  \includegraphics[scale = 0.40]{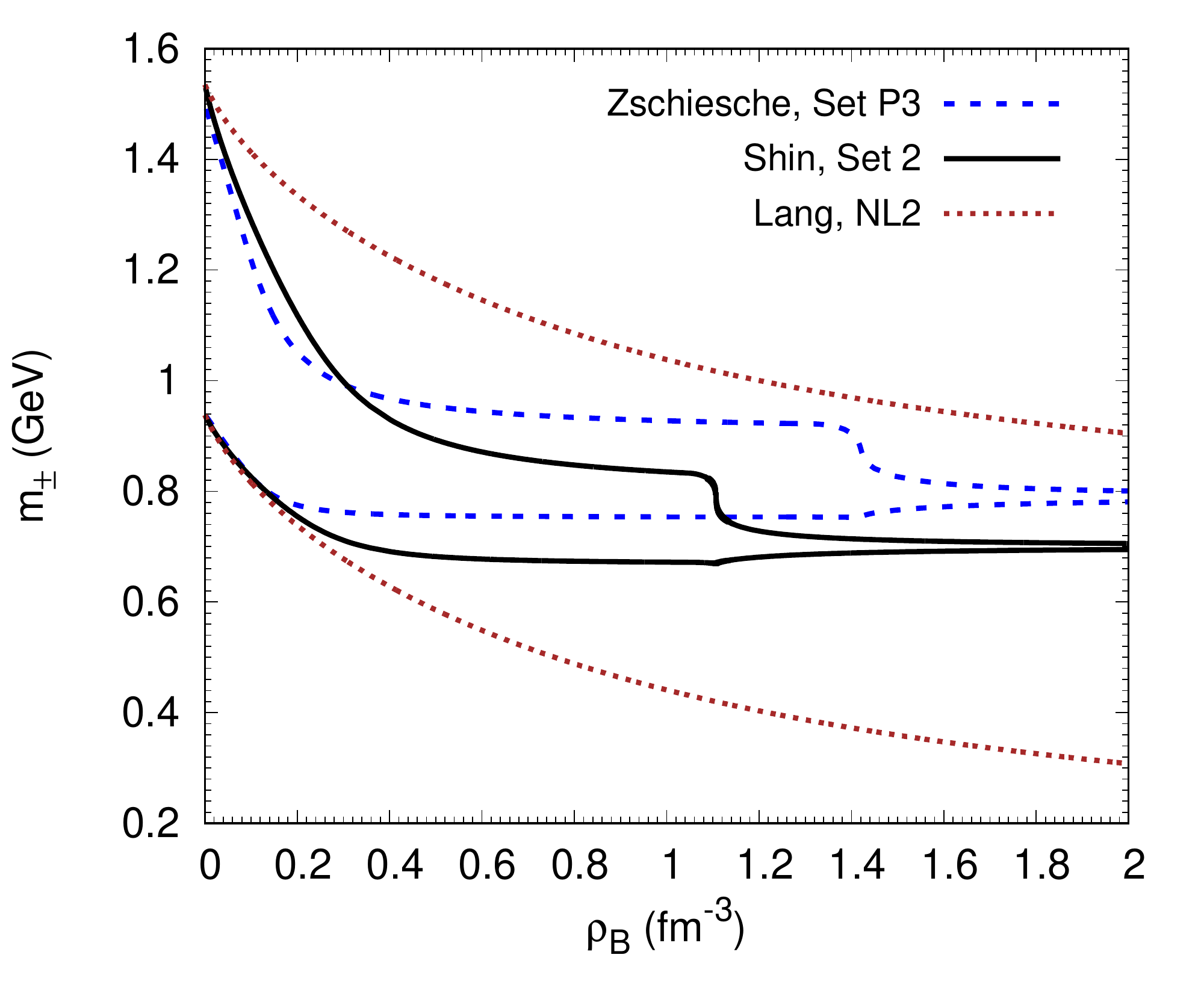}
  \includegraphics[scale = 0.40]{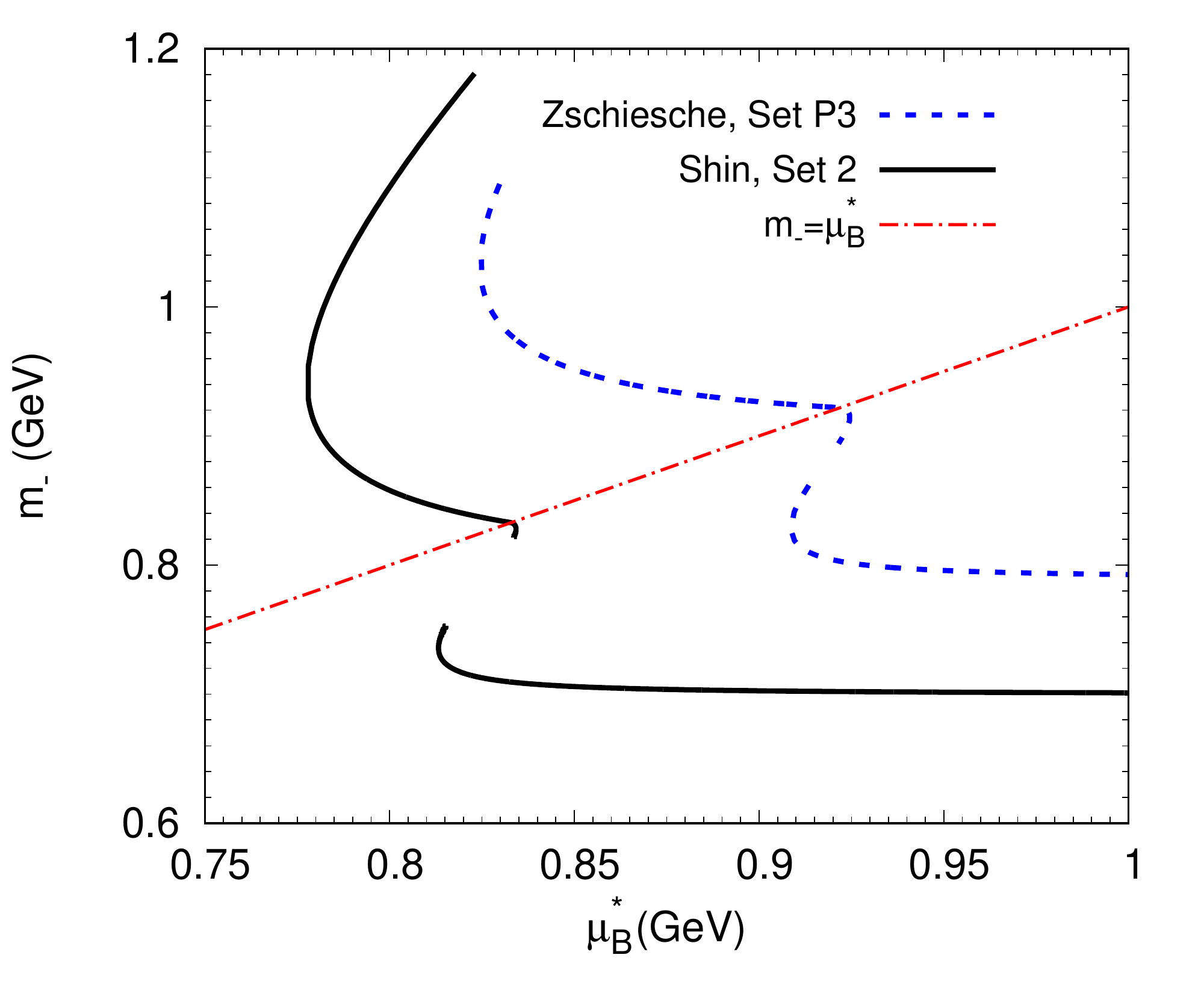}
  \caption{Left panel: the masses of the positive (lower line) and negative (upper line) parity baryons  vs.~baryon density. Right panel: the mass of the negative parity 
  baryon vs.~effective chemical potential. 
  The low-density (nuclear matter) and high-density (chiral) liquid-gas phase transition regions are excluded. Their boundaries are obtained
  via Gibbs conditions of equal pressures and chemical potentials in the corresponding liquid and gas phases.
  Line notations for different EOS's are the same as in Fig.~\ref{fig:eos}.
  The dash-dotted (red) line in the right panel indicates the boundary at $m_-=\mu_B^*$ where
  a Fermi sea of negative parity baryons starts to build up.}
  \label{fig:m_pm} 
\end{figure}

In the left panel of Fig.~\ref{fig:m_pm} we compare the masses of the nucleon and its negative parity partner as functions of the baryon density for the two PDM parameter sets and the same Walecka model EOS.
At small and moderate baryon density the mass of the negative parity baryon is
larger than the effective chemical potential, i.e.\ $m_- > \mu_B^*$, and only the Fermi sea of nucleons gets filled. At $\rho_B \simeq 1-1.4$ fm$^{-3}$, depending on the parameter set, the mass of the negative parity baryons eventually falls below $\mu_B^*$, see the right panel in Fig.~\ref{fig:m_pm}, and their partial density and Fermi sea also start to build up.
Because of the finite binding energy per baryon, their effective mass $m_-$ drops, and one eventually reaches a discontinuous transition.
 The value of the $\sigma$ field decreases discontinuously as well, the masses of nucleons and their negative parity partners  suddenly get very close to one another as a consequence, and their partial baryon densities follow the same pattern. With further increasing $\rho_B$,
beyond this transition, the small residual $\sigma$ field due to explicit chiral  symmetry breaking gradually disappears completely, the parity partner baryons become fully degenerate in mass and their partial baryon densities quickly approach each other as well. This effect is known as chiral symmetry restoration within the PDM. 

In stark contrast, the non-linear Walecka model, which has no chiral symmetry in the first place, predicts that the masses of nucleons and negative parity baryons
both monotonically drop with $\rho_B$ in a way such that their mass splitting remains roughly constant. This is a consequence of a universal dependence $m_\pm(\rho_B) = m_\pm(0) + g_\sigma \Delta\sigma(\rho_B)$
with the coupling constant $g_\sigma$ typically taken to be the same for nucleons and $N^*$ resonances. 

Despite the fact that the baryon density at which the chiral transition is predicted in the PDM mean-field studies is too high to be reached in heavy-ion collisions at SIS18 energies, even with Set 2 where $\rho_B^\mathrm{crit} \sim 6 \rho_0 $,\footnote{Fluctuations beyond mean-field can bring the critical density $\rho_B^\mathrm{crit}$ of the chiral phase transition in the PDM considerably closer to the saturation density $\rho_0$ of normal nuclear matter \cite{Weyrich:2015hha,Tripolt:2021jtp} and therefore have the potential to change this conclusion.} the dependence of $m_\pm$ on the baryon density is drastically different in the PDM and the non-linear Walecka model. For baryon densities up to two to three times $\rho_0$ the most significant difference thereby is the effective mass of the $N^*(1535)$ which drops much faster with increasing baryon density in the PDM than it does in the Walecka model. As explained in the introduction, this might well have observable effects on heavy-ion collision observables at 1-2A GeV.

It is well known that the combination of Fermi motion in colliding nuclei with Lorentz boost results in subthreshold production of hadrons. This basic mechanism is always included in transport calculations. On the top of Fermi motion effects, the mean fields may additionally enhance or suppress subthreshold production depending on the imbalance between potentials of incoming and outgoing particles (cf. Eq.(\ref{sqrts_free})).

To assess these mean field effects, we will study $\eta$, $\rho$, and dilepton production 
in the following sections.
The main mechanism of $\eta$ production is a two-step process $NN \to N N^*(1535),~N^*(1535) \to \eta N$. 
The beam energy threshold of $\eta$ production in $pp$ collisions is 1.255 GeV. Far above threshold,
the available phase space for the intermediate $N^*(1535)$ becomes large and one is less sensitive to mean field
effects. We will therefore focus on beam energies in the 1A GeV region.

\section{Time evolution of Au+Au central collision at 1A GeV}
\label{AuAu1AGeV}

\begin{figure}[ht]
  \includegraphics[scale = 0.48]{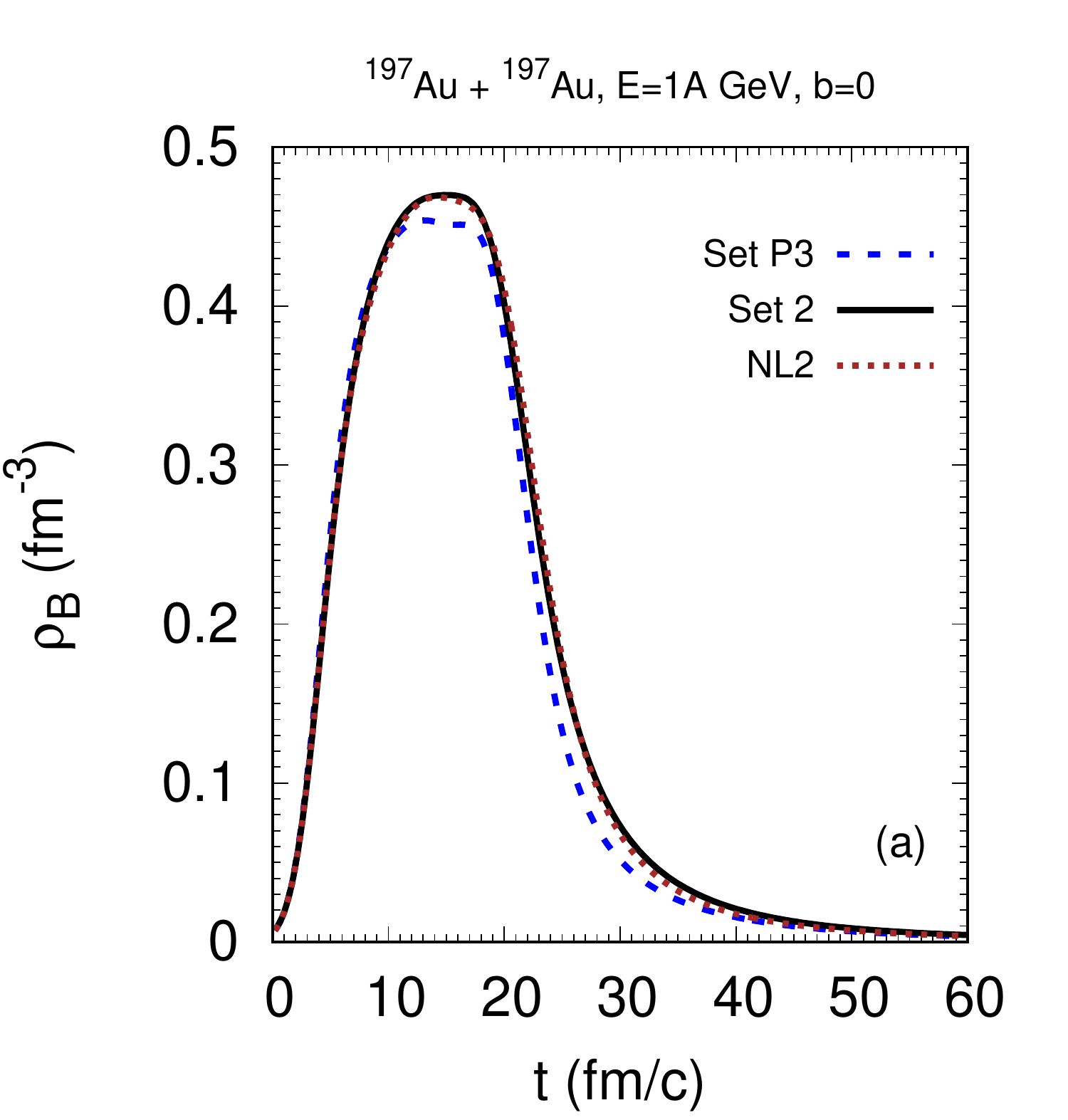}
  \includegraphics[scale = 0.48]{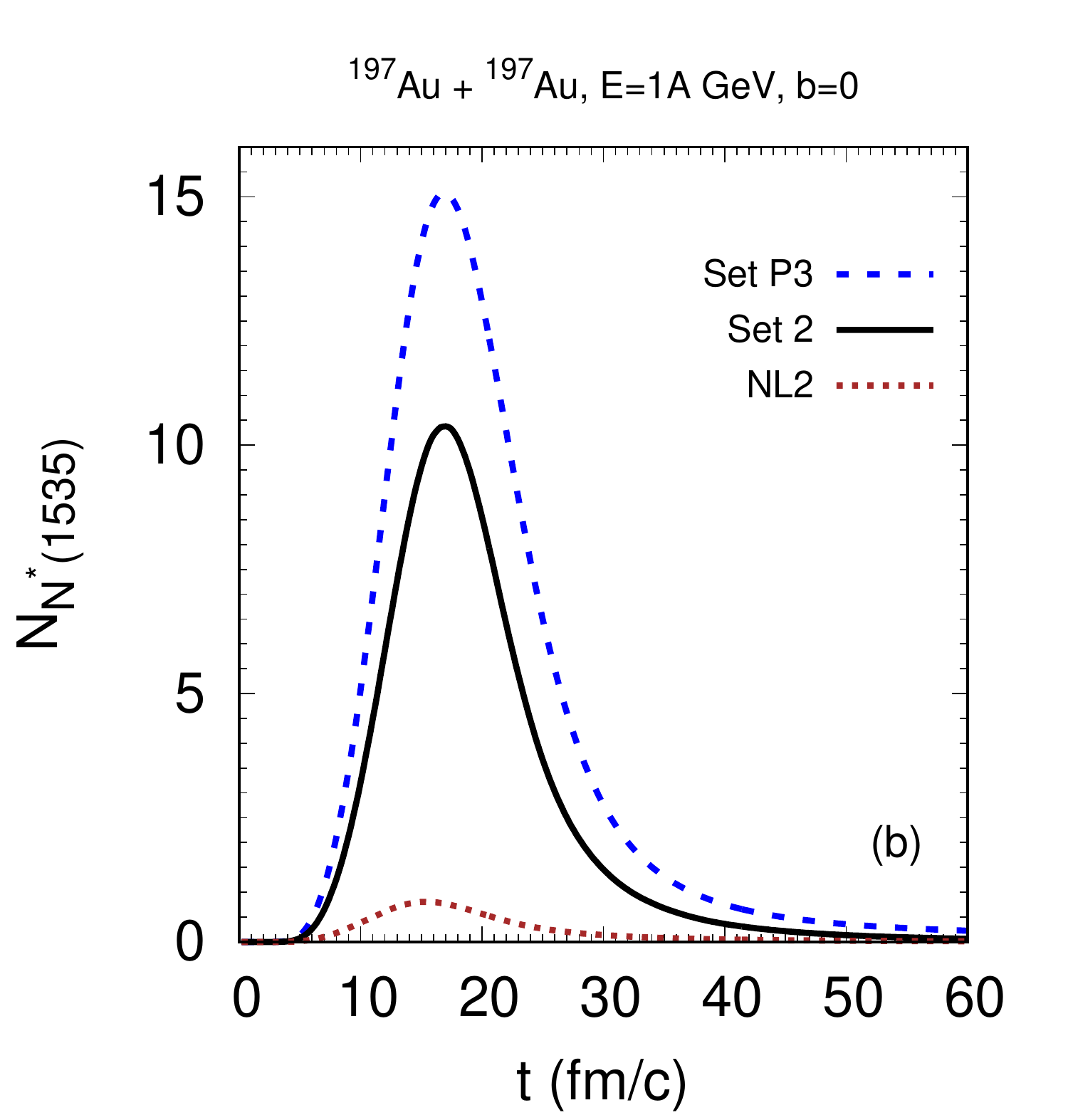}
  \includegraphics[scale = 0.48]{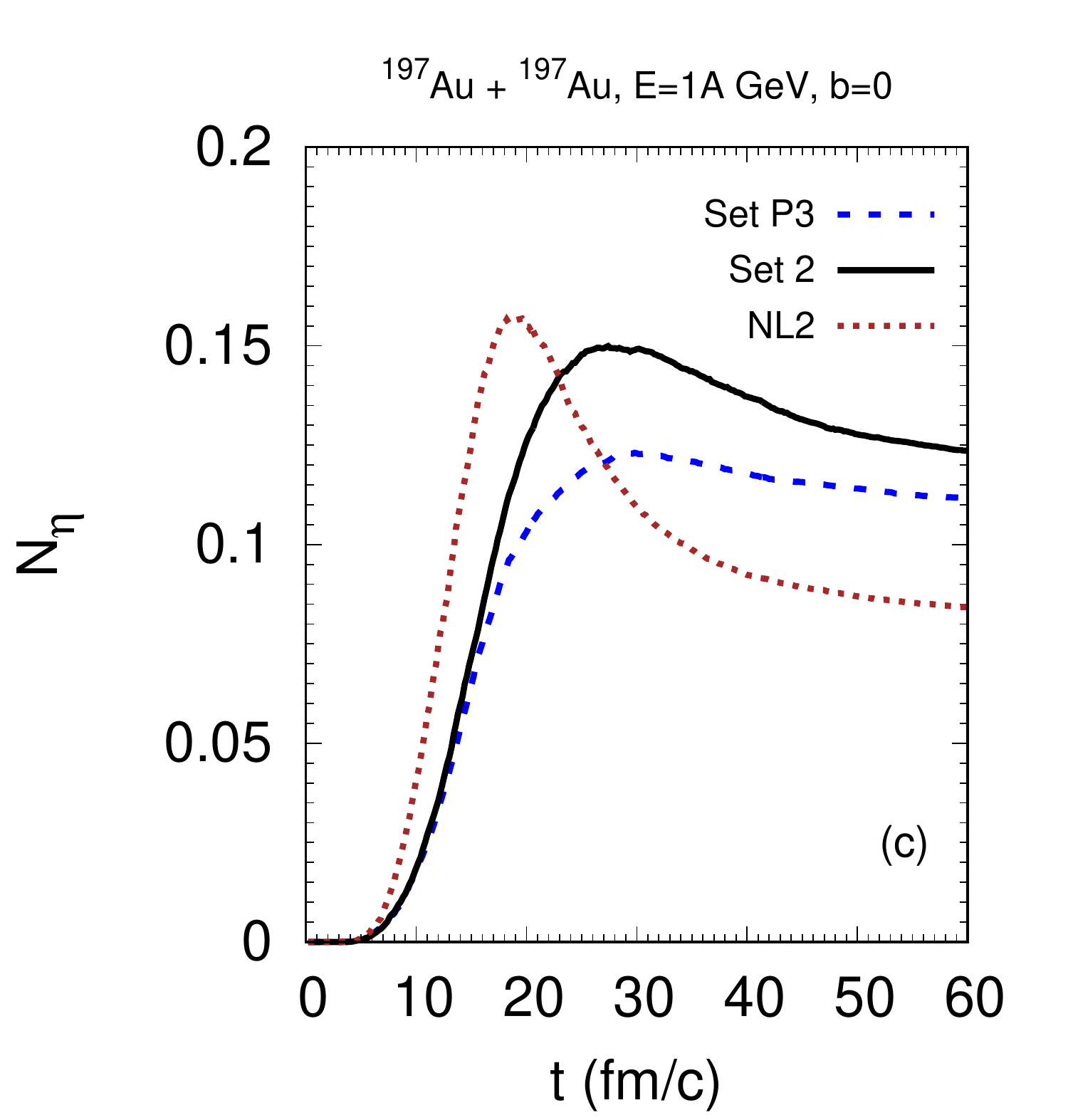}
  \includegraphics[scale = 0.48]{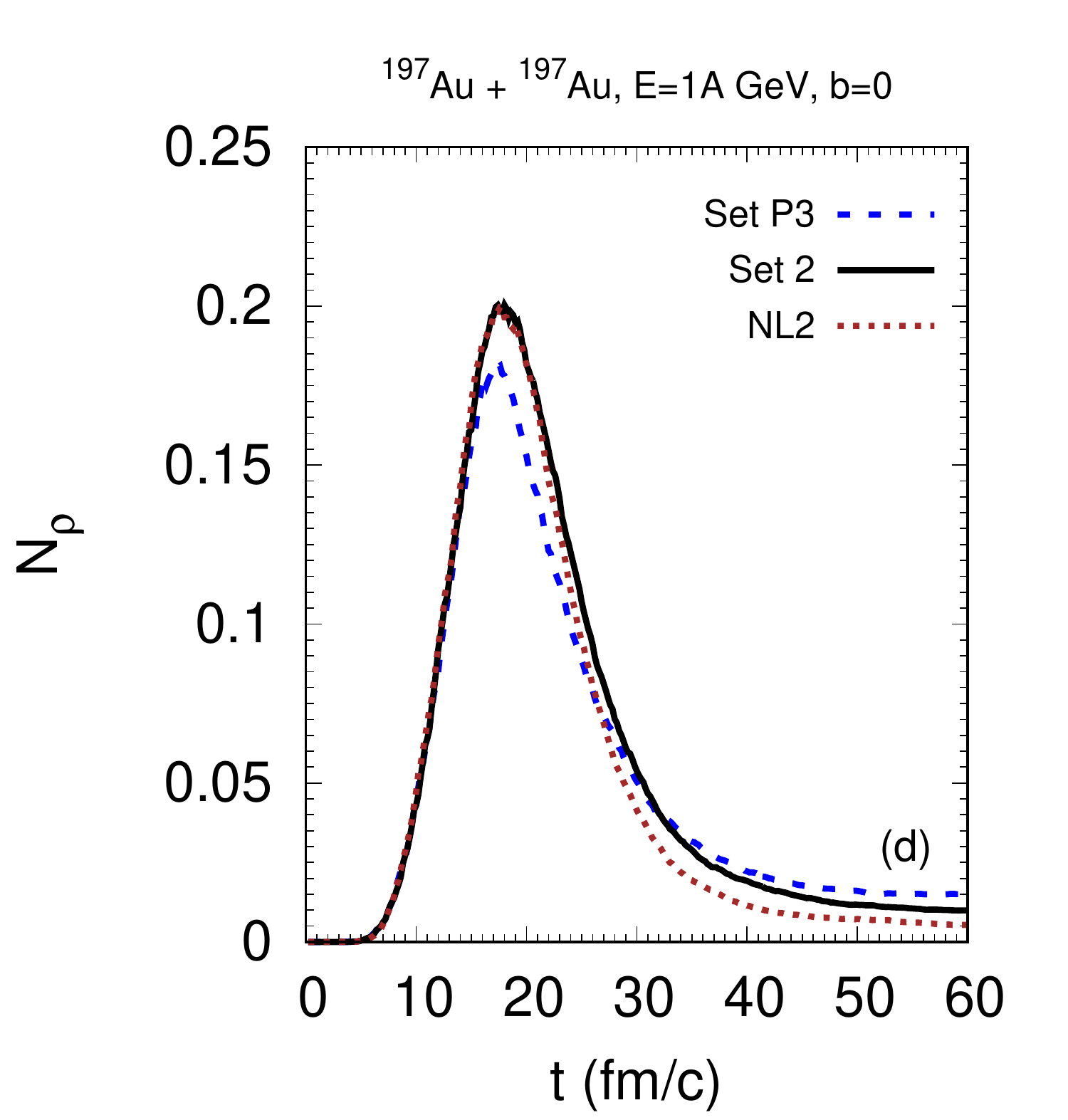}
  \caption{Time evolution of the central baryon density (a), $N^*(1535)$ multiplicity (b), $\eta$ multiplicity (c), and $\rho$ multiplicity (d) for the Au+Au central collision at 1A GeV.
    Line-style labels correspond to those in Fig.~\ref{fig:eos}.} 
    \label{fig:evol} 
\end{figure}

\begin{figure}
   \includegraphics[scale = 0.70]{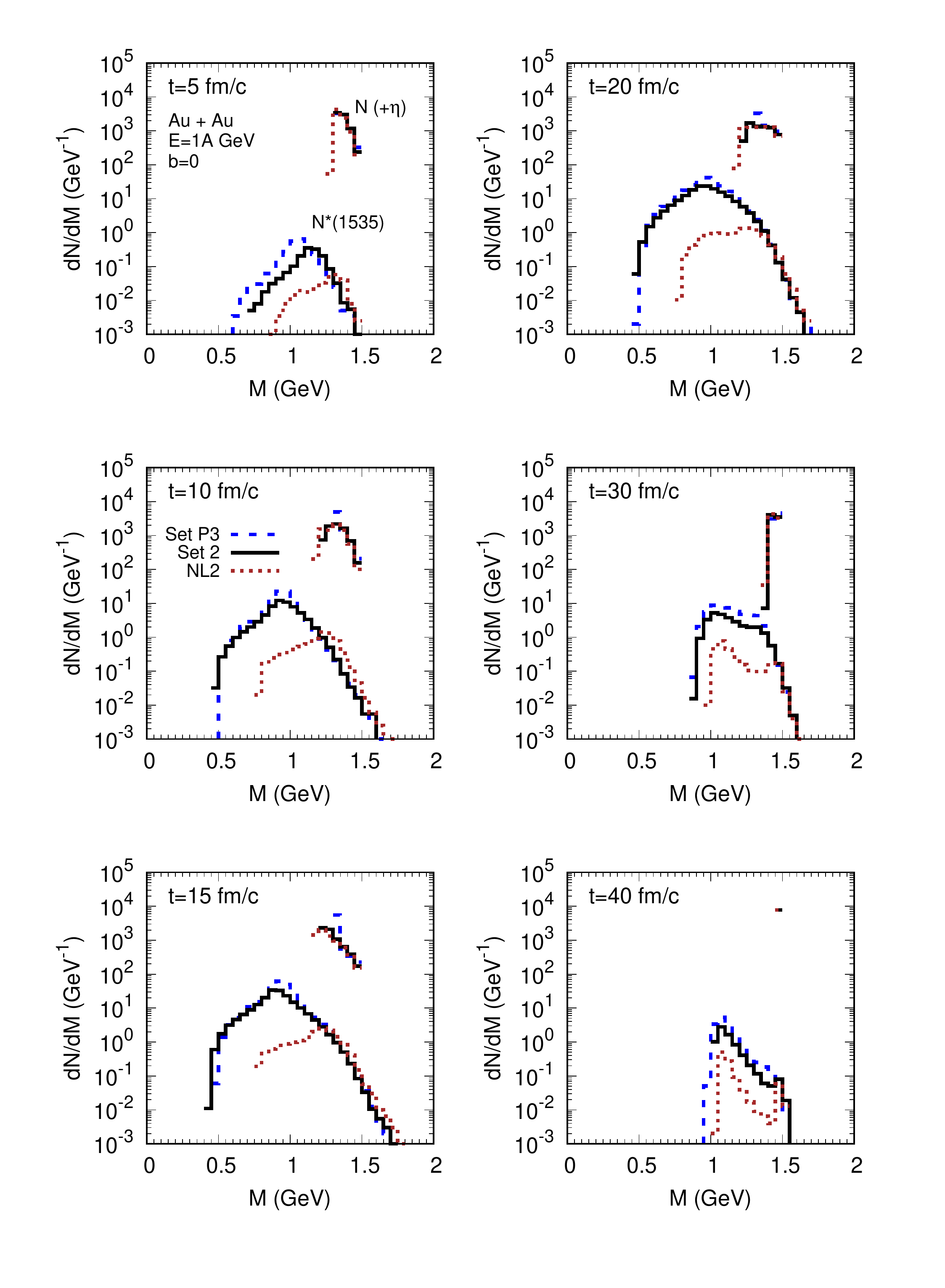}
   \vspace*{-1cm}
   \caption{Invariant mass distributions of $N^*(1535)$ (lower histograms) and nucleons (upper histograms) at different time moments
     for the Au+Au central collision at 1A GeV. The nucleon distributions are right-shifted by the value of $\eta$ mass.
     The contribution of the vector fields is excluded. Line-style labels as in the previous figures.}
     \label{fig:InvMassDist_1535_nuc} 
\end{figure}

The time evolution of the central baryon density together with that of the $N^*(1535)$, $\eta$ and $\rho$ multiplicities is shown in Fig.~\ref{fig:evol}.
The most striking difference is seen in the time dependence of the parity-partner multiplicity in Fig.~\ref{fig:evol}b. We observe an order
of magnitude enhancement of the $N^*(1535)$ maximum multiplicity for calculations with PDM mean fields as compared to the calculation with the non-linear Walecka mean fields. This is the expected consequence of the faster dropping $m_-$ with baryon density relative to the $m_+$, with the PDM sets,
which lowers the threshold $\sqrt{s^*}$ for $N^*(1535)$ production. 
Probably less expected on the other hand is the $\eta$ multiplicity, which is remarkably insensitive to the mean-field effects: it is enhanced by only about 50-60\% with the PDM mean fields.\footnote{Recall that the incompressibilities for the two PDM parameterizations are extremely different, cf.~Tab.~\ref{tab:par}. 
That this has very little influence on the time evolution of the central baryon density must be due to the colliding system being far from the ground
  state where the incompressibility is defined, see Eq.~(\ref{K}).}
This is because the abundant $N^*(1535)$'s of the PDM calculations have too small invariant masses to decay into the $\eta N$ final state. This is demonstrated
in Fig.~\ref{fig:InvMassDist_1535_nuc} which shows the time evolution of the $N^*(1535)$ and nucleon invariant mass distributions.
In the high-density stages of the collision ($t \ltsim 20$ fm/c), the excess of the $N^*(1535)$'s  in the PDM calculations is accumulated at invariant masses below the $N\eta$ in-medium threshold, while the NL2 calculation produces even slightly more $N^*(1535)$'s above the $N\eta$ threshold. This explains the faster initial growth of the $\eta$ production in the NL2 calculation. In the later stages, as the system
expands, the $N^*(1535)$ distributions are shifted towards higher invariant masses. Thus, the low-mass $N^*(1535)$'s move above the $N\eta$ threshold and lead to an increased  $\eta$ production rate in the PDM calculations. Since the rates of $NN \leftrightarrow N N^*(1535)$ processes are small during the expansion stage, the dynamics of the $\eta$ is dominated by the $N^*(1535) \leftrightarrow \eta N$ processes.
However, the $N^*(1535)$ resonance has a large branching ratio also for the $\pi N$ final state.\footnote{The GiBUU resonance parameters adopted from Ref.~\cite{Manley:1992yb} include the following branching ratios for the $S_{11}(1535)$ resonance:
  $\pi N$ 51\%, $\eta N$ 43\%, $\rho N$ S-wave 2\%, $\rho N$ D-wave 1\%, $\sigma N$ 1\%, $\pi P_{11}(1440)$ 2\%.
  The branching ratios for the $D_{13}(1520)$ resonance are: $\pi N$ 59\%, $\pi \Delta$ S-wave 5\%, $\pi \Delta$ D-wave 15\%, $\rho N$ 21\%.}
This results in an overall reduction of the $\eta$ multiplicity due
to the absorption on nucleons $\eta N \to N^*(1535) \to \pi N$. In the NL2 Walecka model calculation $\eta$ absorption clearly wins over the $N^*(1535)$ decays in the expansion stage which is not the case in the PDM calculations
where the $N^*(1535)$ multiplicity is larger. Overall, this results in a somewhat larger final $\eta$ multiplicity with PDM mean fields.

The dynamics of $\rho$ production is more complex.
Here, we have an interplay between $N^*(1520)$ and $N^*(1535)$ decays into the $\rho N$ final state.
As seen in Fig.~\ref{fig:evol}(d), in the intermediate stage of the reaction, for $t \ltsim 30$~fm/c, the $\rho$ multiplicity rather weakly depends on the type of mean field that is used (the somewhat smaller maximum $\rho$ multiplicity for Set P3 can be explained by its stiffer EOS). At later times, however, the $\rho$ multiplicities differ
significantly between PDM and Walecka mean fields.

\begin{figure}
   \includegraphics[scale = 0.70]{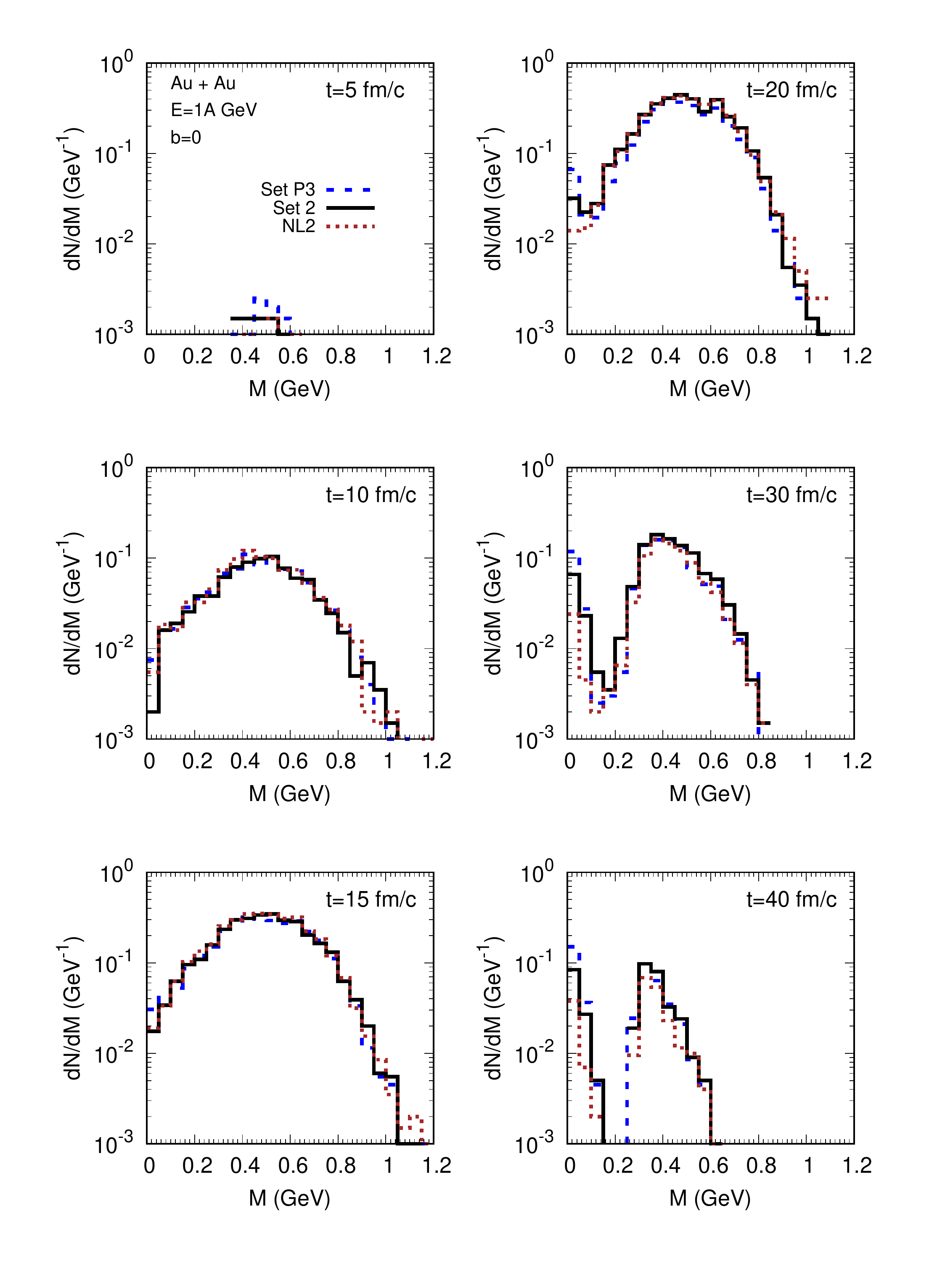}
   \vspace*{-1cm}
   \caption{Invariant mass distribution of $\rho$ mesons at different times in Au+Au central collision at 1A GeV.
     Line-style labels as in the previous figures.}
   \label{fig:InvMassDist_rho}
\end{figure}

These differences can be understood from the time evolution of the $\rho$ invariant mass distributions shown in Fig.~\ref{fig:InvMassDist_rho}.
At $t \leq 15$ fm/c the spectrum of $\rho$ masses is practically independent on the used mean fields. This is because at this early reaction
stage $\rho$ production is dominated by $N^*(1520) \to \rho N$ decays in either case. Recall that in the current PDM calculations,
we set the mean fields acting on all baryonic resonances except the  $N^*(1535)$ equal to the nucleon mean fields.
Therefore, the channels mediated by resonances other than the $N^*(1535)$ are not expected to depend on the mean fields in our calculations.
At $t \geq 20$ fm/c, however, we observe a quickly growing $\rho$ production below the two-pion threshold
where the $\rho$-meson spectral function is supported by the $\rho \to e^+ e^-$ partial width and
by the collisional width $\rho N \to \mbox{resonances}$ (see Ref.~\cite{Larionov:2020fnu} for detail).
This soft part of the $\rho$ invariant-mass spectrum is populated by the decays of low-mass baryon resonances that have a long life time.
(Note that the ``hole'' in the spectrum of $\rho$ masses at $M \simeq 0.2$ GeV is certainly to some extend artificial. It would get at least partially filled by including missing partial widths in the vacuum $\rho$ spectral function such as those from $\rho \to \pi \gamma$ or also  $\rho \to \mu^+ \mu^-$, see the discussion in Ref.~\cite{Larionov:2020fnu}.
At present, there is no experimental identification of the $\rho$
contribution below the $2\pi$ threshold. This part of our results is therefore
largely uncertain and requires further studies in future. For our present study it is irrelevant.)
In PDM based calculations, the multiplicity of low-mass resonances decaying into the $\rho N$ channel is strongly enriched by $N^*(1535)$'s.
This explains the excess of soft $\rho$ production which is also reflected in the overall excess of the $\rho$ multiplicity
at large times in the PDM.
 
\section{Comparison with experimental data}
\label{compExp}

The transverse mass spectra of $\eta$ and $\pi^0$ at midrapidity have been measured by the TAPS Collaboration for the following systems: C+C at 0.8,1.0, and 2.0A~GeV \cite{Averbeck:1997ma},
Ar+Ca at 0.8A~GeV \cite{Marin:1997kj}, and Au+Au at 0.8A GeV \cite{Wolf:1998vn}. For a thermally-equilibrated source the $m_t$-spectra at $y=0$ would be expected to behave exponentially,
\begin{equation}
  \frac{d^2\sigma}{dy m_t^2 dm_t} \propto e^{-m_t/T}~,    \label{d2sigma_dymt^2dmt}
\end{equation}
which is known as the $m_t$-scaling observed for $\eta$ and $\pi^0$ production experimentally. The explanation of $m_t$ scaling purely in terms of thermal equilibrium might seem questionable, however, at beam energies as low as 1-2A GeV especially for light colliding systems such as C+C. Note that our transport calculations do not rely on the assumption of thermal equilibrium.
In an off-equilibrium situation, on the other hand, the $m_t$-spectra should depend on the details of particle production and propagation in the nuclear medium.
It is therefore interesting to study how the $m_t$-spectra are influenced by the PDM description of the mean fields.    

\begin{figure}[ht]
   \includegraphics[scale = 0.48]{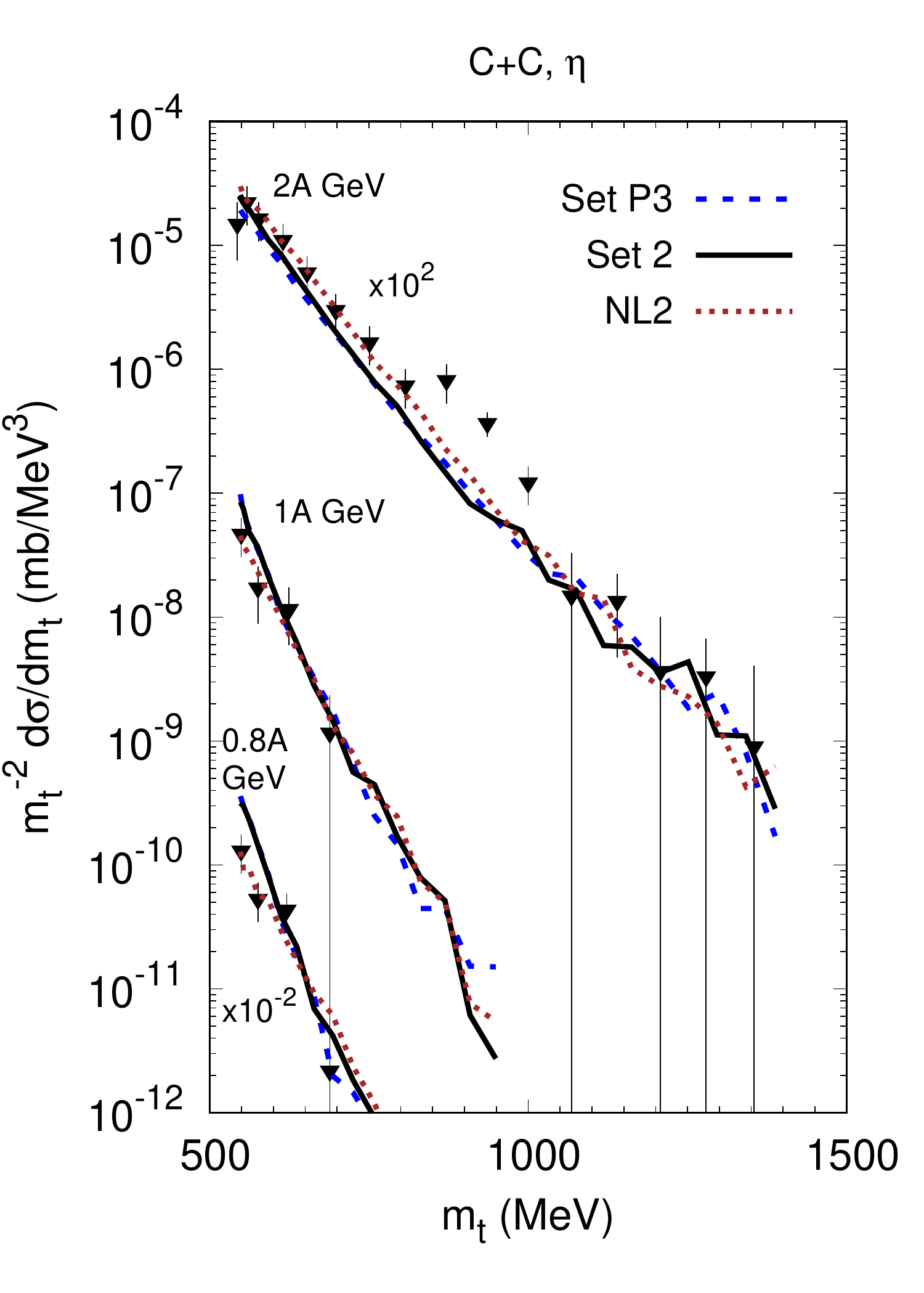}
   \includegraphics[scale = 0.48]{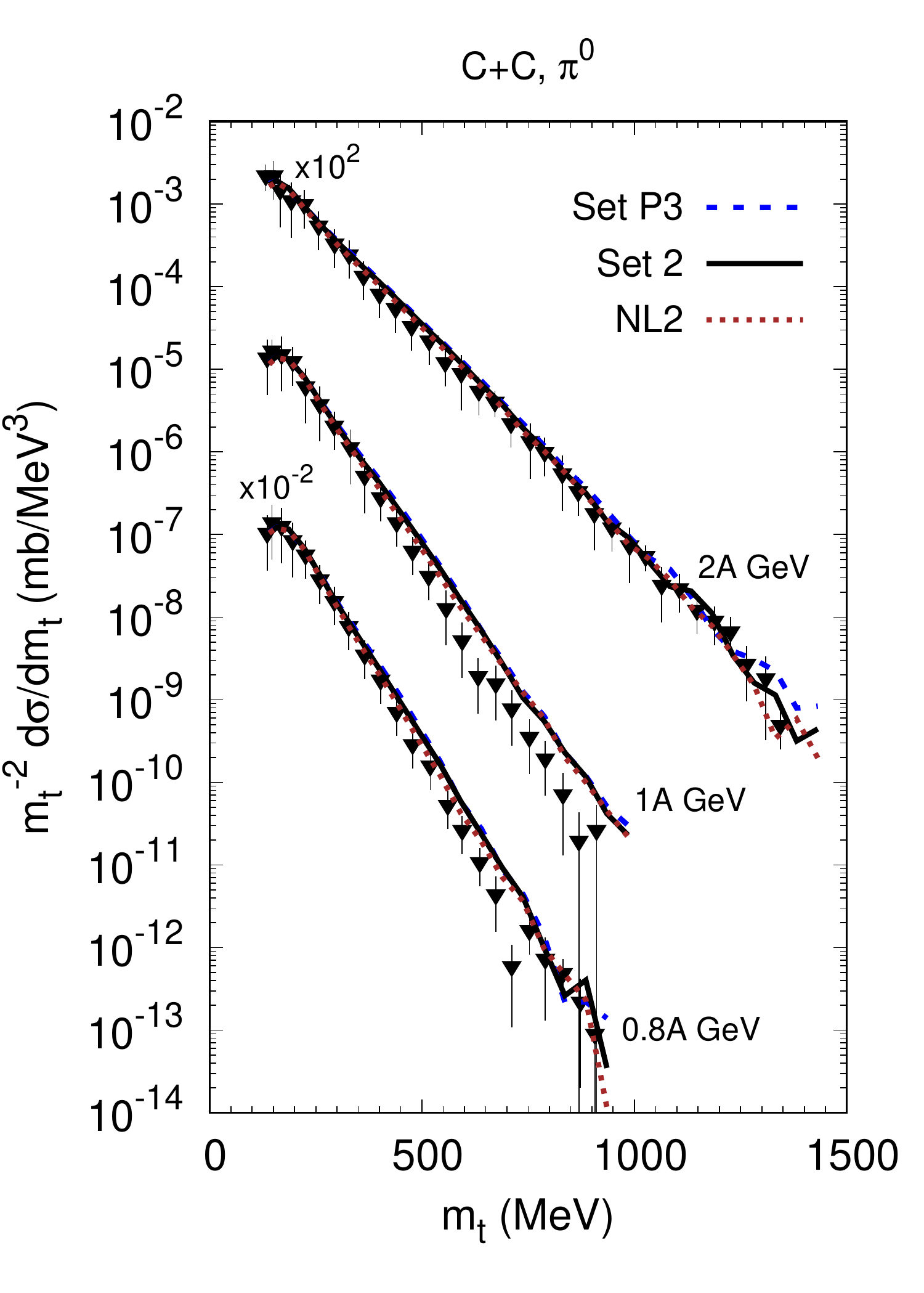}  
   \caption{Transverse-mass differential cross sections for $\eta$ and $\pi^0$ production in C+C collisions at 0.8, 1.0, and 2.0A~GeV.
     The rapidity intervals around the midrapidity values (in parentheses) in the laboratory system are [0.42;0.74] (0.61), [0.42;0.74] (0.68), and [0.80;1.08] (0.90) for 0.8, 1.0, and 2.0A~GeV, respectively.
     Experimental data are from Ref.~\cite{Averbeck:1997ma}.}
   \label{fig:dsig_dmt_CC} 
\end{figure}
Fig.~\ref{fig:dsig_dmt_CC} shows the $m_t$-spectra of $\eta$'s and $\pi^0$'s in C+C collisions.
The $\pi^0$ $m_t$-spectra are in a good agreement with data and are not influenced by the differences in the mean fields. This is expected, since the scalar potential
of the $\Delta(1232)$ resonance is set equal to that of the nucleon in either, for the PDM as well as the non-linear Walecka model. In contrast, for $\eta$ mesons we
see the enhancement at low $m_t$'s in the calculations with the PDM mean fields at the lower beam energies of 0.8 and 1.0A~GeV. At 2.0A GeV this enhancement at low $m_t$'s disappears.
Thus, below the free kinematical threshold, the PDM mean fields tend to enhance the production of slow $\eta$'s in the c.m. frame of the colliding nuclei. 
It might seem  quite surprising at first, that this mean field effect is visible even in the light C+C system where far lower densities are reached than in the heavier colliding nuclei (cf. Fig.~5 in Ref. \cite{Larionov:2020fnu}). The reason of course is that the $m_-$ mass drops so much faster with baryon density at low $\rho_B$ in the PDM description, see Fig.~\ref{fig:m_pm}.

\begin{figure}[ht]
   \includegraphics[scale = 0.48]{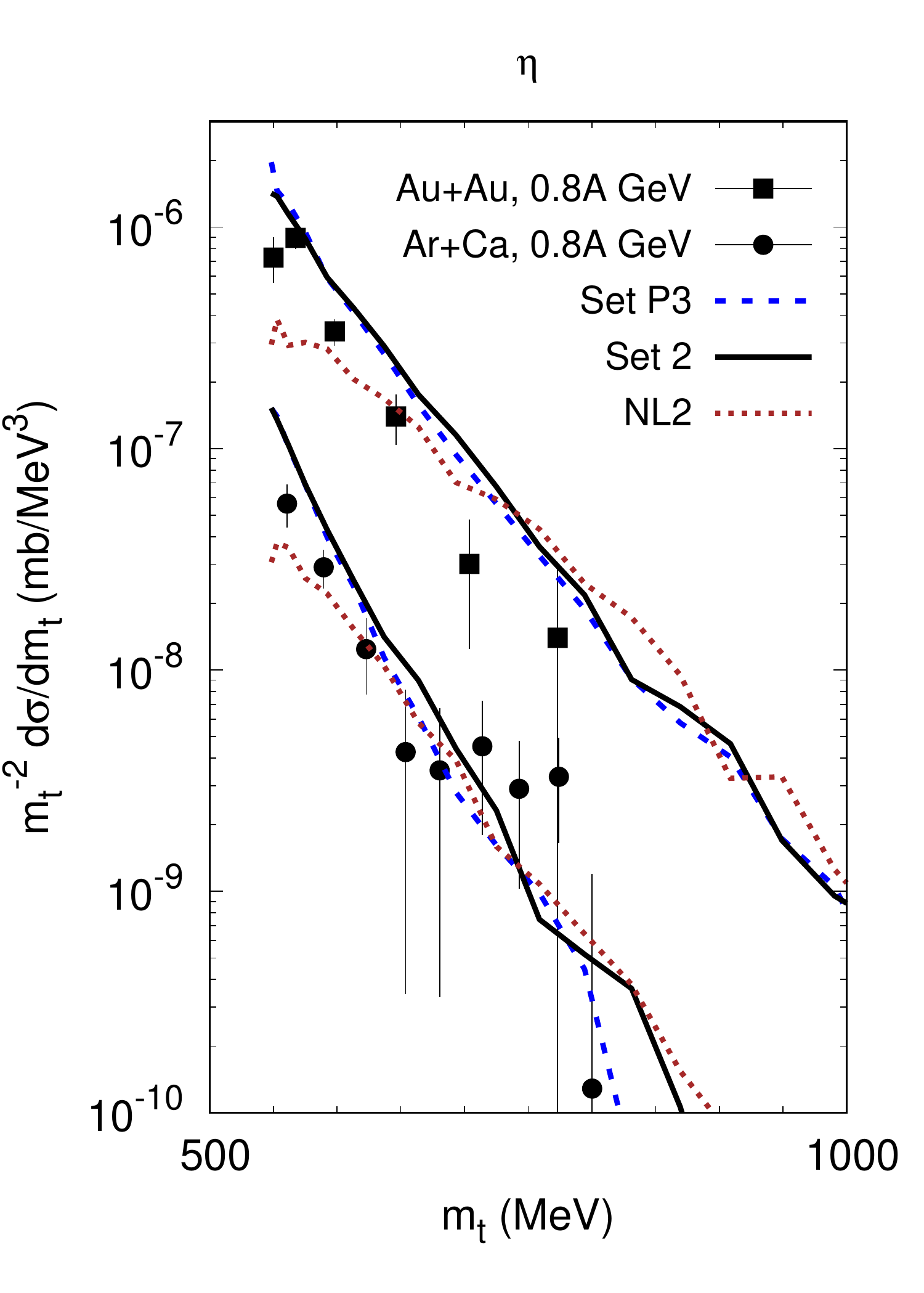}
   \includegraphics[scale = 0.48]{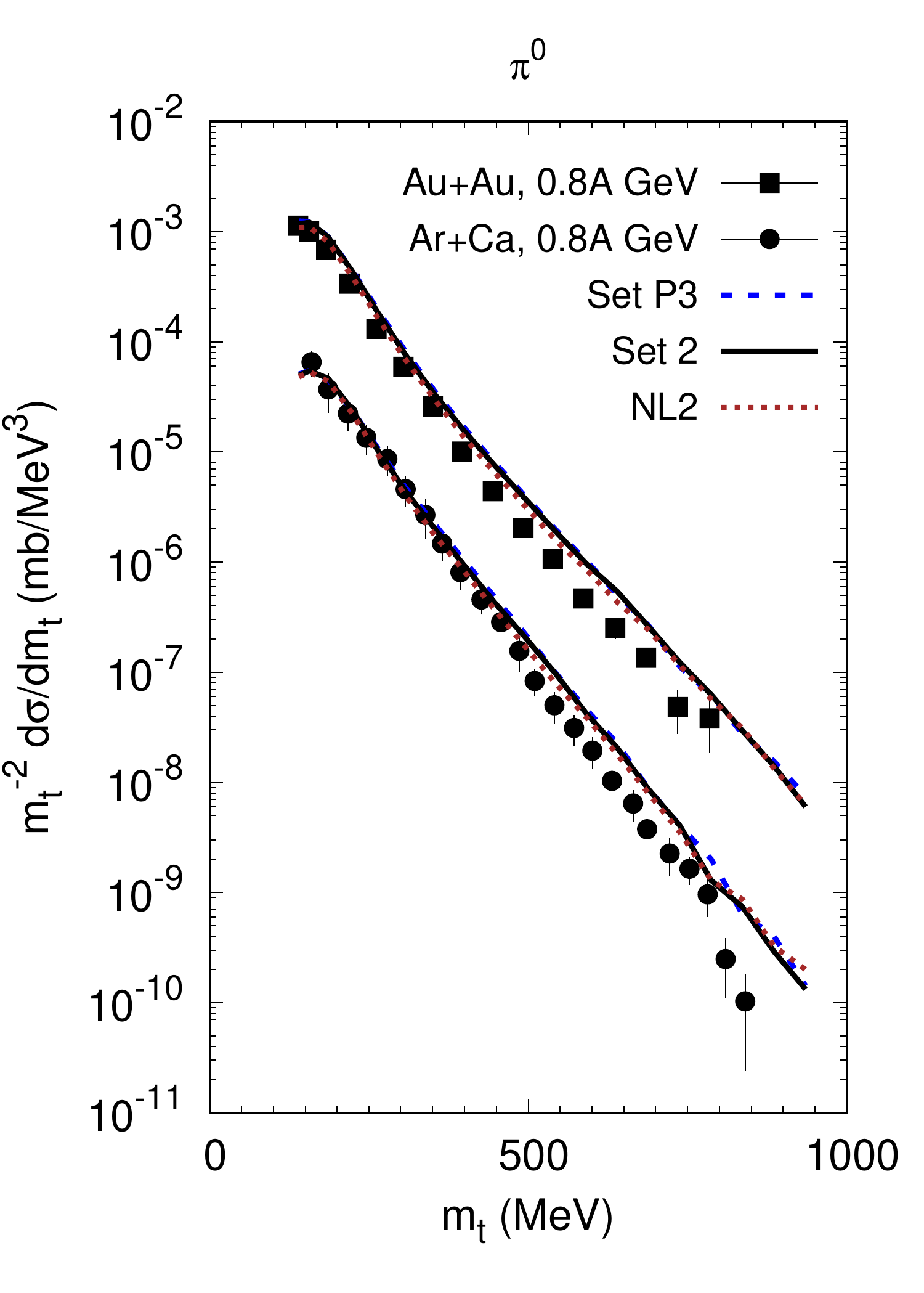}  
   \caption{Transverse-mass differential cross sections for $\eta$ and $\pi^0$ production in Au+Au and Ar+Ca collisions at 0.8A~GeV.
     The rapidity intervals in the laboratory system are [0.510;0.718] and [0.40;0.78] for Ar+Ca and Au+Au, respectively.
     Experimental data are from Refs.~\cite{Marin:1997kj,Wolf:1998vn}.}
     \label{fig:dsig_dmt_AuAu_ArCa}
\end{figure}
In heavier systems at subthreshold energies, the enhancement of $\eta$ production at low $m_t$'s in calculations with PDM becomes more pronounced as demonstrated
in Fig.~\ref{fig:dsig_dmt_AuAu_ArCa}. The variation of the incompressibility $K$ in the PDM sets has practically no effect on the meson spectra.

The TAPS data for C+C do not appear to favor either the PDM or the non-linear Walecka mean fields. However, the PDM calculations better describe the slope of the $m_t$ spectra of $\eta$ mesons at small $m_t$'s for Au+Au and Ar+Ca at 0.8A~GeV.

The $m_t$ spectra of neutral pions in the heavier systems, Au+Au and Ar+Ca, are practically not influenced by the choice of mean field as we also saw for C+C before.
Some overestimation at large transverse masses at 0.8 and 1A~GeV has been also observed in previous  BUU calculations \cite{Larionov:2003av}
and is known to improve when using in-medium $NN \leftrightarrow N \Delta$ cross sections.

\begin{figure}[ht]
   \includegraphics[scale = 0.48]{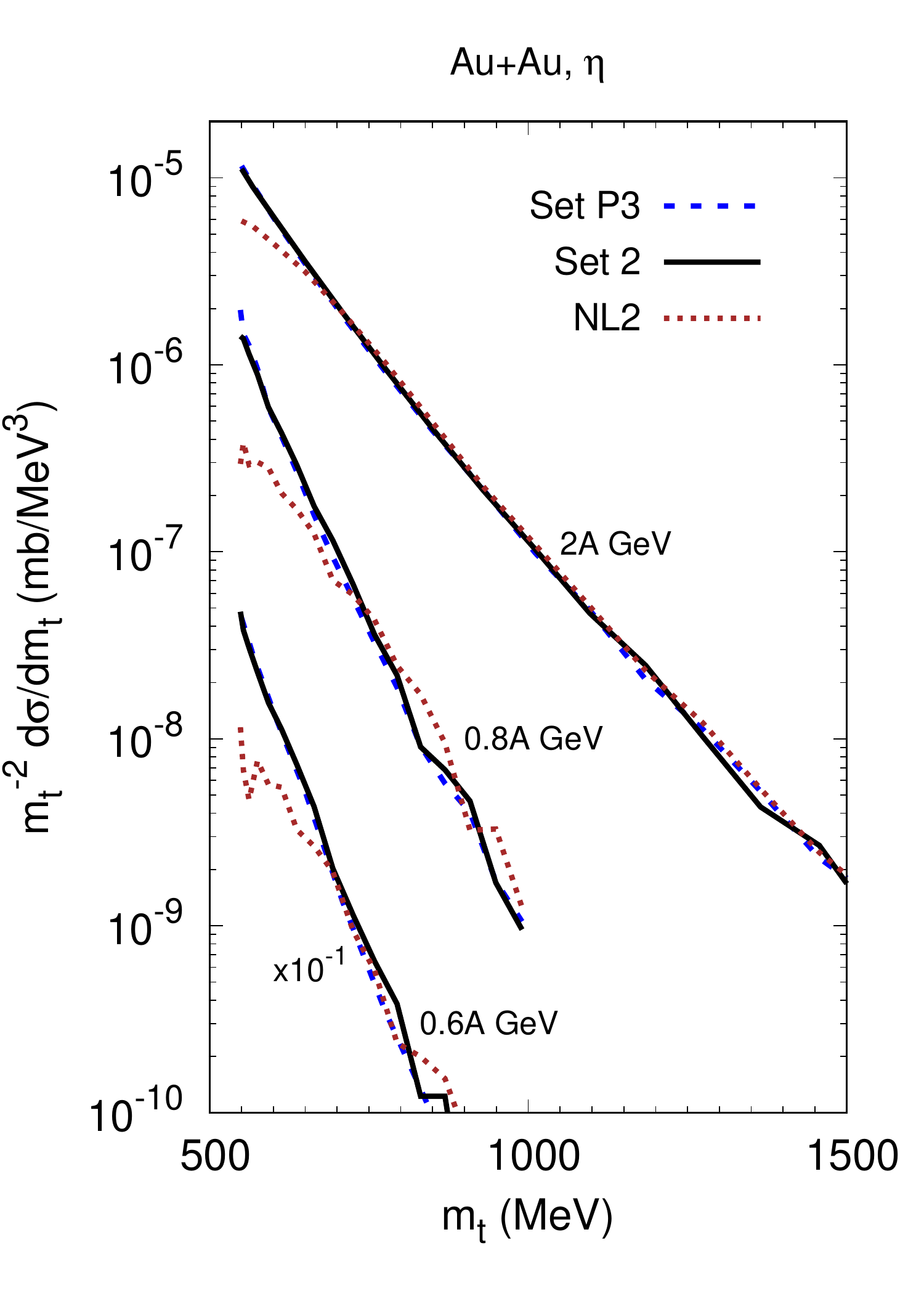}
   \includegraphics[scale = 0.48]{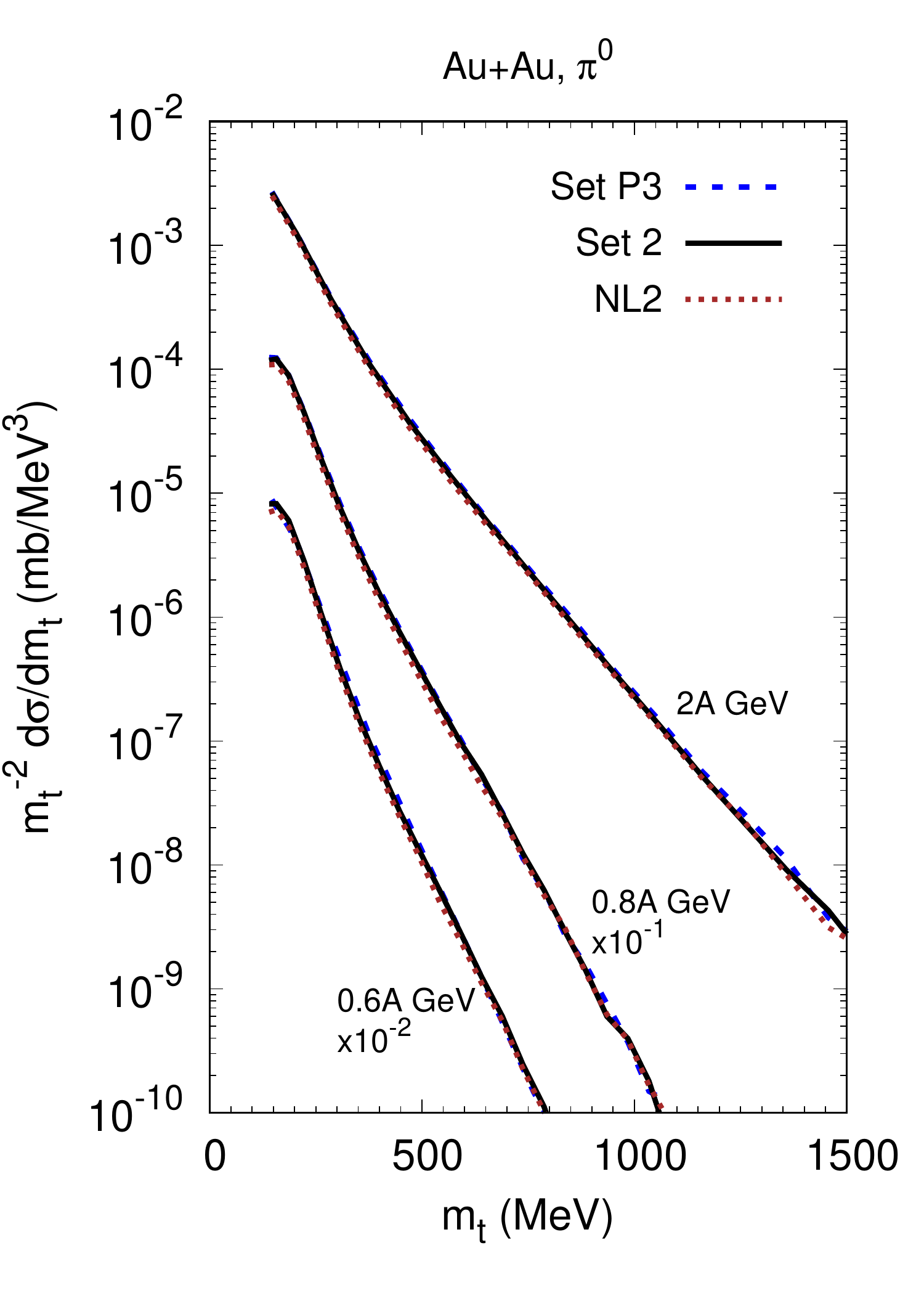}  
   \caption{Transverse-mass differential cross sections predicted for $\eta$ and $\pi^0$ production in Au+Au collisions at 0.6, 0.8, and 2A GeV.
     The $m_t$ spectra at 0.6A GeV and 2A GeV are calculated for the c.m. rapidity intervals [-0.2;0.2] and [-0.15;0.15], respectively,
     while at 0.8A GeV -- in the laboratory rapidity interval [0.40;0.78].}     
     \label{fig:dsig_dmt_AuAu}
\end{figure}

\begin{figure}[ht]
   \includegraphics[scale = 0.48]{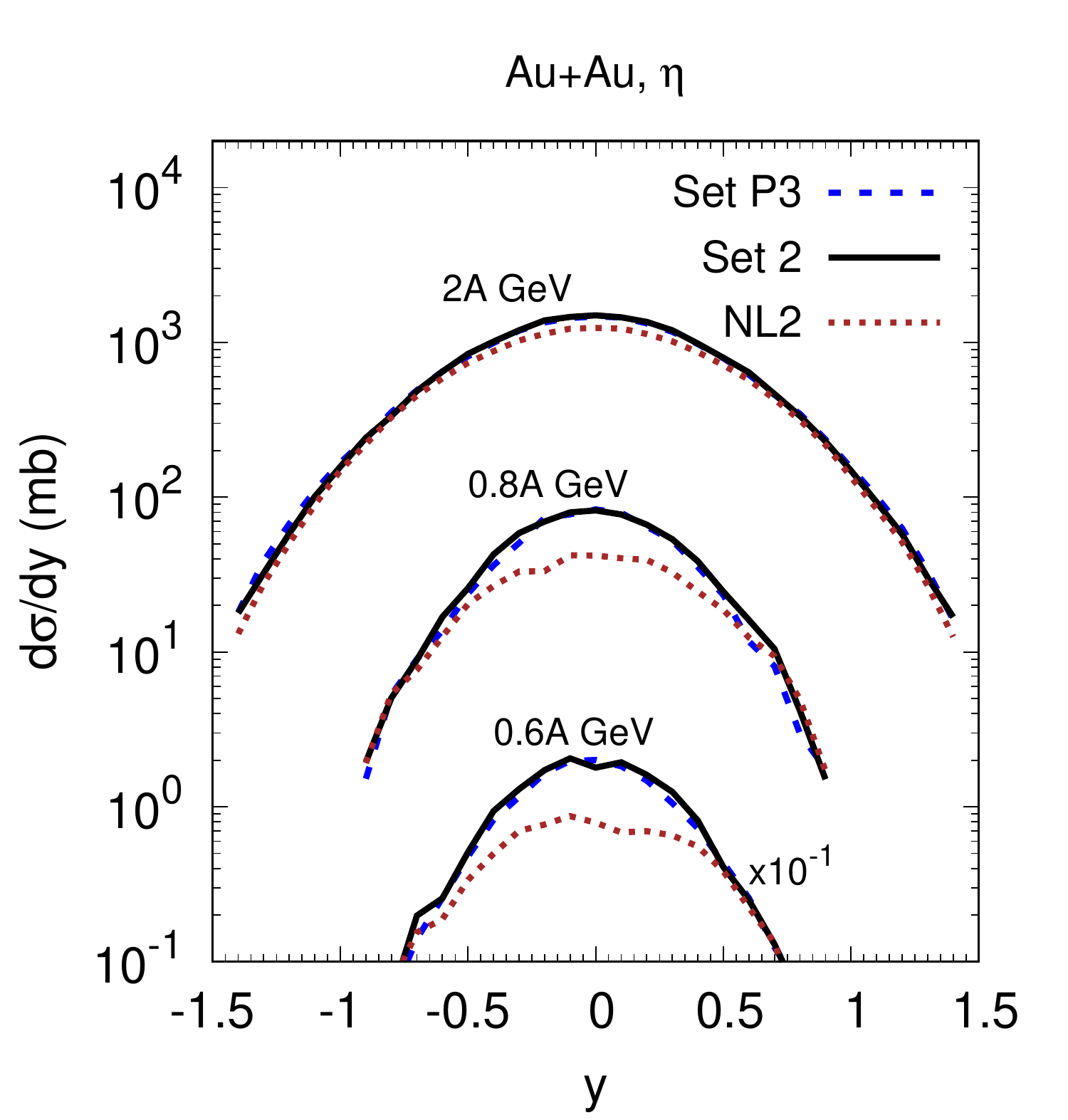}
   \includegraphics[scale = 0.48]{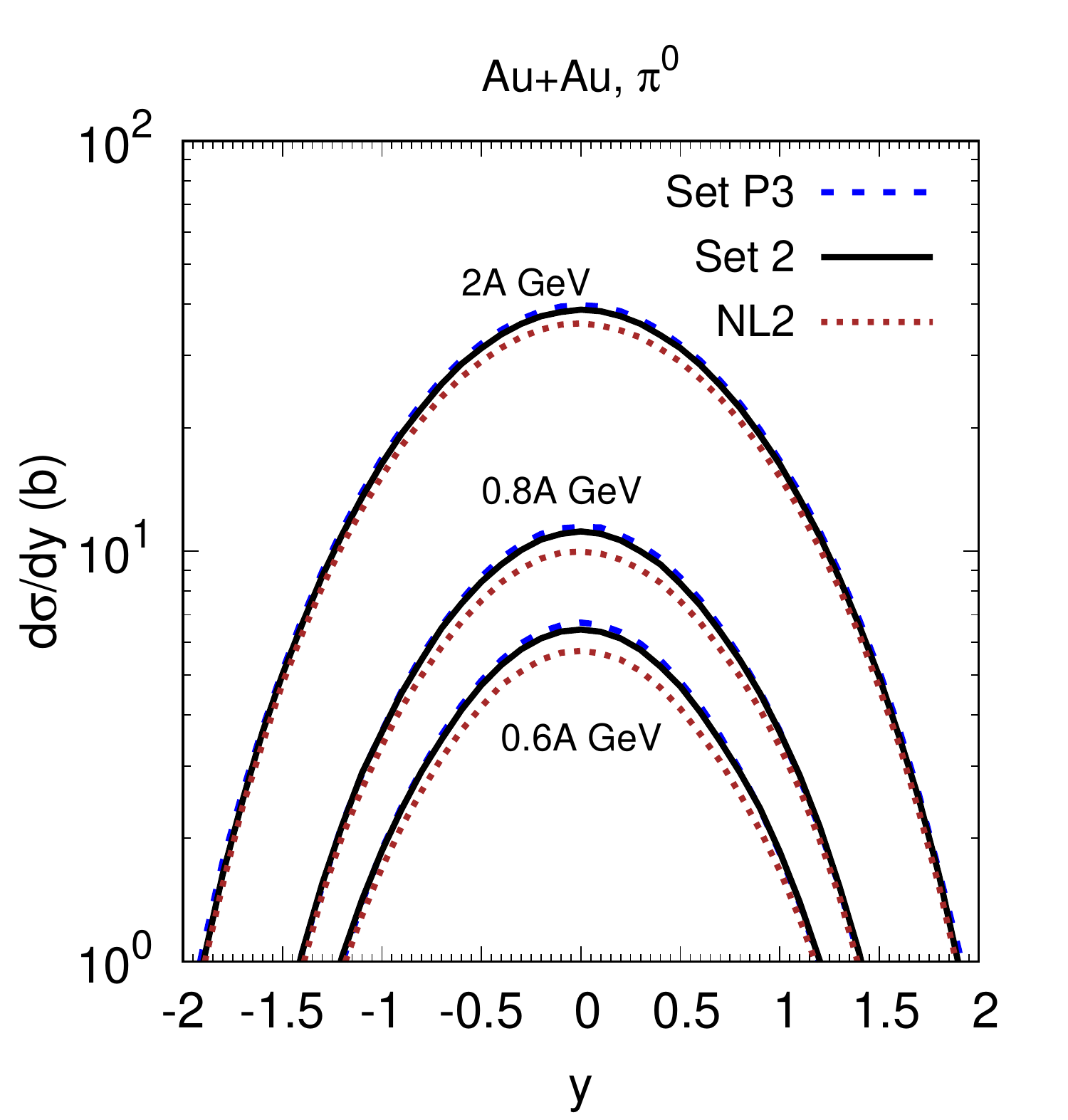}  
   \caption{Center-of-mass rapidity differential cross sections predicted for $\eta$ and $\pi^0$ production in Au+Au collisions at 0.6, 0.8, and 2A GeV.
     Line-style labels refer to chiral PDM (Set P3 and Set 2) and the Walecka model (NL2) mean fields, respectively,  as in the previous figures.}
     \label{fig:dsig_dy_AuAu}
\end{figure}
In Figs.~\ref{fig:dsig_dmt_AuAu}, \ref{fig:dsig_dy_AuAu} we provide our predictions for the $m_t$- and rapidity-spectra of $\eta$'s and $\pi^0$'s
in Au+Au collisions at 0.6, 0.8, and 2A GeV: There is a factor of five enhancement in the $\eta$ production at low transverse masses in the PDM calculations for 0.6 and 0.8A GeV.
Comparing the slopes and the absolute values of the $\eta$ and $\pi^0$ transverse mass spectra we observe that the PDM calculations are much closer
to the $m_t$-scaling regime. The mean-field effects at large transverse masses and/or large absolute values of rapidity generally become small. Overall, the PDM leads to an about a factor of two larger $\eta$ production cross section in Au+Au at 0.6A GeV: $15.4 \pm 0.3$ mb for the PDM (Set 2) versus $7.8 \pm 0.2$ mb with the (NL2) Walecka model mean fields (the errors are due to limited statistics). The production of
$\pi^0$'s at midrapidity is also slightly enhanced in the PDM calculations due to the $N^*(1535)$ resonance contribution.

\begin{figure}[ht]
   \includegraphics[scale = 0.60]{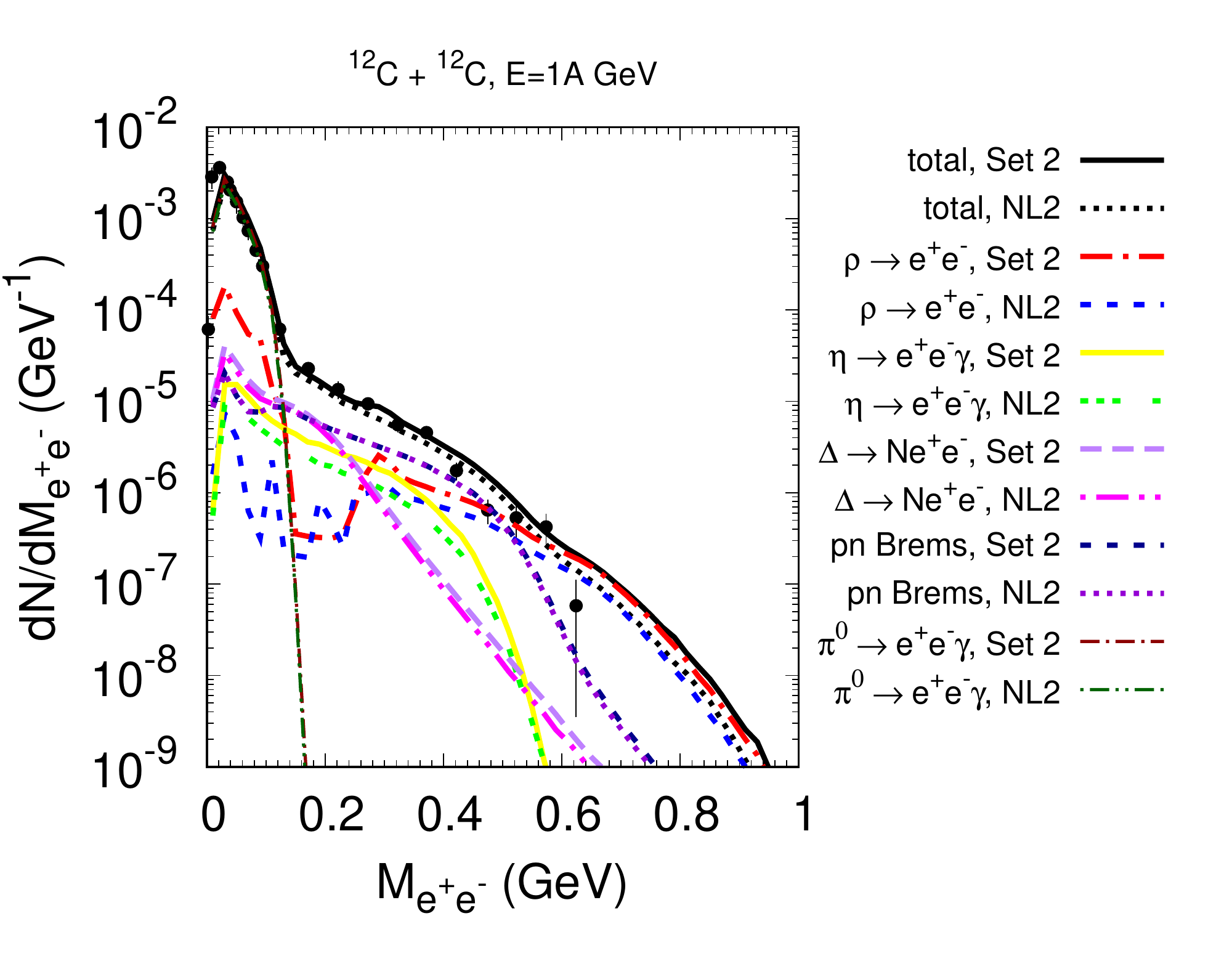}
      \caption{(Color online)
     $e^+e^-$ invariant mass spectrum from C+C collisions at 1A~GeV. Solid (black) and dotted (black) lines denote the total spectrum
     calculated with PDM (Set 2) and Walecka model (NL2) mean fields, respectively.
     Other lines show different partial components of the spectra as indicated.
     The $pp$ and $\pi N$ bremsstrahlung components are included in the total spectra but are not shown.
     Experimental data are from Ref.~\protect\cite{Agakishiev:2007ts}.}
     \label{fig:dNdM_CC1}
\end{figure}

We close our comparison with addressing dilepton production. 
Fig.~\ref{fig:dNdM_CC1} shows the dilepton invariant mass spectrum from C+C at 1A~GeV. The experimental acceptance filter
is taken into account in the calculations. The calculation with Set 2 leads to a larger dilepton yield at $M_{e^+e^-} > 0.15$~GeV.
This is due to enhanced contributions 
from the direct $\rho$ decay (component in red versus blue), the $\eta$ Dalitz component (yellow versus green), and to a lesser extend also from $\Delta$ Dalitz decays (purple versus magenta).
The reason for the enhanced $\eta$ and low-mass $\rho$ components in the PDM calculations has already been discussed in Sec.~\ref{AuAu1AGeV} above.
Moreover, we see from Fig.~\ref{fig:InvMassDist_rho} that some $\rho$ excess appears in the PDM calculations above $2m_\pi$
at large times.
The slightly enhanced pion yield due to the  $N^*(1535) \to \pi N$ decays leads to larger secondary $\Delta(1232)$ production
and thus to slightly larger $\Delta \to e^+ e^- \gamma$ contribution. 
The $pn$ bremsstrahlung component dominating at the intermediate
invariant masses (in the $M_{e^+e^-}=0.2-0.5$~GeV range) is practically independent on the choice of the mean field.

\begin{figure}

\vspace*{-.2cm}
  \includegraphics[scale = 0.60]{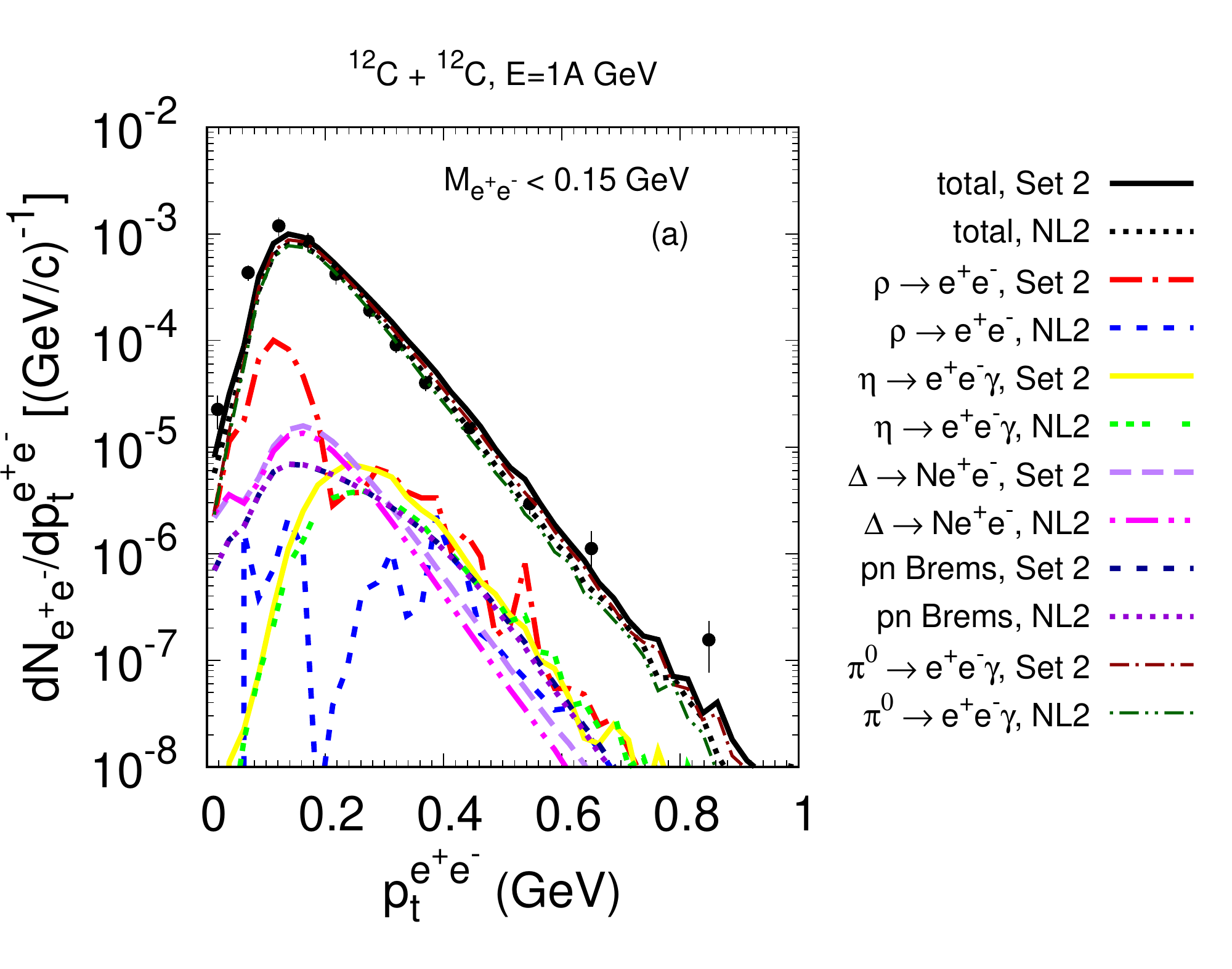}
  
  \vspace*{-.2cm}
  \includegraphics[scale = 0.60]{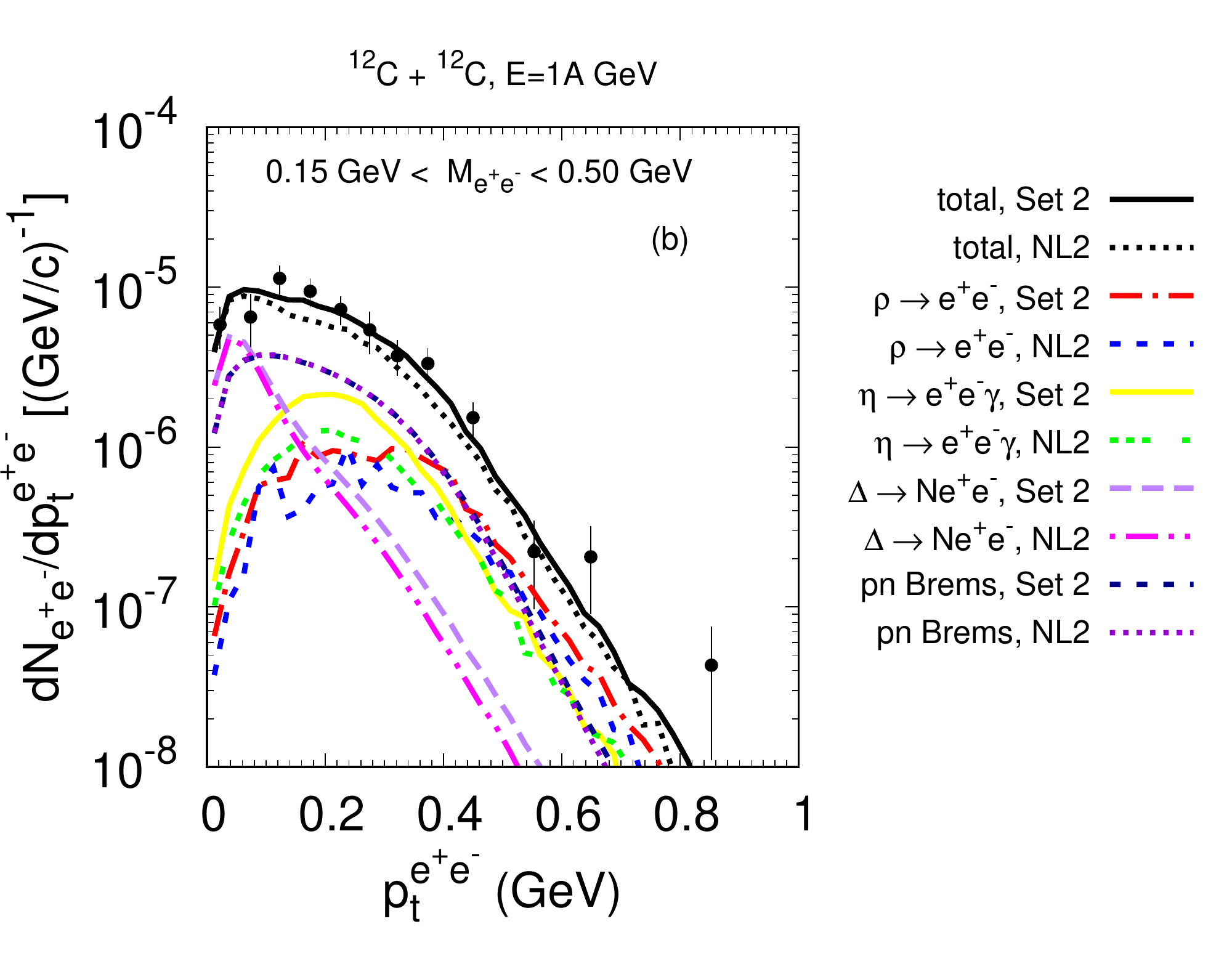}
  
  \vspace{-.4cm}
  \caption{(Color online)
  Transverse momentum distributions of dileptons produced in C+C collisions at $1 A$~GeV
  in the invariant mass intervals $M_{e^+e^-} < 0.15$ GeV (a), $0.15~\mbox{GeV} < M_{e^+e^-} < 0.50$ GeV (b).  
    Solid (black) and dotted (black) lines show the total spectra for PDM (Set 2) 
  and Walecka (NL2) mean fields, respectively. Other lines show partial components as indicated.
     Experimental data are from Ref.~\cite{Pachmayer:2008}.}
  \label{fig:dNdpt_CC1}
\end{figure}

\begin{figure}

\vspace*{-.2cm}
  \includegraphics[scale = 0.60]{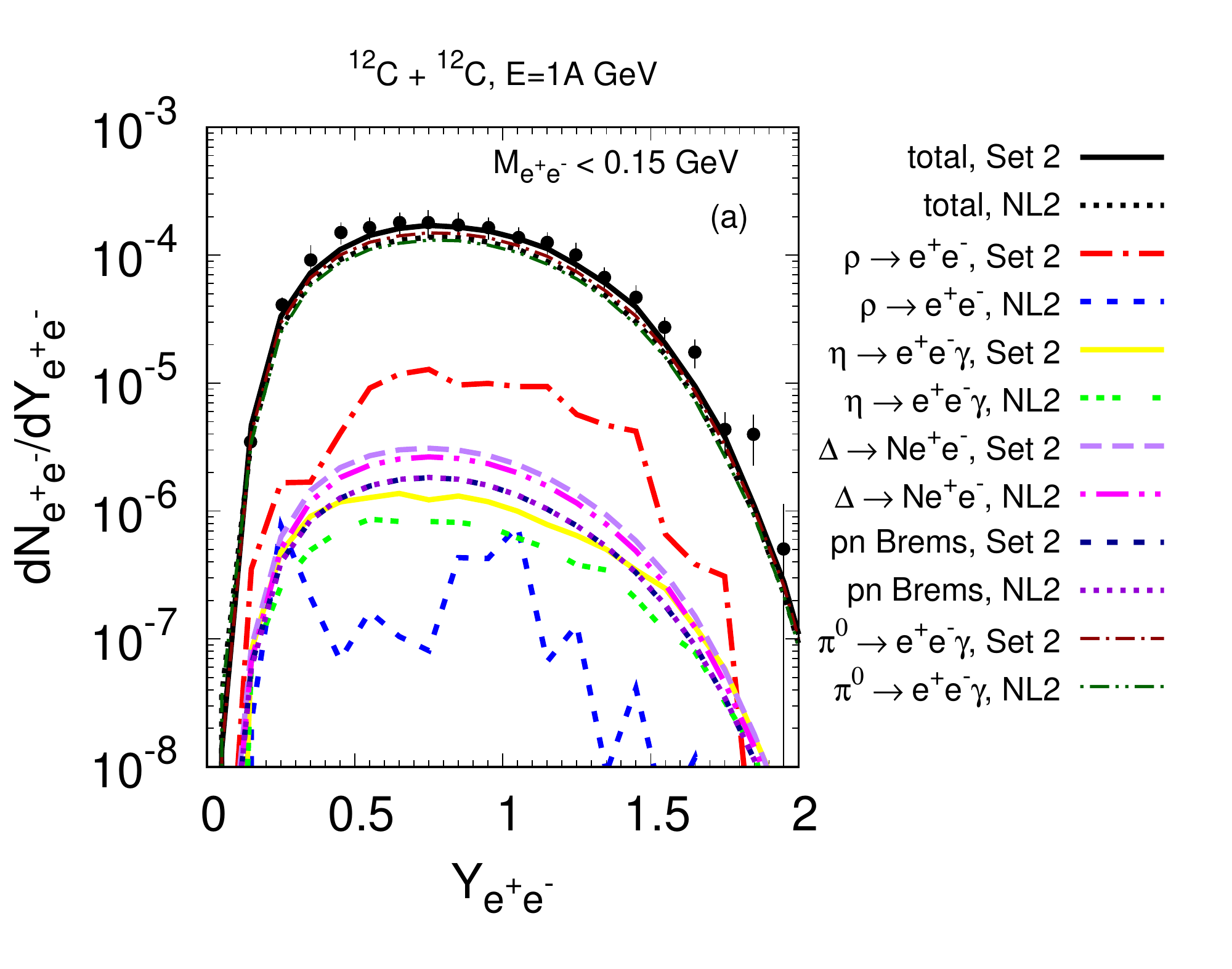}
  
  \vspace*{-.2cm}
  \includegraphics[scale = 0.60]{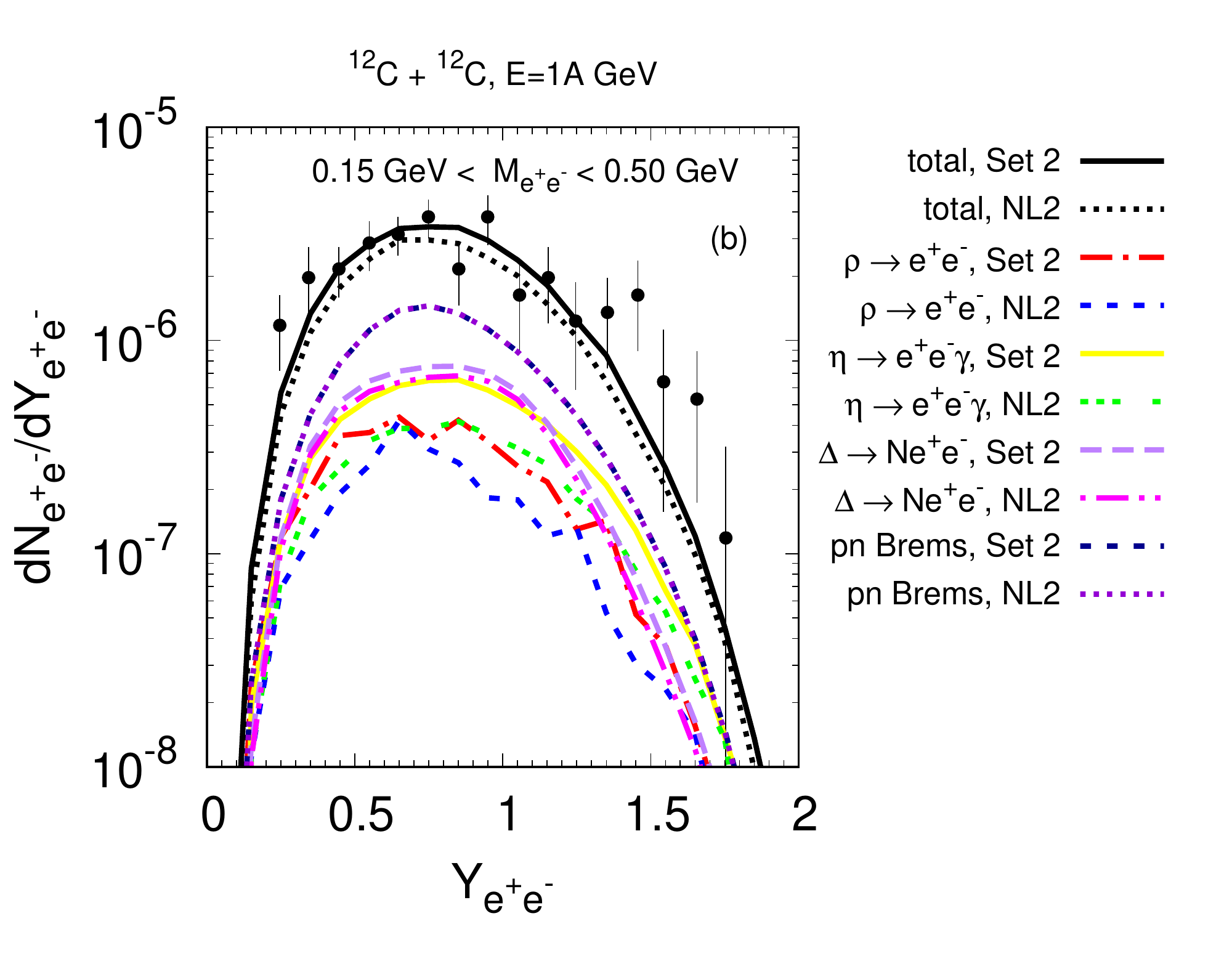}
  
  \vspace{-.4cm}
  \caption{(Color online)
    Rapidity distributions of dileptons produced in C+C collisions at $1 A$~GeV
    in the invariant mass intervals $M_{e^+e^-} < 0.15$ GeV (a), $0.15~\mbox{GeV} < M_{e^+e^-} < 0.50$ GeV (b).
      Solid (black) and dotted (black) lines show the total spectra for PDM (Set 2) 
  and Walecka (NL2) mean fields, respectively. Other lines show partial components as indicated.
    Experimental data are from Ref.~\cite{Pachmayer:2008}.}
    \label{fig:dNdY_CC1}
\end{figure}

Figs.~\ref{fig:dNdpt_CC1} and \ref{fig:dNdY_CC1} show the transverse momentum and rapidity distributions of the dileptons in the low ($M_{e^+e^-} < 0.15~\mbox{GeV}$) and the intermediate ($0.15~\mbox{GeV} < M_{e^+e^-} < 0.50~\mbox{GeV}$) invariant mass regions.
Although the direct $\rho$ decay component in the low-mass region
 is strongly enhanced in the calculation with the PDM (Set 2), this is still hidden 
 under  $\pi^0$ Dalitz decay component which completely dominates this region.
   The enhanced $\eta$ Dalitz component in the intermediate invariant mass region 
for the PDM (Set 2) at $p_t^{e^+e^-} \sim 0.2$ GeV/c almost fills the missing strength there.
Remaining discrepancies in the intermediate invariant mass region are better visible
in the rapidity distribution at small and large rapidities. They are most probably due to the
assumed (for simplicity) isotropic decay of $N^*(1535)$ to the $\eta N$ final state.

\section{Summary and outlook}
\label{summary}

To summarize, we have included parity-doublet model (PDM) mean fields for the nucleon and its parity partner, the $N^*(1535)$ resonance, in simulations based on the GiBUU microscopic transport model.
The modified GiBUU model has been applied to study $\eta$, $\pi^0$ and dilepton production in heavy-ion collisions at SIS18 energies.
In-medium threshold effects have been carefully taken into account which allows us to make quantitative predictions for the $N^*(1535)$ resonance production. The main effect can be described as follows:  
The quickly dropping Dirac mass of the $N^*(1535)$ resonance with baryon density in the PDM leads to an order of magnitude enhancement of
$N^*(1535)$ production in the intermediate stages of central heavy-ion collisions at beam energies of about 1A~GeV. 
Since the baryon density of the expanding nuclear system decreases with time, the in-medium mass shift of the produced $N^*(1535)$ resonances
gradually disappears pushing them above the $N \eta$ threshold. This leads to a significant enhancement of slow $\eta$ production in the c.m. frame of the colliding nuclei. This enhancement effect tends to be stronger for lower beam energies,
i.e.~when going deeper into the subthreshold region.

Since the $N^*(1535)$ resonance is also coupled to the $\pi N$ and $\rho N$ decay channels, pion and intermediate $\rho$ production are also 
influenced by the PDM. Pion production is governed by intermediate $\Delta(1232)$ resonance production, however, and thus changes only slightly.
In contrast, the $\rho$ production at low invariant masses is strongly enhanced. This leads to a slight enhancement in the production of low-invariant-mass dilepton pairs in the PDM calculations as compared to using standard (NL2) Walecka model mean fields.
The $\eta \to e^+ e^- \gamma$ Dalitz decay component is also enhanced in the PDM resulting in a moderate increase in the production of dileptons in the intermediate
invariant mass range. 

We believe that our present work opens the window to further studies of chiral effects on particle production within transport models.
In particular, GiBUU model allows to study the effects of the PDM mean fields in $A(\gamma,\eta)$ reactions where a reduction of $\eta$ production is expected
due to the dropping mass difference between $N^*(1535)$ and nucleon in nuclear medium \cite{Kim:1998upa}.
Another open issue for the future is the chiral description of mean fields for other baryonic resonances, in particular, for the $\Delta(1232)$ that is of utmost importance for pion production. In the present study, for simplicity,  we have assumed that the scalar field acting on the $\Delta(1232)$ is the same as that for the nucleons.
In Ref.~\cite{Jido:1999hd} a quartet scheme has been proposed  to describe the lightest baryons of each spin parity.
In this scheme, the $\Delta(1232)$, $\Delta(1700)$, $N(1520)$, and $N(1720)$ form the chiral quartet in the $J=3/2$ sector.
As a next step it should therefore be interesting to study the effects of the chiral quartet scheme in view of the currently puzzling situation with the theoretical description of recent HADES data on pion production \cite{Adamczewski-Musch:2020vrg}.

\begin{acknowledgments}
We thank Ulrich Mosel for continued interest in our work and stimulating discussions. 
A.L. acknowledges the hospitality of the Institute for Theoretical Physics
at JLU Giessen. This work is supported by the German Federal Ministry of Education and Research (BMBF) through grants No.~05P18RGFCA and No.~05P21RGFCA. 
\end{acknowledgments}

\bibliography{pdm}

\begin{thebibliography}{54}%
\makeatletter
\providecommand \@ifxundefined [1]{%
 \@ifx{#1\undefined}
}%
\providecommand \@ifnum [1]{%
 \ifnum #1\expandafter \@firstoftwo
 \else \expandafter \@secondoftwo
 \fi
}%
\providecommand \@ifx [1]{%
 \ifx #1\expandafter \@firstoftwo
 \else \expandafter \@secondoftwo
 \fi
}%
\providecommand \natexlab [1]{#1}%
\providecommand \enquote  [1]{``#1''}%
\providecommand \bibnamefont  [1]{#1}%
\providecommand \bibfnamefont [1]{#1}%
\providecommand \citenamefont [1]{#1}%
\providecommand \href@noop [0]{\@secondoftwo}%
\providecommand \href [0]{\begingroup \@sanitize@url \@href}%
\providecommand \@href[1]{\@@startlink{#1}\@@href}%
\providecommand \@@href[1]{\endgroup#1\@@endlink}%
\providecommand \@sanitize@url [0]{\catcode `\\12\catcode `\$12\catcode
  `\&12\catcode `\#12\catcode `\^12\catcode `\_12\catcode `\%12\relax}%
\providecommand \@@startlink[1]{}%
\providecommand \@@endlink[0]{}%
\providecommand \url  [0]{\begingroup\@sanitize@url \@url }%
\providecommand \@url [1]{\endgroup\@href {#1}{\urlprefix }}%
\providecommand \urlprefix  [0]{URL }%
\providecommand \Eprint [0]{\href }%
\providecommand \doibase [0]{http://dx.doi.org/}%
\providecommand \selectlanguage [0]{\@gobble}%
\providecommand \bibinfo  [0]{\@secondoftwo}%
\providecommand \bibfield  [0]{\@secondoftwo}%
\providecommand \translation [1]{[#1]}%
\providecommand \BibitemOpen [0]{}%
\providecommand \bibitemStop [0]{}%
\providecommand \bibitemNoStop [0]{.\EOS\space}%
\providecommand \EOS [0]{\spacefactor3000\relax}%
\providecommand \BibitemShut  [1]{\csname bibitem#1\endcsname}%
\let\auto@bib@innerbib\@empty
\bibitem [{\citenamefont {Glozman}\ \emph {et~al.}(2012)\citenamefont
  {Glozman}, \citenamefont {Lang},\ and\ \citenamefont
  {Schrock}}]{Glozman:2012fj}%
  \BibitemOpen
  \bibfield  {author} {\bibinfo {author} {\bibfnamefont {L.~Y.}\ \bibnamefont
  {Glozman}}, \bibinfo {author} {\bibfnamefont {C.~B.}\ \bibnamefont {Lang}}, \
  and\ \bibinfo {author} {\bibfnamefont {M.}~\bibnamefont {Schrock}},\ }\href
  {\doibase 10.1103/PhysRevD.86.014507} {\bibfield  {journal} {\bibinfo
  {journal} {Phys. Rev. D}\ }\textbf {\bibinfo {volume} {86}},\ \bibinfo
  {pages} {014507} (\bibinfo {year} {2012})},\ \Eprint
  {http://arxiv.org/abs/1205.4887} {arXiv:1205.4887 [hep-lat]} \BibitemShut
  {NoStop}%
\bibitem [{\citenamefont {Aarts}\ \emph {et~al.}(2017)\citenamefont {Aarts},
  \citenamefont {Allton}, \citenamefont {De~Boni}, \citenamefont {Hands},
  \citenamefont {J\"ager}, \citenamefont {Praki},\ and\ \citenamefont
  {Skullerud}}]{Aarts:2017rrl}%
  \BibitemOpen
  \bibfield  {author} {\bibinfo {author} {\bibfnamefont {G.}~\bibnamefont
  {Aarts}}, \bibinfo {author} {\bibfnamefont {C.}~\bibnamefont {Allton}},
  \bibinfo {author} {\bibfnamefont {D.}~\bibnamefont {De~Boni}}, \bibinfo
  {author} {\bibfnamefont {S.}~\bibnamefont {Hands}}, \bibinfo {author}
  {\bibfnamefont {B.}~\bibnamefont {J\"ager}}, \bibinfo {author} {\bibfnamefont
  {C.}~\bibnamefont {Praki}}, \ and\ \bibinfo {author} {\bibfnamefont {J.-I.}\
  \bibnamefont {Skullerud}},\ }\href {\doibase 10.1007/JHEP06(2017)034}
  {\bibfield  {journal} {\bibinfo  {journal} {JHEP}\ }\textbf {\bibinfo
  {volume} {06}},\ \bibinfo {pages} {034} (\bibinfo {year} {2017})},\ \Eprint
  {http://arxiv.org/abs/1703.09246} {arXiv:1703.09246 [hep-lat]} \BibitemShut
  {NoStop}%
\bibitem [{\citenamefont {Jido}\ \emph
  {et~al.}(2000{\natexlab{a}})\citenamefont {Jido}, \citenamefont {Nemoto},
  \citenamefont {Oka},\ and\ \citenamefont {Hosaka}}]{Jido:1998av}%
  \BibitemOpen
  \bibfield  {author} {\bibinfo {author} {\bibfnamefont {D.}~\bibnamefont
  {Jido}}, \bibinfo {author} {\bibfnamefont {Y.}~\bibnamefont {Nemoto}},
  \bibinfo {author} {\bibfnamefont {M.}~\bibnamefont {Oka}}, \ and\ \bibinfo
  {author} {\bibfnamefont {A.}~\bibnamefont {Hosaka}},\ }\href {\doibase
  10.1016/S0375-9474(99)00844-1} {\bibfield  {journal} {\bibinfo  {journal}
  {Nucl. Phys. A}\ }\textbf {\bibinfo {volume} {671}},\ \bibinfo {pages} {471}
  (\bibinfo {year} {2000}{\natexlab{a}})},\ \Eprint
  {http://arxiv.org/abs/hep-ph/9805306} {arXiv:hep-ph/9805306} \BibitemShut
  {NoStop}%
\bibitem [{\citenamefont {Jido}\ \emph {et~al.}(2001)\citenamefont {Jido},
  \citenamefont {Oka},\ and\ \citenamefont {Hosaka}}]{Jido:2001nt}%
  \BibitemOpen
  \bibfield  {author} {\bibinfo {author} {\bibfnamefont {D.}~\bibnamefont
  {Jido}}, \bibinfo {author} {\bibfnamefont {M.}~\bibnamefont {Oka}}, \ and\
  \bibinfo {author} {\bibfnamefont {A.}~\bibnamefont {Hosaka}},\ }\href
  {\doibase 10.1143/PTP.106.873} {\bibfield  {journal} {\bibinfo  {journal}
  {Prog. Theor. Phys.}\ }\textbf {\bibinfo {volume} {106}},\ \bibinfo {pages}
  {873} (\bibinfo {year} {2001})},\ \Eprint
  {http://arxiv.org/abs/hep-ph/0110005} {arXiv:hep-ph/0110005} \BibitemShut
  {NoStop}%
\bibitem [{\citenamefont {Detar}\ and\ \citenamefont
  {Kunihiro}(1989)}]{Detar:1988kn}%
  \BibitemOpen
  \bibfield  {author} {\bibinfo {author} {\bibfnamefont {C.~E.}\ \bibnamefont
  {Detar}}\ and\ \bibinfo {author} {\bibfnamefont {T.}~\bibnamefont
  {Kunihiro}},\ }\href {\doibase 10.1103/PhysRevD.39.2805} {\bibfield
  {journal} {\bibinfo  {journal} {Phys. Rev. D}\ }\textbf {\bibinfo {volume}
  {39}},\ \bibinfo {pages} {2805} (\bibinfo {year} {1989})}\BibitemShut
  {NoStop}%
\bibitem [{\citenamefont {Hatsuda}\ and\ \citenamefont
  {Prakash}(1989)}]{Hatsuda:1988mv}%
  \BibitemOpen
  \bibfield  {author} {\bibinfo {author} {\bibfnamefont {T.}~\bibnamefont
  {Hatsuda}}\ and\ \bibinfo {author} {\bibfnamefont {M.}~\bibnamefont
  {Prakash}},\ }\href {\doibase 10.1016/0370-2693(89)91040-X} {\bibfield
  {journal} {\bibinfo  {journal} {Phys. Lett. B}\ }\textbf {\bibinfo {volume}
  {224}},\ \bibinfo {pages} {11} (\bibinfo {year} {1989})}\BibitemShut
  {NoStop}%
\bibitem [{\citenamefont {Zschiesche}\ \emph {et~al.}(2007)\citenamefont
  {Zschiesche}, \citenamefont {Tolos}, \citenamefont {Schaffner-Bielich},\ and\
  \citenamefont {Pisarski}}]{Zschiesche:2006zj}%
  \BibitemOpen
  \bibfield  {author} {\bibinfo {author} {\bibfnamefont {D.}~\bibnamefont
  {Zschiesche}}, \bibinfo {author} {\bibfnamefont {L.}~\bibnamefont {Tolos}},
  \bibinfo {author} {\bibfnamefont {J.}~\bibnamefont {Schaffner-Bielich}}, \
  and\ \bibinfo {author} {\bibfnamefont {R.~D.}\ \bibnamefont {Pisarski}},\
  }\href {\doibase 10.1103/PhysRevC.75.055202} {\bibfield  {journal} {\bibinfo
  {journal} {Phys. Rev. C}\ }\textbf {\bibinfo {volume} {75}},\ \bibinfo
  {pages} {055202} (\bibinfo {year} {2007})},\ \Eprint
  {http://arxiv.org/abs/nucl-th/0608044} {arXiv:nucl-th/0608044} \BibitemShut
  {NoStop}%
\bibitem [{\citenamefont {Sasaki}\ and\ \citenamefont
  {Mishustin}(2010)}]{Sasaki:2010bp}%
  \BibitemOpen
  \bibfield  {author} {\bibinfo {author} {\bibfnamefont {C.}~\bibnamefont
  {Sasaki}}\ and\ \bibinfo {author} {\bibfnamefont {I.}~\bibnamefont
  {Mishustin}},\ }\href {\doibase 10.1103/PhysRevC.82.035204} {\bibfield
  {journal} {\bibinfo  {journal} {Phys. Rev. C}\ }\textbf {\bibinfo {volume}
  {82}},\ \bibinfo {pages} {035204} (\bibinfo {year} {2010})},\ \Eprint
  {http://arxiv.org/abs/1005.4811} {arXiv:1005.4811 [hep-ph]} \BibitemShut
  {NoStop}%
\bibitem [{\citenamefont {Weyrich}\ \emph {et~al.}(2015)\citenamefont
  {Weyrich}, \citenamefont {Strodthoff},\ and\ \citenamefont {von
  Smekal}}]{Weyrich:2015hha}%
  \BibitemOpen
  \bibfield  {author} {\bibinfo {author} {\bibfnamefont {J.}~\bibnamefont
  {Weyrich}}, \bibinfo {author} {\bibfnamefont {N.}~\bibnamefont {Strodthoff}},
  \ and\ \bibinfo {author} {\bibfnamefont {L.}~\bibnamefont {von Smekal}},\
  }\href {\doibase 10.1103/PhysRevC.92.015214} {\bibfield  {journal} {\bibinfo
  {journal} {Phys. Rev. C}\ }\textbf {\bibinfo {volume} {92}},\ \bibinfo
  {pages} {015214} (\bibinfo {year} {2015})},\ \Eprint
  {http://arxiv.org/abs/1504.02697} {arXiv:1504.02697 [nucl-th]} \BibitemShut
  {NoStop}%
\bibitem [{\citenamefont {Tripolt}\ \emph {et~al.}(2021)\citenamefont
  {Tripolt}, \citenamefont {Jung}, \citenamefont {von Smekal},\ and\
  \citenamefont {Wambach}}]{Tripolt:2021jtp}%
  \BibitemOpen
  \bibfield  {author} {\bibinfo {author} {\bibfnamefont {R.-A.}\ \bibnamefont
  {Tripolt}}, \bibinfo {author} {\bibfnamefont {C.}~\bibnamefont {Jung}},
  \bibinfo {author} {\bibfnamefont {L.}~\bibnamefont {von Smekal}}, \ and\
  \bibinfo {author} {\bibfnamefont {J.}~\bibnamefont {Wambach}},\ }\href@noop
  {} {\  (\bibinfo {year} {2021})},\ \Eprint {http://arxiv.org/abs/2105.00861}
  {arXiv:2105.00861 [hep-ph]} \BibitemShut {NoStop}%
\bibitem [{\citenamefont {Jung}\ \emph {et~al.}(2017)\citenamefont {Jung},
  \citenamefont {Rennecke}, \citenamefont {Tripolt}, \citenamefont {von
  Smekal},\ and\ \citenamefont {Wambach}}]{Jung:2016yxl}%
  \BibitemOpen
  \bibfield  {author} {\bibinfo {author} {\bibfnamefont {C.}~\bibnamefont
  {Jung}}, \bibinfo {author} {\bibfnamefont {F.}~\bibnamefont {Rennecke}},
  \bibinfo {author} {\bibfnamefont {R.-A.}\ \bibnamefont {Tripolt}}, \bibinfo
  {author} {\bibfnamefont {L.}~\bibnamefont {von Smekal}}, \ and\ \bibinfo
  {author} {\bibfnamefont {J.}~\bibnamefont {Wambach}},\ }\href {\doibase
  10.1103/PhysRevD.95.036020} {\bibfield  {journal} {\bibinfo  {journal} {Phys.
  Rev. D}\ }\textbf {\bibinfo {volume} {95}},\ \bibinfo {pages} {036020}
  (\bibinfo {year} {2017})},\ \Eprint {http://arxiv.org/abs/1610.08754}
  {arXiv:1610.08754 [hep-ph]} \BibitemShut {NoStop}%
\bibitem [{\citenamefont {Jung}\ and\ \citenamefont {von
  Smekal}(2019)}]{Jung:2019nnr}%
  \BibitemOpen
  \bibfield  {author} {\bibinfo {author} {\bibfnamefont {C.}~\bibnamefont
  {Jung}}\ and\ \bibinfo {author} {\bibfnamefont {L.}~\bibnamefont {von
  Smekal}},\ }\href {\doibase 10.1103/PhysRevD.100.116009} {\bibfield
  {journal} {\bibinfo  {journal} {Phys. Rev. D}\ }\textbf {\bibinfo {volume}
  {100}},\ \bibinfo {pages} {116009} (\bibinfo {year} {2019})},\ \Eprint
  {http://arxiv.org/abs/1909.13712} {arXiv:1909.13712 [hep-ph]} \BibitemShut
  {NoStop}%
\bibitem [{\citenamefont {Sasaki}(2020)}]{Sasaki:2019jyh}%
  \BibitemOpen
  \bibfield  {author} {\bibinfo {author} {\bibfnamefont {C.}~\bibnamefont
  {Sasaki}},\ }\href {\doibase 10.1016/j.physletb.2019.135172} {\bibfield
  {journal} {\bibinfo  {journal} {Phys. Lett. B}\ }\textbf {\bibinfo {volume}
  {801}},\ \bibinfo {pages} {135172} (\bibinfo {year} {2020})},\ \Eprint
  {http://arxiv.org/abs/1906.05077} {arXiv:1906.05077 [hep-ph]} \BibitemShut
  {NoStop}%
\bibitem [{\citenamefont {Zyla}\ \emph {et~al.}(2020)\citenamefont {Zyla} \emph
  {et~al.}}]{Zyla:2020zbs}%
  \BibitemOpen
  \bibfield  {author} {\bibinfo {author} {\bibfnamefont {P.}~\bibnamefont
  {Zyla}} \emph {et~al.} (\bibinfo {collaboration} {Particle Data Group}),\
  }\href {\doibase 10.1093/ptep/ptaa104} {\bibfield  {journal} {\bibinfo
  {journal} {PTEP}\ }\textbf {\bibinfo {volume} {2020}},\ \bibinfo {pages}
  {083C01} (\bibinfo {year} {2020})}\BibitemShut {NoStop}%
\bibitem [{\citenamefont {R\"obig-Landau}\ \emph {et~al.}(1996)\citenamefont
  {R\"obig-Landau} \emph {et~al.}}]{RoebigLandau:1996xa}%
  \BibitemOpen
  \bibfield  {author} {\bibinfo {author} {\bibfnamefont {M.}~\bibnamefont
  {R\"obig-Landau}} \emph {et~al.},\ }\href {\doibase
  10.1016/0370-2693(96)00125-6} {\bibfield  {journal} {\bibinfo  {journal}
  {Phys. Lett. B}\ }\textbf {\bibinfo {volume} {373}},\ \bibinfo {pages} {45}
  (\bibinfo {year} {1996})}\BibitemShut {NoStop}%
\bibitem [{\citenamefont {Lehr}\ \emph {et~al.}(2003)\citenamefont {Lehr},
  \citenamefont {Post},\ and\ \citenamefont {Mosel}}]{Lehr:2003km}%
  \BibitemOpen
  \bibfield  {author} {\bibinfo {author} {\bibfnamefont {J.}~\bibnamefont
  {Lehr}}, \bibinfo {author} {\bibfnamefont {M.}~\bibnamefont {Post}}, \ and\
  \bibinfo {author} {\bibfnamefont {U.}~\bibnamefont {Mosel}},\ }\href
  {\doibase 10.1103/PhysRevC.68.044601} {\bibfield  {journal} {\bibinfo
  {journal} {Phys. Rev. C}\ }\textbf {\bibinfo {volume} {68}},\ \bibinfo
  {pages} {044601} (\bibinfo {year} {2003})},\ \Eprint
  {http://arxiv.org/abs/nucl-th/0306024} {arXiv:nucl-th/0306024} \BibitemShut
  {NoStop}%
\bibitem [{\citenamefont {Jido}\ \emph {et~al.}(2008)\citenamefont {Jido},
  \citenamefont {Kolomeitsev}, \citenamefont {Nagahiro},\ and\ \citenamefont
  {Hirenzaki}}]{Jido:2008ng}%
  \BibitemOpen
  \bibfield  {author} {\bibinfo {author} {\bibfnamefont {D.}~\bibnamefont
  {Jido}}, \bibinfo {author} {\bibfnamefont {E.~E.}\ \bibnamefont
  {Kolomeitsev}}, \bibinfo {author} {\bibfnamefont {H.}~\bibnamefont
  {Nagahiro}}, \ and\ \bibinfo {author} {\bibfnamefont {S.}~\bibnamefont
  {Hirenzaki}},\ }\href {\doibase 10.1016/j.nuclphysa.2008.07.012} {\bibfield
  {journal} {\bibinfo  {journal} {Nucl. Phys. A}\ }\textbf {\bibinfo {volume}
  {811}},\ \bibinfo {pages} {158} (\bibinfo {year} {2008})},\ \Eprint
  {http://arxiv.org/abs/0801.4834} {arXiv:0801.4834 [nucl-th]} \BibitemShut
  {NoStop}%
\bibitem [{\citenamefont {Kim}\ \emph {et~al.}(1998)\citenamefont {Kim},
  \citenamefont {Jido},\ and\ \citenamefont {Oka}}]{Kim:1998upa}%
  \BibitemOpen
  \bibfield  {author} {\bibinfo {author} {\bibfnamefont {H.-c.}\ \bibnamefont
  {Kim}}, \bibinfo {author} {\bibfnamefont {D.}~\bibnamefont {Jido}}, \ and\
  \bibinfo {author} {\bibfnamefont {M.}~\bibnamefont {Oka}},\ }\href {\doibase
  10.1016/S0375-9474(98)00451-5} {\bibfield  {journal} {\bibinfo  {journal}
  {Nucl. Phys. A}\ }\textbf {\bibinfo {volume} {640}},\ \bibinfo {pages} {77}
  (\bibinfo {year} {1998})},\ \Eprint {http://arxiv.org/abs/hep-ph/9806275}
  {arXiv:hep-ph/9806275} \BibitemShut {NoStop}%
\bibitem [{\citenamefont {Suenaga}(2018)}]{Suenaga:2017wbb}%
  \BibitemOpen
  \bibfield  {author} {\bibinfo {author} {\bibfnamefont {D.}~\bibnamefont
  {Suenaga}},\ }\href {\doibase 10.1103/PhysRevC.97.045203} {\bibfield
  {journal} {\bibinfo  {journal} {Phys. Rev. C}\ }\textbf {\bibinfo {volume}
  {97}},\ \bibinfo {pages} {045203} (\bibinfo {year} {2018})},\ \Eprint
  {http://arxiv.org/abs/1704.03630} {arXiv:1704.03630 [nucl-th]} \BibitemShut
  {NoStop}%
\bibitem [{\citenamefont {Zhang}\ and\ \citenamefont
  {Ko}(2018)}]{Zhang:2018ool}%
  \BibitemOpen
  \bibfield  {author} {\bibinfo {author} {\bibfnamefont {Z.}~\bibnamefont
  {Zhang}}\ and\ \bibinfo {author} {\bibfnamefont {C.~M.}\ \bibnamefont {Ko}},\
  }\href {\doibase 10.1103/PhysRevC.98.054614} {\bibfield  {journal} {\bibinfo
  {journal} {Phys. Rev. C}\ }\textbf {\bibinfo {volume} {98}},\ \bibinfo
  {pages} {054614} (\bibinfo {year} {2018})}\BibitemShut {NoStop}%
\bibitem [{\citenamefont {Buss}\ \emph {et~al.}(2012)\citenamefont {Buss},
  \citenamefont {Gaitanos}, \citenamefont {Gallmeister}, \citenamefont {van
  Hees}, \citenamefont {Kaskulov}, \citenamefont {Lalakulich}, \citenamefont
  {Larionov}, \citenamefont {Leitner}, \citenamefont {Weil},\ and\
  \citenamefont {Mosel}}]{Buss:2011mx}%
  \BibitemOpen
  \bibfield  {author} {\bibinfo {author} {\bibfnamefont {O.}~\bibnamefont
  {Buss}}, \bibinfo {author} {\bibfnamefont {T.}~\bibnamefont {Gaitanos}},
  \bibinfo {author} {\bibfnamefont {K.}~\bibnamefont {Gallmeister}}, \bibinfo
  {author} {\bibfnamefont {H.}~\bibnamefont {van Hees}}, \bibinfo {author}
  {\bibfnamefont {M.}~\bibnamefont {Kaskulov}}, \bibinfo {author}
  {\bibfnamefont {O.}~\bibnamefont {Lalakulich}}, \bibinfo {author}
  {\bibfnamefont {A.~B.}\ \bibnamefont {Larionov}}, \bibinfo {author}
  {\bibfnamefont {T.}~\bibnamefont {Leitner}}, \bibinfo {author} {\bibfnamefont
  {J.}~\bibnamefont {Weil}}, \ and\ \bibinfo {author} {\bibfnamefont
  {U.}~\bibnamefont {Mosel}},\ }\href {\doibase 10.1016/j.physrep.2011.12.001}
  {\bibfield  {journal} {\bibinfo  {journal} {Phys. Rept.}\ }\textbf {\bibinfo
  {volume} {512}},\ \bibinfo {pages} {1} (\bibinfo {year} {2012})},\ \Eprint
  {http://arxiv.org/abs/1106.1344} {arXiv:1106.1344 [hep-ph]} \BibitemShut
  {NoStop}%
\bibitem [{\citenamefont {Shin}\ \emph {et~al.}(2018)\citenamefont {Shin},
  \citenamefont {Paeng}, \citenamefont {Harada},\ and\ \citenamefont
  {Kim}}]{Shin:2018axs}%
  \BibitemOpen
  \bibfield  {author} {\bibinfo {author} {\bibfnamefont {I.~J.}\ \bibnamefont
  {Shin}}, \bibinfo {author} {\bibfnamefont {W.-G.}\ \bibnamefont {Paeng}},
  \bibinfo {author} {\bibfnamefont {M.}~\bibnamefont {Harada}}, \ and\ \bibinfo
  {author} {\bibfnamefont {Y.}~\bibnamefont {Kim}},\ }\href@noop {} {\
  (\bibinfo {year} {2018})},\ \Eprint {http://arxiv.org/abs/1805.03402}
  {arXiv:1805.03402 [nucl-th]} \BibitemShut {NoStop}%
\bibitem [{\citenamefont {Kim}\ \emph {et~al.}(2020)\citenamefont {Kim},
  \citenamefont {Jeon}, \citenamefont {Kim}, \citenamefont {Kim},\ and\
  \citenamefont {Lee}}]{Kim:2020sjy}%
  \BibitemOpen
  \bibfield  {author} {\bibinfo {author} {\bibfnamefont {M.}~\bibnamefont
  {Kim}}, \bibinfo {author} {\bibfnamefont {S.}~\bibnamefont {Jeon}}, \bibinfo
  {author} {\bibfnamefont {Y.-M.}\ \bibnamefont {Kim}}, \bibinfo {author}
  {\bibfnamefont {Y.}~\bibnamefont {Kim}}, \ and\ \bibinfo {author}
  {\bibfnamefont {C.-H.}\ \bibnamefont {Lee}},\ }\href {\doibase
  10.1103/PhysRevC.101.064614} {\bibfield  {journal} {\bibinfo  {journal}
  {Phys. Rev. C}\ }\textbf {\bibinfo {volume} {101}},\ \bibinfo {pages}
  {064614} (\bibinfo {year} {2020})},\ \Eprint
  {http://arxiv.org/abs/2006.02023} {arXiv:2006.02023 [nucl-th]} \BibitemShut
  {NoStop}%
\bibitem [{\citenamefont {Larionov}\ \emph {et~al.}(2008)\citenamefont
  {Larionov}, \citenamefont {Mishustin}, \citenamefont {Satarov},\ and\
  \citenamefont {Greiner}}]{Larionov:2008wy}%
  \BibitemOpen
  \bibfield  {author} {\bibinfo {author} {\bibfnamefont {A.~B.}\ \bibnamefont
  {Larionov}}, \bibinfo {author} {\bibfnamefont {I.~N.}\ \bibnamefont
  {Mishustin}}, \bibinfo {author} {\bibfnamefont {L.~M.}\ \bibnamefont
  {Satarov}}, \ and\ \bibinfo {author} {\bibfnamefont {W.}~\bibnamefont
  {Greiner}},\ }\href {\doibase 10.1103/PhysRevC.78.014604} {\bibfield
  {journal} {\bibinfo  {journal} {Phys. Rev. C}\ }\textbf {\bibinfo {volume}
  {78}},\ \bibinfo {pages} {014604} (\bibinfo {year} {2008})},\ \Eprint
  {http://arxiv.org/abs/0802.1845} {arXiv:0802.1845 [nucl-th]} \BibitemShut
  {NoStop}%
\bibitem [{\citenamefont {Blaettel}\ \emph {et~al.}(1993)\citenamefont
  {Blaettel}, \citenamefont {Koch},\ and\ \citenamefont
  {Mosel}}]{Blaettel:1993uz}%
  \BibitemOpen
  \bibfield  {author} {\bibinfo {author} {\bibfnamefont {B.}~\bibnamefont
  {Blaettel}}, \bibinfo {author} {\bibfnamefont {V.}~\bibnamefont {Koch}}, \
  and\ \bibinfo {author} {\bibfnamefont {U.}~\bibnamefont {Mosel}},\ }\href
  {\doibase 10.1088/0034-4885/56/1/001} {\bibfield  {journal} {\bibinfo
  {journal} {Rept. Prog. Phys.}\ }\textbf {\bibinfo {volume} {56}},\ \bibinfo
  {pages} {1} (\bibinfo {year} {1993})}\BibitemShut {NoStop}%
\bibitem [{\citenamefont {Larionov}\ \emph {et~al.}(2021)\citenamefont
  {Larionov}, \citenamefont {Mosel},\ and\ \citenamefont {von
  Smekal}}]{Larionov:2020fnu}%
  \BibitemOpen
  \bibfield  {author} {\bibinfo {author} {\bibfnamefont {A.~B.}\ \bibnamefont
  {Larionov}}, \bibinfo {author} {\bibfnamefont {U.}~\bibnamefont {Mosel}}, \
  and\ \bibinfo {author} {\bibfnamefont {L.}~\bibnamefont {von Smekal}},\
  }\href {\doibase 10.1103/PhysRevC.102.064913} {\bibfield  {journal} {\bibinfo
   {journal} {Phys. Rev. C}\ }\textbf {\bibinfo {volume} {102}},\ \bibinfo
  {pages} {064913} (\bibinfo {year} {2021})},\ \Eprint
  {http://arxiv.org/abs/2009.11702} {arXiv:2009.11702 [nucl-th]} \BibitemShut
  {NoStop}%
\bibitem [{\citenamefont {Ferini}\ \emph {et~al.}(2005)\citenamefont {Ferini},
  \citenamefont {Colonna}, \citenamefont {Gaitanos},\ and\ \citenamefont
  {Di~Toro}}]{Ferrini:2005jw}%
  \BibitemOpen
  \bibfield  {author} {\bibinfo {author} {\bibfnamefont {G.}~\bibnamefont
  {Ferini}}, \bibinfo {author} {\bibfnamefont {M.}~\bibnamefont {Colonna}},
  \bibinfo {author} {\bibfnamefont {T.}~\bibnamefont {Gaitanos}}, \ and\
  \bibinfo {author} {\bibfnamefont {M.}~\bibnamefont {Di~Toro}},\ }\href
  {\doibase 10.1016/j.nuclphysa.2005.08.007} {\bibfield  {journal} {\bibinfo
  {journal} {Nucl. Phys. A}\ }\textbf {\bibinfo {volume} {762}},\ \bibinfo
  {pages} {147} (\bibinfo {year} {2005})},\ \Eprint
  {http://arxiv.org/abs/nucl-th/0504032} {arXiv:nucl-th/0504032} \BibitemShut
  {NoStop}%
\bibitem [{\citenamefont {Arnaldi}\ \emph {et~al.}(2006)\citenamefont {Arnaldi}
  \emph {et~al.}}]{NA60:2006ymb}%
  \BibitemOpen
  \bibfield  {author} {\bibinfo {author} {\bibfnamefont {R.}~\bibnamefont
  {Arnaldi}} \emph {et~al.} (\bibinfo {collaboration} {NA60}),\ }\href
  {\doibase 10.1103/PhysRevLett.96.162302} {\bibfield  {journal} {\bibinfo
  {journal} {Phys. Rev. Lett.}\ }\textbf {\bibinfo {volume} {96}},\ \bibinfo
  {pages} {162302} (\bibinfo {year} {2006})},\ \Eprint
  {http://arxiv.org/abs/nucl-ex/0605007} {arXiv:nucl-ex/0605007} \BibitemShut
  {NoStop}%
\bibitem [{\citenamefont {van Hees}\ and\ \citenamefont
  {Rapp}(2006)}]{vanHees:2006ng}%
  \BibitemOpen
  \bibfield  {author} {\bibinfo {author} {\bibfnamefont {H.}~\bibnamefont {van
  Hees}}\ and\ \bibinfo {author} {\bibfnamefont {R.}~\bibnamefont {Rapp}},\
  }\href {\doibase 10.1103/PhysRevLett.97.102301} {\bibfield  {journal}
  {\bibinfo  {journal} {Phys. Rev. Lett.}\ }\textbf {\bibinfo {volume} {97}},\
  \bibinfo {pages} {102301} (\bibinfo {year} {2006})},\ \Eprint
  {http://arxiv.org/abs/hep-ph/0603084} {arXiv:hep-ph/0603084} \BibitemShut
  {NoStop}%
\bibitem [{\citenamefont {Bratkovskaya}\ and\ \citenamefont
  {Cassing}(2008)}]{Bratkovskaya:2007jk}%
  \BibitemOpen
  \bibfield  {author} {\bibinfo {author} {\bibfnamefont {E.}~\bibnamefont
  {Bratkovskaya}}\ and\ \bibinfo {author} {\bibfnamefont {W.}~\bibnamefont
  {Cassing}},\ }\href {\doibase 10.1016/j.nuclphysa.2008.04.004} {\bibfield
  {journal} {\bibinfo  {journal} {Nucl. Phys. A}\ }\textbf {\bibinfo {volume}
  {807}},\ \bibinfo {pages} {214} (\bibinfo {year} {2008})},\ \Eprint
  {http://arxiv.org/abs/0712.0635} {arXiv:0712.0635 [nucl-th]} \BibitemShut
  {NoStop}%
\bibitem [{\citenamefont {Endres}\ \emph {et~al.}(2015)\citenamefont {Endres},
  \citenamefont {van Hees}, \citenamefont {Weil},\ and\ \citenamefont
  {Bleicher}}]{Endres:2015fna}%
  \BibitemOpen
  \bibfield  {author} {\bibinfo {author} {\bibfnamefont {S.}~\bibnamefont
  {Endres}}, \bibinfo {author} {\bibfnamefont {H.}~\bibnamefont {van Hees}},
  \bibinfo {author} {\bibfnamefont {J.}~\bibnamefont {Weil}}, \ and\ \bibinfo
  {author} {\bibfnamefont {M.}~\bibnamefont {Bleicher}},\ }\href {\doibase
  10.1103/PhysRevC.92.014911} {\bibfield  {journal} {\bibinfo  {journal} {Phys.
  Rev. C}\ }\textbf {\bibinfo {volume} {92}},\ \bibinfo {pages} {014911}
  (\bibinfo {year} {2015})},\ \Eprint {http://arxiv.org/abs/1505.06131}
  {arXiv:1505.06131 [nucl-th]} \BibitemShut {NoStop}%
\bibitem [{\citenamefont {Staudenmaier}\ \emph {et~al.}(2018)\citenamefont
  {Staudenmaier}, \citenamefont {Weil}, \citenamefont {Steinberg},
  \citenamefont {Endres},\ and\ \citenamefont
  {Petersen}}]{Staudenmaier:2017vtq}%
  \BibitemOpen
  \bibfield  {author} {\bibinfo {author} {\bibfnamefont {J.}~\bibnamefont
  {Staudenmaier}}, \bibinfo {author} {\bibfnamefont {J.}~\bibnamefont {Weil}},
  \bibinfo {author} {\bibfnamefont {V.}~\bibnamefont {Steinberg}}, \bibinfo
  {author} {\bibfnamefont {S.}~\bibnamefont {Endres}}, \ and\ \bibinfo {author}
  {\bibfnamefont {H.}~\bibnamefont {Petersen}},\ }\href {\doibase
  10.1103/PhysRevC.98.054908} {\bibfield  {journal} {\bibinfo  {journal} {Phys.
  Rev. C}\ }\textbf {\bibinfo {volume} {98}},\ \bibinfo {pages} {054908}
  (\bibinfo {year} {2018})},\ \Eprint {http://arxiv.org/abs/1711.10297}
  {arXiv:1711.10297 [nucl-th]} \BibitemShut {NoStop}%
\bibitem [{\citenamefont {Schmidt}\ \emph {et~al.}(2021)\citenamefont
  {Schmidt}, \citenamefont {Bratkovskaya}, \citenamefont {Gumberidze},\ and\
  \citenamefont {Holzmann}}]{Schmidt:2021hhs}%
  \BibitemOpen
  \bibfield  {author} {\bibinfo {author} {\bibfnamefont {I.}~\bibnamefont
  {Schmidt}}, \bibinfo {author} {\bibfnamefont {E.}~\bibnamefont
  {Bratkovskaya}}, \bibinfo {author} {\bibfnamefont {M.}~\bibnamefont
  {Gumberidze}}, \ and\ \bibinfo {author} {\bibfnamefont {R.}~\bibnamefont
  {Holzmann}},\ }\href {\doibase 10.1103/PhysRevD.104.015008} {\bibfield
  {journal} {\bibinfo  {journal} {Phys. Rev. D}\ }\textbf {\bibinfo {volume}
  {104}},\ \bibinfo {pages} {015008} (\bibinfo {year} {2021})},\ \Eprint
  {http://arxiv.org/abs/2105.00569} {arXiv:2105.00569 [hep-ph]} \BibitemShut
  {NoStop}%
\bibitem [{\citenamefont {Shyam}\ and\ \citenamefont
  {Mosel}(2010)}]{Shyam:2010vr}%
  \BibitemOpen
  \bibfield  {author} {\bibinfo {author} {\bibfnamefont {R.}~\bibnamefont
  {Shyam}}\ and\ \bibinfo {author} {\bibfnamefont {U.}~\bibnamefont {Mosel}},\
  }\href {\doibase 10.1103/PhysRevC.82.062201} {\bibfield  {journal} {\bibinfo
  {journal} {Phys. Rev. C}\ }\textbf {\bibinfo {volume} {82}},\ \bibinfo
  {pages} {062201} (\bibinfo {year} {2010})},\ \Eprint
  {http://arxiv.org/abs/1006.3873} {arXiv:1006.3873 [hep-ph]} \BibitemShut
  {NoStop}%
\bibitem [{\citenamefont {Agakishiev}\ \emph {et~al.}(2010)\citenamefont
  {Agakishiev} \emph {et~al.}}]{Agakishiev:2009yf}%
  \BibitemOpen
  \bibfield  {author} {\bibinfo {author} {\bibfnamefont {G.}~\bibnamefont
  {Agakishiev}} \emph {et~al.} (\bibinfo {collaboration} {HADES}),\ }\href
  {\doibase 10.1016/j.physletb.2010.05.010} {\bibfield  {journal} {\bibinfo
  {journal} {Phys. Lett.}\ }\textbf {\bibinfo {volume} {B690}},\ \bibinfo
  {pages} {118} (\bibinfo {year} {2010})},\ \Eprint
  {http://arxiv.org/abs/0910.5875} {arXiv:0910.5875 [nucl-ex]} \BibitemShut
  {NoStop}%
\bibitem [{\citenamefont {Blaizot}(1980)}]{Blaizot:1980tw}%
  \BibitemOpen
  \bibfield  {author} {\bibinfo {author} {\bibfnamefont {J.~P.}\ \bibnamefont
  {Blaizot}},\ }\href {\doibase 10.1016/0370-1573(80)90001-0} {\bibfield
  {journal} {\bibinfo  {journal} {Phys. Rept.}\ }\textbf {\bibinfo {volume}
  {64}},\ \bibinfo {pages} {171} (\bibinfo {year} {1980})}\BibitemShut
  {NoStop}%
\bibitem [{\citenamefont {Garg}\ and\ \citenamefont
  {Col\`o}(2018)}]{Garg:2018uam}%
  \BibitemOpen
  \bibfield  {author} {\bibinfo {author} {\bibfnamefont {U.}~\bibnamefont
  {Garg}}\ and\ \bibinfo {author} {\bibfnamefont {G.}~\bibnamefont {Col\`o}},\
  }\href {\doibase 10.1016/j.ppnp.2018.03.001} {\bibfield  {journal} {\bibinfo
  {journal} {Prog. Part. Nucl. Phys.}\ }\textbf {\bibinfo {volume} {101}},\
  \bibinfo {pages} {55} (\bibinfo {year} {2018})},\ \Eprint
  {http://arxiv.org/abs/1801.03672} {arXiv:1801.03672 [nucl-ex]} \BibitemShut
  {NoStop}%
\bibitem [{\citenamefont {Harakeh}\ and\ \citenamefont {van~der
  Woude}(2001)}]{HW01}%
  \BibitemOpen
  \bibfield  {author} {\bibinfo {author} {\bibfnamefont {M.}~\bibnamefont
  {Harakeh}}\ and\ \bibinfo {author} {\bibfnamefont {A.}~\bibnamefont {van~der
  Woude}},\ }\href@noop {} {\emph {\bibinfo {title} {Giant Resonances:
  Fundamental High-Frequency Modes of Nuclear Excitation}}}\ (\bibinfo
  {publisher} {Oxford University Press},\ \bibinfo {year} {2001})\BibitemShut
  {NoStop}%
\bibitem [{\citenamefont {Bonasera}\ \emph {et~al.}(2018)\citenamefont
  {Bonasera}, \citenamefont {Anders},\ and\ \citenamefont
  {Shlomo}}]{Bonasera:2018onp}%
  \BibitemOpen
  \bibfield  {author} {\bibinfo {author} {\bibfnamefont {G.}~\bibnamefont
  {Bonasera}}, \bibinfo {author} {\bibfnamefont {M.~R.}\ \bibnamefont
  {Anders}}, \ and\ \bibinfo {author} {\bibfnamefont {S.}~\bibnamefont
  {Shlomo}},\ }\href {\doibase 10.1103/PhysRevC.98.054316} {\bibfield
  {journal} {\bibinfo  {journal} {Phys. Rev. C}\ }\textbf {\bibinfo {volume}
  {98}},\ \bibinfo {pages} {054316} (\bibinfo {year} {2018})}\BibitemShut
  {NoStop}%
\bibitem [{\citenamefont {Bonasera}\ \emph {et~al.}(2021)\citenamefont
  {Bonasera}, \citenamefont {1+}, \citenamefont {Youngblood}, \citenamefont
  {Lui}, \citenamefont {Button},\ and\ \citenamefont
  {Chen}}]{Bonasera:2020twz}%
  \BibitemOpen
  \bibfield  {author} {\bibinfo {author} {\bibfnamefont {G.}~\bibnamefont
  {Bonasera}}, \bibinfo {author} {\bibfnamefont {S.~S.}\ \bibnamefont {1+}},
  \bibinfo {author} {\bibfnamefont {D.~H.}\ \bibnamefont {Youngblood}},
  \bibinfo {author} {\bibfnamefont {Y.~W.}\ \bibnamefont {Lui}}, \bibinfo
  {author} {\bibfnamefont {J.}~\bibnamefont {Button}}, \ and\ \bibinfo {author}
  {\bibfnamefont {X.}~\bibnamefont {Chen}},\ }\href {\doibase
  10.1016/j.nuclphysa.2021.122159} {\bibfield  {journal} {\bibinfo  {journal}
  {Nucl. Phys. A}\ }\textbf {\bibinfo {volume} {1010}},\ \bibinfo {pages}
  {122159} (\bibinfo {year} {2021})},\ \Eprint
  {http://arxiv.org/abs/2009.00451} {arXiv:2009.00451 [nucl-th]} \BibitemShut
  {NoStop}%
\bibitem [{\citenamefont {Danielewicz}\ \emph {et~al.}(2002)\citenamefont
  {Danielewicz}, \citenamefont {Lacey},\ and\ \citenamefont
  {Lynch}}]{Danielewicz:2002pu}%
  \BibitemOpen
  \bibfield  {author} {\bibinfo {author} {\bibfnamefont {P.}~\bibnamefont
  {Danielewicz}}, \bibinfo {author} {\bibfnamefont {R.}~\bibnamefont {Lacey}},
  \ and\ \bibinfo {author} {\bibfnamefont {W.~G.}\ \bibnamefont {Lynch}},\
  }\href {\doibase 10.1126/science.1078070} {\bibfield  {journal} {\bibinfo
  {journal} {Science}\ }\textbf {\bibinfo {volume} {298}},\ \bibinfo {pages}
  {1592} (\bibinfo {year} {2002})},\ \Eprint
  {http://arxiv.org/abs/nucl-th/0208016} {arXiv:nucl-th/0208016} \BibitemShut
  {NoStop}%
\bibitem [{\citenamefont {Ghosh}\ \emph {et~al.}(2021)\citenamefont {Ghosh},
  \citenamefont {Chatterjee},\ and\ \citenamefont
  {Schaffner-Bielich}}]{Ghosh:2021bvw}%
  \BibitemOpen
  \bibfield  {author} {\bibinfo {author} {\bibfnamefont {S.}~\bibnamefont
  {Ghosh}}, \bibinfo {author} {\bibfnamefont {D.}~\bibnamefont {Chatterjee}}, \
  and\ \bibinfo {author} {\bibfnamefont {J.}~\bibnamefont
  {Schaffner-Bielich}},\ }\href@noop {} {\  (\bibinfo {year} {2021})},\ \Eprint
  {http://arxiv.org/abs/2107.09371} {arXiv:2107.09371 [astro-ph.HE]}
  \BibitemShut {NoStop}%
\bibitem [{\citenamefont {Gaitanos}\ \emph {et~al.}(2010)\citenamefont
  {Gaitanos}, \citenamefont {Larionov}, \citenamefont {Lenske},\ and\
  \citenamefont {Mosel}}]{Gaitanos:2010fd}%
  \BibitemOpen
  \bibfield  {author} {\bibinfo {author} {\bibfnamefont {T.}~\bibnamefont
  {Gaitanos}}, \bibinfo {author} {\bibfnamefont {A.~B.}\ \bibnamefont
  {Larionov}}, \bibinfo {author} {\bibfnamefont {H.}~\bibnamefont {Lenske}}, \
  and\ \bibinfo {author} {\bibfnamefont {U.}~\bibnamefont {Mosel}},\ }\href
  {\doibase 10.1103/PhysRevC.81.054316} {\bibfield  {journal} {\bibinfo
  {journal} {Phys. Rev. C}\ }\textbf {\bibinfo {volume} {81}},\ \bibinfo
  {pages} {054316} (\bibinfo {year} {2010})},\ \Eprint
  {http://arxiv.org/abs/1003.4863} {arXiv:1003.4863 [nucl-th]} \BibitemShut
  {NoStop}%
\bibitem [{\citenamefont {Zhang}\ and\ \citenamefont
  {Li}(2021)}]{Zhang:2021xdt}%
  \BibitemOpen
  \bibfield  {author} {\bibinfo {author} {\bibfnamefont {N.-B.}\ \bibnamefont
  {Zhang}}\ and\ \bibinfo {author} {\bibfnamefont {B.-A.}\ \bibnamefont {Li}},\
  }\href {\doibase 10.3847/1538-4357/ac1e8c} {\bibfield  {journal} {\bibinfo
  {journal} {Astrophys. J.}\ }\textbf {\bibinfo {volume} {921}},\ \bibinfo
  {pages} {111} (\bibinfo {year} {2021})},\ \Eprint
  {http://arxiv.org/abs/2105.11031} {arXiv:2105.11031 [nucl-th]} \BibitemShut
  {NoStop}%
\bibitem [{\citenamefont {Lang}\ \emph {et~al.}(1992)\citenamefont {Lang},
  \citenamefont {Cassing}, \citenamefont {Mosel},\ and\ \citenamefont
  {Weber}}]{Lang:1992jz}%
  \BibitemOpen
  \bibfield  {author} {\bibinfo {author} {\bibfnamefont {A.}~\bibnamefont
  {Lang}}, \bibinfo {author} {\bibfnamefont {W.}~\bibnamefont {Cassing}},
  \bibinfo {author} {\bibfnamefont {U.}~\bibnamefont {Mosel}}, \ and\ \bibinfo
  {author} {\bibfnamefont {K.}~\bibnamefont {Weber}},\ }\href {\doibase
  10.1016/0375-9474(92)90189-Q} {\bibfield  {journal} {\bibinfo  {journal}
  {Nucl. Phys. A}\ }\textbf {\bibinfo {volume} {541}},\ \bibinfo {pages} {507}
  (\bibinfo {year} {1992})}\BibitemShut {NoStop}%
\bibitem [{\citenamefont {Manley}\ and\ \citenamefont
  {Saleski}(1992)}]{Manley:1992yb}%
  \BibitemOpen
  \bibfield  {author} {\bibinfo {author} {\bibfnamefont {D.~M.}\ \bibnamefont
  {Manley}}\ and\ \bibinfo {author} {\bibfnamefont {E.~M.}\ \bibnamefont
  {Saleski}},\ }\href {\doibase 10.1103/PhysRevD.45.4002} {\bibfield  {journal}
  {\bibinfo  {journal} {Phys. Rev.}\ }\textbf {\bibinfo {volume} {D45}},\
  \bibinfo {pages} {4002} (\bibinfo {year} {1992})}\BibitemShut {NoStop}%
\bibitem [{\citenamefont {Averbeck}\ \emph {et~al.}(1997)\citenamefont
  {Averbeck} \emph {et~al.}}]{Averbeck:1997ma}%
  \BibitemOpen
  \bibfield  {author} {\bibinfo {author} {\bibfnamefont {R.}~\bibnamefont
  {Averbeck}} \emph {et~al.} (\bibinfo {collaboration} {TAPS}),\ }\href
  {\doibase 10.1007/s002180050368} {\bibfield  {journal} {\bibinfo  {journal}
  {Z. Phys. A}\ }\textbf {\bibinfo {volume} {359}},\ \bibinfo {pages} {65}
  (\bibinfo {year} {1997})}\BibitemShut {NoStop}%
\bibitem [{\citenamefont {Mar\'\i{}n}\ \emph {et~al.}(1997)\citenamefont
  {Mar\'\i{}n} \emph {et~al.}}]{Marin:1997kj}%
  \BibitemOpen
  \bibfield  {author} {\bibinfo {author} {\bibfnamefont {A.}~\bibnamefont
  {Mar\'\i{}n}} \emph {et~al.},\ }\href {\doibase
  10.1016/S0370-2693(97)00937-4} {\bibfield  {journal} {\bibinfo  {journal}
  {Phys. Lett. B}\ }\textbf {\bibinfo {volume} {409}},\ \bibinfo {pages} {77}
  (\bibinfo {year} {1997})}\BibitemShut {NoStop}%
\bibitem [{\citenamefont {Wolf}\ \emph {et~al.}(1998)\citenamefont {Wolf} \emph
  {et~al.}}]{Wolf:1998vn}%
  \BibitemOpen
  \bibfield  {author} {\bibinfo {author} {\bibfnamefont {A.~R.}\ \bibnamefont
  {Wolf}} \emph {et~al.},\ }\href {\doibase 10.1103/PhysRevLett.80.5281}
  {\bibfield  {journal} {\bibinfo  {journal} {Phys. Rev. Lett.}\ }\textbf
  {\bibinfo {volume} {80}},\ \bibinfo {pages} {5281} (\bibinfo {year}
  {1998})}\BibitemShut {NoStop}%
\bibitem [{\citenamefont {Larionov}\ and\ \citenamefont
  {Mosel}(2003)}]{Larionov:2003av}%
  \BibitemOpen
  \bibfield  {author} {\bibinfo {author} {\bibfnamefont {A.~B.}\ \bibnamefont
  {Larionov}}\ and\ \bibinfo {author} {\bibfnamefont {U.}~\bibnamefont
  {Mosel}},\ }\href {\doibase 10.1016/j.nuclphysa.2003.08.005} {\bibfield
  {journal} {\bibinfo  {journal} {Nucl. Phys. A}\ }\textbf {\bibinfo {volume}
  {728}},\ \bibinfo {pages} {135} (\bibinfo {year} {2003})},\ \Eprint
  {http://arxiv.org/abs/nucl-th/0307035} {arXiv:nucl-th/0307035} \BibitemShut
  {NoStop}%
\bibitem [{\citenamefont {Agakishiev}\ \emph {et~al.}(2008)\citenamefont
  {Agakishiev} \emph {et~al.}}]{Agakishiev:2007ts}%
  \BibitemOpen
  \bibfield  {author} {\bibinfo {author} {\bibfnamefont {G.}~\bibnamefont
  {Agakishiev}} \emph {et~al.} (\bibinfo {collaboration} {HADES}),\ }\href
  {\doibase 10.1016/j.physletb.2008.03.062} {\bibfield  {journal} {\bibinfo
  {journal} {Phys. Lett.}\ }\textbf {\bibinfo {volume} {B663}},\ \bibinfo
  {pages} {43} (\bibinfo {year} {2008})},\ \Eprint
  {http://arxiv.org/abs/0711.4281} {arXiv:0711.4281 [nucl-ex]} \BibitemShut
  {NoStop}%
\bibitem [{\citenamefont {Pachmayer}(2008)}]{Pachmayer:2008}%
  \BibitemOpen
  \bibfield  {author} {\bibinfo {author} {\bibfnamefont {Y.}~\bibnamefont
  {Pachmayer}},\ }\emph {\bibinfo {title} {{Dielektronenproduktion in $^{12}$C
  + $^{12}$C Kollisionen bei 1 GeV pro Nukleon}}},\ \href@noop {} {Ph.D.
  thesis},\ \bibinfo  {school} {Frankfurt U.} (\bibinfo {year} {2008}),\
  \bibinfo {note}
  {\url{http://publikationen.ub.uni-frankfurt.de/frontdoor/index/index/docId/5895}}\BibitemShut
  {NoStop}%
\bibitem [{\citenamefont {Jido}\ \emph
  {et~al.}(2000{\natexlab{b}})\citenamefont {Jido}, \citenamefont {Hatsuda},\
  and\ \citenamefont {Kunihiro}}]{Jido:1999hd}%
  \BibitemOpen
  \bibfield  {author} {\bibinfo {author} {\bibfnamefont {D.}~\bibnamefont
  {Jido}}, \bibinfo {author} {\bibfnamefont {T.}~\bibnamefont {Hatsuda}}, \
  and\ \bibinfo {author} {\bibfnamefont {T.}~\bibnamefont {Kunihiro}},\ }\href
  {\doibase 10.1103/PhysRevLett.84.3252} {\bibfield  {journal} {\bibinfo
  {journal} {Phys. Rev. Lett.}\ }\textbf {\bibinfo {volume} {84}},\ \bibinfo
  {pages} {3252} (\bibinfo {year} {2000}{\natexlab{b}})},\ \Eprint
  {http://arxiv.org/abs/hep-ph/9910375} {arXiv:hep-ph/9910375} \BibitemShut
  {NoStop}%
\bibitem [{\citenamefont {Adamczewski-Musch}\ \emph {et~al.}(2020)\citenamefont
  {Adamczewski-Musch} \emph {et~al.}}]{Adamczewski-Musch:2020vrg}%
  \BibitemOpen
  \bibfield  {author} {\bibinfo {author} {\bibfnamefont {J.}~\bibnamefont
  {Adamczewski-Musch}} \emph {et~al.} (\bibinfo {collaboration} {HADES}),\
  }\href {\doibase 10.1140/epja/s10050-020-00237-2} {\bibfield  {journal}
  {\bibinfo  {journal} {Eur. Phys. J. A}\ }\textbf {\bibinfo {volume} {56}},\
  \bibinfo {pages} {259} (\bibinfo {year} {2020})},\ \Eprint
  {http://arxiv.org/abs/2005.08774} {arXiv:2005.08774 [nucl-ex]} \BibitemShut
  {NoStop}%
\end{thebibliography}%

\end{document}